\documentclass[aps,prd,showpacs,amsmath,10pt,twocolumn,superscriptaddress,floatfix,nofootinbib]{revtex4-2} 

\usepackage[utf8]{inputenc}
\usepackage[english]{babel}
\usepackage{bm,color,xcolor,amsfonts,amsmath,amssymb,epsfig,esint,orcidlink}

\usepackage{lipsum}
\DeclareMathOperator\arctanh{arctanh}
\allowdisplaybreaks
\usepackage{slashed}
\usepackage{subcaption}
\usepackage{tikz}
\usepackage{float}
\usepackage{mathtools}
\usepackage{statmath}
\usepackage{adjustbox}
\newcommand\scalemath[2]{\scalebox{#1}{\mbox{\ensuremath{\displaystyle #2}}}}

\RequirePackage[compat=1.1.0]{tikz-feynman}
\RequirePackage[compat=1.1.0]{tikz-feynhand}

\usepackage{hyperref}
\usepackage{comment}

\newcommand{\be}{\begin{equation}}
\newcommand{\ee}{\end{equation}}
\newcommand{\bea}{\begin{eqnarray}}
\newcommand{\eea}{\end{eqnarray}}
\newcommand{\beas}{\begin{eqnarray*}}
\newcommand{\eeas}{\end{eqnarray*}}

\newcommand{\zb}{Z_b^{\pm}(10610)}
\newcommand{\zbp}{Z_b^{\pm}(10650)}
\newcommand{\zc}{Z_{c}^{\pm}(3900)}
\newcommand{\zcp}{Z_{c}^{\pm}(4020)}
\def\vec#1{\boldsymbol{#1}}

\newcommand{\fzj}{\affiliation{Institute for Advanced Simulation, Institut f\"ur Kernphysik and J\"ulich Center for Hadron Physics, Forschungszentrum J\"ulich, D-52425 J\"ulich, Germany}}

\newcommand{\rub}{\affiliation{Institut f\"ur Theoretische Physik II, Ruhr-Universit\"at Bochum, D-44780 Bochum, Germany}}

\newcommand{\berlin}{\affiliation{Machine Learning Group,
Technische Universität Berlin,
10587 Charlottenburg, Germany}}

\graphicspath{{figs.dir/}}

\begin{document}

\title{Two-pion exchange for coupled-channel scattering of two heavy mesons
}

\author{J.T.~Chacko\orcidlink{0009-0007-1006-5923}}
\fzj

\author{V.~Baru\orcidlink{0000-0001-6472-1008}}
\rub

\author{C.~Hanhart\orcidlink{0000-0002-7603-451X}}
\fzj

\author{S.L.~Krug\orcidlink{0000-0002-9821-8280}}\footnote{Present address: Department of Chemistry and Applied
Biosciences, ETH Zurich, 8093 Zurich, Switzerland.}
\fzj
\berlin

\begin{abstract}

  To improve the theoretical understanding of multiquark states like $Z_b(10610)$ and $Z_b(10650)$, we calculate the heavy-meson heavy-(anti)meson scattering potential up to next-to-leading order, $\mathcal{O}(Q^2)$, within chiral effective field theory ($\chi$EFT) employing a power counting scheme that explicitly keeps track with the large momentum scale  $Q \sim \sqrt{2\mu \delta}$ (where $\delta = m_V - m_P$ is the vector-pseudoscalar mass difference and $\mu$ their reduced mass)
 introduced by the coupled channel dynamics.
 We provide expressions for the two-pion exchange (TPE) terms up to $\mathcal{O}(Q^2)$ and their partial-wave decomposition. We show that these potentials are well-approximated by contact terms at $\mathcal{O}(Q^2)$, 
 with minor residual non-analytic TPE contributions, supporting $\chi$EFT convergence  in the theoretical predictions for $Z_b(10610)$ and $Z_b(10650)$, as well as their spin partners. These findings are also relevant for $D^{(*)}D^{(*)}$ scattering, especially for the $T_{cc}$ state, for both physical and lattice QCD data with moderately larger pion masses. We further demonstrate that the differences between isovector and isoscalar potentials for heavy mesons are naturally explained by the TPE contributions.

\end{abstract}


\maketitle

\section{Introduction}
Understanding exotic hadrons within  Quantum Chromodynamics (QCD) holds the promise for deeper insights into the strong force. 
In addition to the well-established quark model states, that describe mesons as quark-antiquark and baryons as three-quark systems, 
exotic hadrons encompass a large variety of states, including  multiquark configurations, glueballs, hybrids, etc., as documented
 in a large number of review articles ~\cite{Lebed:2016hpi,Esposito:2016noz,Ali:2017jda,Guo:2017jvc,Olsen:2017bmm,Liu:2019zoy,Brambilla:2019esw,Guo:2019twa,Chen:2022asf,Meng:2022ozq}. 
Exotic mesons with heavy quarks, often denoted as XYZ states, challenge the predictions of the conventional quark model, giving rise to intriguing questions about their internal structure.
 Some of these states are manifestly exotic -- notable examples include the 
 charged states of the $Z$ family, namely, 
$\zb$, $\zbp$~\cite{Belle:2011aa}, $\zc$~\cite{BESIII:2013ris,Belle:2013yex}, $\zcp$~\cite{BESIII:2013ouc}, $Z_c^\pm(4430)$~\cite{Belle:2007hrb,Belle:2009lvn,Belle:2013shl,LHCb:2014zfx}, 
which decay into final states containing  a heavy quarkonium accompanied by a single light hadron.

The states $Z_b(10610)$ and $Z_b(10650)$ discovered by the Belle collaboration~\cite{Adachi:2011mks}, commonly denoted as $Z_b$ and $Z_b'$ for brevity,  provide an excellent playground for gaining deeper insights into exotic states.  Both have $ J^{PC}=1^{+-} $~\cite{Adachi:2011mks}
and manifest themselves as
 two narrow peaks, separated by approximately 45 MeV, in the invariant mass distributions of the $\pi^\pm\Upsilon(nS)$ ($n=1,2,3$) and $\pi^\pm h_b(mP)$ ($m=1,2$) subsystems in the dipion transitions from the vector bottomonium $\Upsilon(10860)$ \cite{Belle:2011aa} (hereafter referred to as inelastic channels). 
Moreover, these states have been observed in the $B\bar{B}^{*}$\footnote{Here, a properly normalized $C$-odd combination of the $B\bar{B}^*$ and $\bar{B}B^*$ components is understood.} and $B^{*}\bar{B}^{*}$ invariant mass distributions (hereafter referred to as elastic channels) in the decays $\Upsilon(10860)\to\pi B^{(*)}\bar{B}^*$ with dominant branching fractions \cite{Belle:2012koo,Belle:2015upu}.
The proximity of $Z_b(10610)$ and $Z_b(10650)$ to the $B \bar B^*$ and $B^* \bar B^*$ thresholds, respectively,  
along with the predominance of the open-flavor branching fractions, strongly support their molecular interpretation~\cite{Bondar:2011ev}. However, despite this observation, the two primary explanations for the $Z_b$ states, consistent with the data, are a tetraquark model and a hadronic molecule picture, see, e.g.,  
Refs.~\cite{Esposito:2016noz,Ali:2017jda,Guo:2017jvc,Brambilla:2019esw,Albuquerque:2018jkn}  for review articles and references therein.

Understanding the nature of near-threshold exotic states hinges on accurately extracting their properties from data, typically found in the pole position and residue of the elastic scattering amplitude or probed using low-energy hadronic parameters like scattering length and effective range. 
Then, Weinberg's compositeness criterion can be employed to assess their internal structure. This criterion establishes a relationship between the pole position and residue of a state and the molecular component within its complete wave function. Although originally formulated for shallow bound states \cite{Weinberg:1965zz, Morgan:1992ge, Baru:2003qq}, it was recently extended to virtual states and resonances~\cite{Baru:2003qq,Matuschek:2020gqe,Bruns:2019xgo,Oller:2017alp,Kang:2016jxw,Sekihara:2016xnq,Kamiya:2016oao,Guo:2015daa,Sekihara:2014kya,Aceti:2012dd,Gamermann:2009uq,Kinugawa:2024kwb},  provided their constituents are narrow \cite{Filin:2010se}.
Since the states of interest are located very close to the production thresholds
of particle pairs to which they couple strongly, 
 systematic theoretical analyses, respecting analyticity and unitarity principles, are compulsory.
 Especially, simplistic Breit-Wigner parameterizations
 are to be avoided for the parameters extracted in this way are reaction dependent.
An Effective Field Theory (EFT) approach serves as a suitable framework for this purpose, offering model independence in the relevant energy range. Furthermore, it enables the formulation of testable predictions for various observables, facilitating the observation of potential molecular candidates and their heavy-quark (HQ) partner states.

Recently, a chiral  EFT-based approach was formulated to address  experimental data for all measured production and decay channels of the bottomonium-like states $Z_b(10610)$ and $Z_b(10650)$~\cite{Wang:2018jlv,Baru:2019xnh}. 
The EFT approach is constructed based on an effective Lagrangian that respects both chiral and heavy-quark spin symmetry (HQSS) of QCD. The essential aspects of this approach can be summarized as follows:

\begin{itemize}
\item The coupled-channel hadron-hadron EFT is formulated using the Weinberg counting~\cite{Weinberg:1990rz}, initially designed for treating few-nucleon systems. 
Furthermore, it was demonstrated in~\cite{Beane:2001bc} that at least in channels where the pion tensor force is operative, a non-perturbative treatment of  the one-pion exchange is necessary
--- for a modern discussion on the subject see Ref.~\cite{Epelbaum:2008ga}. The potential is built to a specific order in the chiral expansion $Q/\Lambda_\chi$, with the hard scale of the chiral EFT being represented by $\Lambda_\chi\approx 1$ GeV, and then nonperturbatively resummed through Lippmann-Schwinger type equations. 

\item Simultaneously with the chiral   expansion, the potential undergoes an expansion around the spin symmetry limit. 
At the leading order, this involves incorporating the mass difference of the spin partners $B$-$B^*$
 along with all interaction vertices constructed in accordance with HQSS. In the loops at next-to-leading order
 the $B$-$B^*$ mass difference can be dropped.

\item The binding momenta, pion mass, and the momentum scale resulting from the splitting between open-flavor partner thresholds  $B\bar B^{(*)}$ and $B^{*}\bar B^{*}$ are considered as soft scales of the system, collectively denoted as $Q$. The explicit inclusion of the coupled-channel scale extends the  energy range, where the theory is applicable. This is crucial for analyzing experimental data around the two elastic thresholds and in between. 
Note that  for energies near the $B^{*}\bar B^{*}$ threshold,  the on-shell relative momentum in the $B\bar B^{*}$  channel can be as large as $p_{\rm typ}= \sqrt{m_B \delta} \simeq  500$ MeV, where $\delta = m_{B^*}-m_B$, with $m_{B^*}$ and  $m_{B}$
being the $B^*$ and $B$ meson mass, respectively. 
The expansion parameter can be therefore as large as 
\be\label{eq:chi_MCS}
\chi = Q/\Lambda_{\chi}\sim p_{\rm typ}/\Lambda_{\chi}\simeq 1/2.
\ee
Thus the convergence of the chiral expansion needs to be investigated.

\item The elastic coupled-channel effective potential $V$  constructed in chiral   EFT up to ${\cal O}(Q^2)$ reads
\begin{eqnarray}\nonumber
V_{\text{EFT}}
 	&{=}&V_{\text{OPE}}^{(0)}{+}V_{\text{cont}}^{(0)}{+}V_{\text{OEE}}^{(0)}\\
& &\qquad {+}V_{\text{cont}}^{(2)}{+}V_{\text{TPE}}^{(2)}+
{\mathcal O}(\chi^4) \ .\label{eq:veft}
\end{eqnarray}

The potential at leading order (LO) includes two momentum-independent, ${\cal O}(Q^0)$, 
contact interactions, consistent with HQSS, 
while its long-ranged component is attributed to the pseudoscalar Goldstone boson exchange   represented here by the one-pion (OPE) and the one-$\eta$-meson (OEE) exchanges. 
Note that the Goldstone-boson exchange  potential  is well defined in the sense of an EFT only in combination with the pertinent contact operators \cite{Baru:2015nea}. 
To tame the strong regulator dependence arising from higher-momentum OPE contributions, especially when multiple open-flavor coupled channels are considered, a formally ${\cal O}(Q^2)$ $S$-wave-to-$D$-wave counter term is 
promoted to leading order, as detailed in \cite{Wang:2018jlv,Baru:2019xnh}. At next-to-leading order (NLO),  
two momentum-dependent
 ${\cal O}(Q^2)$ S-wave-to-S-wave contact terms  appear. The intermediate range contributions are represented by 
 the two-pion exchanges (TPE).  However, the current calculations~\cite{Wang:2018jlv,Baru:2019xnh} have, until now, omitted the contribution from TPE. In this work we provide the potentials
 that are necessary to overcome this 
 shortcoming.

\item The potential incorporates the contributions of  inelastic channels by enabling their coupling to the $S$-wave open-flavor thresholds. This inclusion results in effective elastic open-flavor potentials with imaginary components driven by unitarity. Meanwhile, the contributions to the real parts of the elastic potentials from inelastic channels can be absorbed through a redefinition of the momentum-independent ${\cal O}(Q^0)$ contact interactions~\cite{Hanhart:2015cua,Baru:2019xnh}.

\item All low-energy constants, including the two elastic couplings, effective couplings to inelastic channels, and the $S$-$S$ and $S$-$D$ contact interactions, are determined through a combined fit to all available experimental line shapes.

\item The leading contribution to the production operator   typically comes from the   open-flavor channels, unless notable structures in line shapes, like a dip near the threshold, suggest otherwise.  
If a dip is present, production might occur through more distant inelastic channels, as discussed in \cite{Dong:2020hxe}, see also \cite{Baru:2024ptl} for a recent application to the $X(3872)$. Another possibility in this case is the existence of a Castillejo-Dalitz-Dyson (CDD) zero near the threshold, challenging the effective range expansion of the scattering amplitude~\cite{Baru:2021ldu,Baru:2010ww,Kang:2016jxw}. For insights into the impact of triangle singularities on near-threshold line shapes, we  refer to the review article~\cite{Guo:2019twa}.  
However, none of these structures  play a role in   production of  $Z_b$  states from the decay of $\Upsilon(10860)$.

\item To be able to analyse data in the final states involving a quarkonium and two pions, namely the transitions $\Upsilon(10860)\to \pi\pi h_b(mP)$ and $\Upsilon(10860)\to \pi\pi \Upsilon(nS)$ (n=1,2,3),   the $\pi\pi$ final state interaction (FSI)     in an $S$-wave including its coupling to the $K\bar K$ channel  has to be taken into account. 
This was achieved by employing dispersion theory in Refs.~\cite{Baru:2020ywb,Baru:2022xne}.
\end{itemize}

The EFT approach formulated above has been employed to analyse all available data from the decay of $\Upsilon(10860)$. 
Utilizing data on the decays of $\Upsilon(10860)\to\pi^\pm h_b(mP)$ ($m=1,2$) and $\Upsilon(10860)\to\pi B^{(*)}\bar{B}^*$ to fix  unknown low-energy constants 
from the best $\chi^2$ fits, 
the approach revealed a very good understanding of these line shapes and resulted in the extraction of the pole positions and residues of the $Z_b$ states~\cite{Wang:2018jlv,Baru:2019xnh}.  Based on these results,  the line shapes for the HQSS partner states of the   $Z_b$'s,  namely the positive $P$- and $C$-parity states 
$W_{bJ}^{++} (J=0,1,2)$ were predicted parameter free  in the radiative decays $\Upsilon(10860)\to \gamma B^{(*)}\bar{B}^{(*)}$,  
$\Upsilon(10860)\to \gamma \pi \chi_{bJ}(mP) (m=1,2; J=0,1,2)$ and $\Upsilon(10860)\to \gamma \pi \eta_{b0}(nS) (n=1,2)$~\cite{Baru:2019xnh}.
Furthermore,  a version of the EFT amplitudes corresponding to a contact fit from Refs.~\cite{Wang:2018jlv,Baru:2019xnh}, augmented with the   $\pi\pi/K\bar K$ FSI, was 
used to show consistency with  the two-dimensional Dalitz plots for the $\Upsilon(10860)\to\pi^+\pi^-\Upsilon(nS)$ ($n=1,2,3$), revealing the importance of these FSI effects especially for  $\Upsilon(1S)$ and $\Upsilon(2S)$ final bottomonium states.  On the other hand,  $\pi\pi/K\bar K$ FSI in the process 
$\Upsilon(10860)\to\pi^\pm h_b(mP)$ ($m=1,2$)  was shown to be strongly suppressed by HQSS \cite{Baru:2022xne}.  

The effect of the OPE on the results of Refs.~\cite{Wang:2018jlv,Baru:2019xnh} can be formulated as follows. 
First, we stress once again that the OPE potential  is well defined in the sense of an EFT only in connection with  contact operators, 
which implies that the true effect of the OPE on observables can only be seen after the potential is  renormalized. 
Second, the effect from the OPE depends on   isospin and $C$-parity \cite{Baru:2017gwo}.    At large distances the OPE contribution  to the $Z_b$'s  
(isovector  $B^{(*)}\bar B^*$ scattering with $C=-1$) is repulsive, and the OPE cannot go on shell since the 
decay $B^*\to B\pi$ is not possible at physical pion masses.  Without the OPE,  the 
$Z_b(10610)$ and $Z_b(10650)$ were found to be virtual states with respect 
to the nearby $B \bar B^*$ and $B^* \bar B^*$ thresholds, respectively~\cite{Wang:2018jlv,Baru:2019xnh}. However,
the nontrivial interplay of the repulsive OPE and attractive contact potentials resulted in shifting poles into the complex plane, rendering them resonance states located just below the corresponding thresholds.
The same pattern was also observed for their predicted HQSS partners that in some channels can even move above threshold. 
This picture is entirely consistent with recent findings for $DD^*$ scattering at unphysically large pion masses in Ref.~\cite{Meng:2023bmz},
where it was shown that the inclusion of the repulsive OPE in the analysis of 
  lattice finite-volume spectra from Ref.~\cite{Padmanath:2022cvl} shifts the  $T_{cc}$ pole to the complex plane (see also Refs.~\cite{Du:2023hlu,Collins:2024sfi} for related studies). 
  As a result, the repulsion generated by the OPE has a pronounced effect on the $T_{cc}$  pole trajectory  as a function of the pion mass, by pushing it into the complex energy plane for  $m_{\pi}\ge 230$ MeV~\cite{Abolnikov:2024key}.

In this work, we provide the contributions to the
$B^{(*)}\bar B^{(*)}$ potential that are so far missing in the analysis to
NLO. Those comprise
two-pion exchange contributions to one loop. 
  The inclusion of these
  contributions is necessary to test the convergence of the chiral EFT formalism (which is especially important given the rather large expansion parameter $\chi\sim 1/2$), provide a systematic uncertainty estimate of the theoretical results, and finally extract the pole positions reliably. It is worth noting that 
we are aiming at the calculation of the TPE terms in the so-called momentum counting scheme (MCS),  which treats the momentum between 
$B \bar B^{(*)}$ and $B^{*}\bar B^{*}$ as a leading soft scale -- see Eq.~\eqref{eq:chi_MCS}.  As will become clear below, the MCS selects a specific subclass
of the one-loop contributions, which are enhanced due to the presence of the large momentum scale $p$.  
This kind of power counting was introduced originally to
provide a convergent EFT for pion production in nucleon-nucleon collisions near the threshold, where the large momentum scale,
$p\simeq \sqrt{m_{\pi}M_N}$ with  $m_{\pi}(M_N)$ being the pion(nucleon) mass,
was introduced through the large momentum (relative to the pion mass) necessary in the
initial state.
This approach indeed led to a successful understanding of the non-trivial pion production mechanisms in $NN\to NN\pi$, which was not achievable within the original Weinberg's counting where momenta and the pion mass
where treated at the same order -- see 
Refs.~\cite{Hanhart:2003pg,Baru:2013zpa} for reviews. The TPE contributions for pion production in two nucleon collisions were calculated in this power counting in Refs.~\cite{Filin:2013uma,Filin:2012za,Baru:2016kru}.

It should be stressed that the rate of convergence of the chiral expansion is a crucial diagnostic tool to understand the nature of the multi-quark states under investigation here:
a compact state driving the experimental signals would call for a pole term in the scattering potential. Its presence should then lead to a bad convergence of the series of local contact terms included in the EFT. After all,  in a low-energy EFT organized according to an expansion in powers of the soft momenta, the presence of a compact state can only influence the rate of convergence of the expansion, or in other words, the power counting. However, it cannot modify the individual contributions as such, since those are built solely on the symmetries of the underlying theory. While the data existing up to date for the $Z_b$ states can be well described with higher-order contact terms naturally suppressed in accordance with the power counting (up to the need to promote the $S$-$D$ counter term called for by renormalizability of the OPE in the presence of several coupled hadronic channels), data with improved statistics expected from the Belle II experiment will call for a refined theoretical effort.

As the results for TPE diagrams derived here up to order ${\cal O}(Q^2)$ do not depend on the heavy-meson mass, they can also be applied to other heavy-meson heavy- (anti)meson systems, such as $D^{(*)}D^* - D^{(*)}D^*$ scattering, particularly in the context of the $T_{cc}$ and its possible partner states.
It should be noted that for the physical pion mass, the three-body cuts in the TPE diagrams will provide some contributions to the $T_{cc}$ width, which  need to be included if one aims at the high-accuracy calculation of this quantity.  On the other hand, the effect of the cuts on the the real part of the TPE diagrams should be very small, and the TPE contributions can be still largely absorbed by the contact operators.  This conclusion should also hold for not too large  unphysical pion masses,  $m_{\pi}^{\rm ph}< m_{\pi}< p_{\rm typ}$, in the context of lattice QCD data analyses.

The TPE contributions to heavy-meson–heavy-(anti)meson scattering have already been addressed in the literature. For instance, phenomenological calculations in Refs.~\cite{Dias:2014pva,Aceti:2014uea} considered a particular subclass of TPE operators. Additionally, Refs.\cite{Wang:2018atz,Wang:2020} (see also \cite{Wang:2022jop,Xu:2017tsr}) provide investigations within the framework of EFT. These TPE operators, however, were derived ignoring all coupled-channel transitions based on the original Weinberg's counting and no attempt was made to check for the renormalisation of those to the given order. 
 In this work we overcome those short comings. Moreover,
  we will provide a   comparison of our results with the earlier EFT works.

The paper is organized as follows. In Sec.~\ref{lag}, the Lagrangian and the vertices for our approach are provided. The power counting scheme is discussed in detail in Sec.~\ref{pc}.
In Sec.~\ref{sec:effpots}, the effective potential of one representative channel is presented at $\mathcal{O}(\chi^0)$ and $\mathcal{O}(\chi^2)$. Appendix \ref{alleffpotentials} contains the effective potentials of all the channels.
Sec.~\ref{sec:pwd} provides the partial wave decomposed potentials with Appendix \ref{PWP} containing a
complete set of the relevant projection
operators. 
Sec.~\ref{sec:check} summarizes the various checks conducted on our PWD potentials and Sec.~\ref{sec:comparison} gives the comparison our potentials to those of previous works.
Sec.~\ref{summary} gives a summary and outlook of this paper.
Additionally, Appendix \ref{sec:pertintegrals} provides details about the evaluation of the pertinent 
loop integrals of the TPE potentials.

\section{Lagrangian and Vertices}
\label{lag}
The effective Lagrangian describing  $B^{(*)}$ $\bar{B}^{(*)}$ scattering at low energies reads~\cite{Mehen:2011yh,Mehen:2013,Wang:2018jlv,Baru:2019xnh}:
\begin{widetext}
    \begin{multline}
    \begin{split} 
    \label{eq:lagrangian}
    \mathcal{L} &= \mathrm{Tr}[H^{\dag}_a \, (i D_0)_{ba} \, H_b] + \frac{\delta}{4}  \mathrm{Tr}[H^{\dag}_a \,\sigma^i \, H_a \, \sigma^i] + \mathrm{Tr}[\Bar{H}^{\dag}_a \, (i D_0)_{ab} \, \Bar{H}_b] + \frac{\delta}{4}\mathrm{Tr}[\Bar{H}^{\dag}_a \, \sigma^i \, \Bar{H}_b \sigma^i]
    - \frac{g_{Q}}{2} \mathrm{Tr}[\boldsymbol{\sigma} \, \mathbf{\cdot} \, \mathbf{u}_{ab} \, H^{\dag}_a H_b   ] \\ &+ \frac{g_Q}{2} \mathrm{Tr}[\Bar{H}_a \Bar{H}^{\dag}_b \, \boldsymbol{\sigma} \, \mathbf{\cdot} \, \mathbf{u}_{ab}  ] 
    - \frac{C_{10}}{8} \mathrm{Tr}[\Bar{H}^{\dag}_a \, \tau^{A}_{aa'} H^{\dag}_{a'} H_b \tau^{A}_{bb'} \Bar{H}_{b'}]
    -\frac{C_{11}}{8} \mathrm{Tr}[\Bar{H}^{\dag}_a \, \tau^{A}_{aa'} \sigma^i H^{\dag}_{a'} H_b \tau^{A}_{bb'} \sigma^i \Bar{H}_{b'}]\\
    &- \frac{D_{10}}{8} \Big\{ \mathrm{Tr} [\nabla^i \Bar{H}^{\dag}_a \, \tau^{A}_{aa'} \nabla^i  H^{\dag}_{a'} H_b \tau^{A}_{bb'} \, \Bar{H}_{b'}] + \mathrm{Tr} [ \Bar{H}^{\dag}_a \, \tau^{A}_{aa'}  H^{\dag}_{a'} \nabla^i H_b  \tau^{A}_{bb'} \nabla^i \Bar{H}_{b'}] \Big\}  \\
    &- \frac{D_{11}}{8}\Big\{ \mathrm{Tr} [\nabla^i \Bar{H}^{\dag}_a \, \tau^{A}_{aa'} \sigma^j \nabla^i  H^{\dag}_{a'} H_b \tau^{A}_{bb'} \, \sigma^j \Bar{H}_{b'}] + \mathrm{Tr} [ \Bar{H}^{\dag}_a \, \tau^{A}_{aa'} \, \sigma^j  H^{\dag}_{a'} \nabla^i H_b  \tau^{A}_{bb'}\, \sigma^j \nabla^i \Bar{H}_{b'}] \Big\}\\
    &-\frac{D_{12}}{8} \Bigg\{  \mathrm{Tr} \bigg[ \Big( \nabla^i \Bar{H}^{\dag}_a \, \tau^{A}_{aa'} \, \sigma^i \nabla^j  H^{\dag}_{a'} + \nabla^j \Bar{H}^{\dag}_a \, \tau^{A}_{aa'} \, \sigma^i \nabla^i  H^{\dag}_{a'} - \frac{2}{3} \delta^{ij} \nabla^k \Bar{H}^{\dag}_a \, \tau^{A}_{aa'} \sigma^i \nabla^k  H^{\dag}_{a'} \Big) 
    H_b \tau^{A}_{bb'} \, \sigma^j \Bar{H}_{b'} \bigg] \\
   &+ \mathrm{Tr} \bigg[ \Bar{H}^{\dag}_a \, \tau^{A}_{aa'} \, \sigma^i  H^{\dag}_{a'} \Big( \nabla^i H_b  \tau^{A}_{bb'}\, \sigma^j \nabla^j \Bar{H}_{b'} + \nabla^j H_b  \tau^{A}_{bb'}\, \sigma^j \nabla^i \Bar{H}_{b'} - \frac{2}{3} \delta^{ij} \nabla^k H_b  \tau^{A}_{bb'}\, \sigma^j \nabla^k \Bar{H}_{b'} \Big) \bigg] \Bigg\}  + \dots,  
    \end{split}
\end{multline}
\end{widetext}
where $a$ and $b$ are isospin indices, $\sigma$'s and $\tau$'s are the spin and isospin Pauli matrices, respectively. The isospin matrices are normalized as $ \tau^{A}_{ab} \tau^{B}_{ba} =2 \delta^{AB}$ and the trace (Tr) is taken over spin space. 
The contact terms $C_{1i}$ ($i=0,1$) and 
$D_{1i}$ ($i=0,1,2$) represent short-range interactions, while the ellipsis in Eq.~\eqref{eq:lagrangian} denotes similar terms $C_{0i}$ and $D_{0i}$  which are not shown explicitly and appear without the $\tau$ matrices. 
The terms proportional to $\delta = m_{B^*}-m_B \approx 45$ MeV are the leading terms that violate spin symmetry. To the order we are working (NLO in the chiral expansion
and in the heavy quark expansion) they do not
contribute to the loops in the potentials but only
to the two-body propagators in the LS-equation.

The $H_a$ and $\Bar{H}_a$ are super-fields which contain the $B^{(*)}$ and $\Bar{B}^{(*)}$ fields,
respectively, with $H_a= B_a+B^{*\, i }_a \sigma^{i}$ and $\Bar{H}_a= (\Bar{B}\tau_2)_a - ( \Bar{B}^{* \, i}\tau_2)_a \sigma^{i}$, where $B_a$($\Bar{B}_a$) and $B^{*\, i }_a$($ \Bar{B}^{* \, i}_a$) are the pseudoscalar and vector $B$ mesons (antimesons), respectively.
The $\tau_2$, acting as the charge conjugation matrix in isospin space, appears in the expressions for the anti-B-mesons, since they contain light antiquarks.
$H_1$ contains $B^0$ and $(B^{0})^*$ and $H_2$ contains $B^+$ and  $(B^{+})^*$, while $\Bar{H}_1$ and $\Bar{H}_2$ contain the respective antiparticles \cite{Mehen:2013}.
The zeroth component of the 
chiral covariant derivative is given by $D_0= \partial_0+ \Gamma_0$  with
\begin{equation}
        \Gamma_0
        = \frac{i}{4 f^2_{\pi}}
        \left(\vec \pi\times\partial_0 \vec \pi \right)\cdot \vec \tau 
        + \mathcal{O}(\boldsymbol{\pi}^4) \ ,
\end{equation}
where $f_{\pi}=92.4$ MeV denotes the pion decay constant.
The 
 spatial components of the axial current
 read
 \begin{equation}
     \mathbf{\mathbf{u}}= -\vec \nabla (\vec{\tau}\cdot \vec{\pi}) /f_{\pi}+ \mathcal{O}(\boldsymbol{\pi}^3) \ .
     \end{equation}
In both cases $\vec{\tau}$ and $\vec{\pi}$ are 3-dimensional vectors made of the Pauli matrices and the pions fields, respectively.  
Employing heavy quark spin symmetry, the pion-heavy meson coupling constant is fixed to
$$g_Q\approx g_b \approx g_c \approx g = 0.57 \ ,$$ extracted
from the partial decay width $D^*\to D\pi$ provided in Ref.~\cite{ParticleDataGroup:2022pth} (this value agrees within 10\% with that extracted in lattice QCD for static sources~\cite{Bernardoni:2014kla}). 
The terms proportional to the low-energy constants (LECs) $C_{10}$ and $C_{11}$ correspond to the $\mathcal{O}(p^0)$ $S$-wave contact interactions, whereas the terms proportional to $D_{10}$ and $D_{11}$ correspond to $\mathcal{O}(p^2)$ $S$-wave contact interactions. 
The term $D_{12}$ gives rise to $S$-$D$  transitions --- this is the counter term
formally appearing at NLO, however, promoted
to LO as detailed above.
As we are only interested in $S$-$S$ and $S$-$D$ transitions, terms proportional to $\nabla^i H^{\dagger} \nabla^j H$, leading to 
$P$-wave interactions, are ignored \cite{Wang:2018jlv}. 
\par
From the Lagrangian provided in Eq.~(\ref{eq:lagrangian}) we now derive the vertex structures relevant for this work.
The $B$-meson--pion interactions are  
\begin{equation}\label{eq:LpiBB}
 \mathcal{L}_{B^{(*)}B^{(*)}\pi }   {=}  \frac{g}{2 f_{\pi}} \partial_i \Vec{\pi}\Big[
B^{\dag} \Vec{\tau} B_i  {+} B_i^{\dag} \Vec{\tau}  B {+} 
i\epsilon_{jki}  B_j^{\dag} \Vec{\tau} B_k  \Big] \, ,
\end{equation}

where the indices $i,j,k$ refer to the 
spacial index of the derivative or the 
$B^*$ polarisation vectors---summation over those
is assumed. The factor of $(1/2)$ comes from the normalisation of the heavy-meson field as shown in Appendix A of Ref.~\cite{Fleming:2007}.
If we define $\vec k$, the momentum of the pion, as outgoing, we can replace the spacial  derivative in Eq.~(\ref{eq:LpiBB}) by $-i \vec k$ and get for the various vertices
\begin{eqnarray} \nonumber
{v}_{B \rightarrow B\pi_a\phantom{^*}\phantom{^*} } &=& 0 \ , \\   \nonumber 
{v}_{B^{*} \rightarrow B\pi_a\phantom{^*} }   &=&  \frac{g }{2f_{\pi}} (\Vec{\epsilon} \cdot \Vec{k}) \tau_a
\ ,
\\  \nonumber
{v}_{B \rightarrow B^{*}\pi_a\phantom{^*} }   &=&  \frac{g }{2f_{\pi}} 
(\vec{\epsilon}^* \cdot \vec{k}) \tau_a \ ,
\\  \nonumber
{v}_{B^{*} \rightarrow B^{*}\pi_a }   &=& -i \frac{g }{2f_{\pi}}    
\tau_a  ( {\vec \epsilon}{\times} {{\vec \epsilon}^*}) \vec k \ .
\end{eqnarray}
In all expressions $a$ denotes the isospin index of the pion. The corresponding vertices for the
anti-mesons are near identical to the meson case with the exception that the charge-conjugated Pauli matrix,
related to the antifundamental representation of the isospin group,
is to be used which reads $\boldsymbol{\tau^c}=\tau_2 \, \boldsymbol{\tau}^T \, \tau_2= \boldsymbol{-\tau}$ \cite{Baru:2019xnh}.

The leading two-pion $B^{(*)}B^{(*)}$ couplings arise from the chiral covariant derivative acting on the heavy fields, namely the Weinberg-Tomozawa (WT) vertex.
In particular we have
\begin{eqnarray}\nonumber
 \mathcal{L}_{B^{(*)}B^{(*)}\pi \pi   }   &=& -\frac{1}{4 f^2_{\pi}}\epsilon_{abc} \pi_{a} \partial_0 \pi_{b} \\
& & \qquad \times \Big[ B^{\dag} \tau_c B  +
B^{* j\, \dag} \tau_c B^{* j}  \Big] \ .
\end{eqnarray}
From this the pertinent vertices
are 
\begin{eqnarray}\nonumber
v_{B \pi_a  \rightarrow B\pi_b }\phantom{^*}\phantom{^*}  
&=& \frac{ 1}{4 f^2_{\pi}}  \epsilon_{abc}\tau_c (  k'_0+   k_0) \ ,
\\ \nonumber
v_{B^* \pi_a  \rightarrow B^*\pi_b }  
&=&  \frac{ 1}{4 f^2_{\pi}}  \epsilon_{abc} \tau_c
(\vec{\epsilon} \cdot \vec{\epsilon}\,^*) (  k'_0+   k_0) \ ,
\end{eqnarray}
where $k_0$ ($k_0'$) denotes the
zeroth component of the incoming (outgoing) pion four-momentum.
Again, to switch to the corresponding vertices for the 
anti-B-mesons the $\tau$ matrices need to be replaced by their charge conjugate counterparts.

\section{Power counting }
\label{pc}

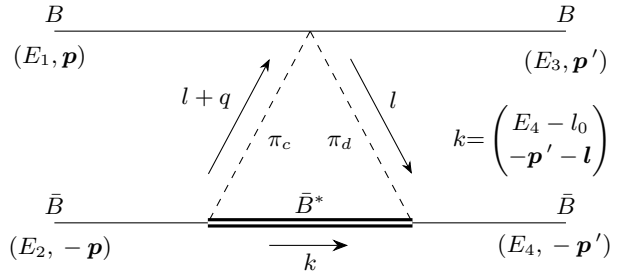
\begin{figure}[h]
\centering
     \begin{tikzpicture}[scale=1.7]
     \begin{feynman}[]
     \vertex [](a) at (0,0);
     \vertex [] (b) at (-0.8, -1.5);
     \vertex [] (c) at (0.8, -1.5);
     \vertex [label =left: $(E_1{,} \, \vec{p}) \quad \quad $](p1) at (-1.3,-0.2);
      \vertex [label =left: $(E_2{,} \, -\vec{p}) \quad \quad $](p2) at (-1.1,-1.7);
     \vertex [label = $B$](p1) at (-2,0);
      \vertex [label = $B$](p'1) at (2,0);
     \vertex [label = $\Bar{B}$] (p2) at (-2,-1.5);
     \vertex [label = $\Bar{B}$] (p'2) at (2,-1.5) ;
      \node[yshift=-0.4cm,xshift=0cm] at (p'1) {$(E_3{,} \, \vec{p}\,') $};
      \node[yshift=-0.3cm,xshift=-0.15cm] at (p'2) {$(E_4{,} \, -\vec{p}\,') $};
     \node[yshift=1.1cm,xshift=0.2cm] at (p'2) {$k{=} \begin{pmatrix}
     E_4-l_0 \\
     -\vec{p}\,'-\vec{l}  
     \end{pmatrix}  \qquad  \qquad $};
     \diagram*{
         (p1) -- [plain  ] 
         (a) -- [ plain] (p'1) ,
         (a) -- [scalar, reversed momentum' = {[arrow shorten=0.2]$l+q$}, edge label=$\pi_c$](b) ,
         (a) -- [scalar,  momentum = {[arrow shorten=0.2]\(l\)}, edge label'=$\pi_d$](c) ,
         (p2)  -- [plain]
         (b) -- [double,double distance=0.3ex,very thick, momentum' = {[arrow shorten=0.3]\(k\)},edge label=$\Bar{B}^*$] (c),
         (c) -- [plain](p'2)
     };
     \end{feynman}

     \end{tikzpicture}
   \caption{Typical one loop diagram that appears
    at NLO in
    the momentum expansion as well as the standard power counting. }
    \label{fig:tripowercounting}
\end{figure}

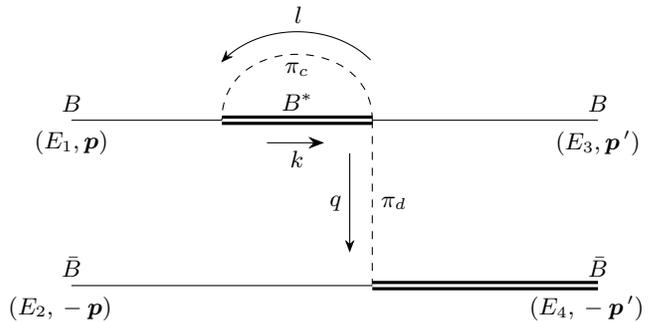
\begin{figure}[h]
    \centering
    \begin{tikzpicture}[scale=2]
     \begin{feynman}[]
     \vertex [](a) at (-1,-0.7);
     \vertex [] (b) at (0, -0.7);
     \vertex [] (c) at (1, -0.7);
     \vertex [] (d) at (0,-1.8) ;
     \vertex  [label = $B$](p1) at (-2,-0.7);
     \vertex [label = $B$] (p'1) at (1.5, -0.7);
      \vertex [label = $\Bar{B}$] (p2) at (-2,-1.8) ;
     \vertex  [label = $\Bar{B}$] (p'2) at (1.5,-1.8) ;
      \node[yshift=-0.3cm,xshift=0cm] at (p1) {$(E_1{,} \, \vec{p}) $};
      \node[yshift=-0.3cm,xshift=-0.15cm] at (p2) {$(E_2{,} \, -\vec{p}) $};
       \node[yshift=-0.3cm,xshift=0cm] at (p'1) {$(E_3{,} \, \vec{p}\,') $};
      \node[yshift=-0.3cm,xshift=-0.15cm] at (p'2) {$(E_4{,} \, -\vec{p}\,') $};

     \diagram*{
          (p1) -- [plain] (a) -- [double,double distance=0.3ex,very thick, momentum' = {[arrow shorten=0.3]\(k\)},edge label=$B^*$ ] (b) -- [scalar,momentum' = {[arrow shorten=0.2]\(q\)}, edge label=$\pi_d$ ] (d) ,
         (b) -- [scalar, half right,momentum' = {[arrow shorten=0.2]\(l\)}, edge label=$\pi_c$ ](a) ,
         (b)  -- [plain](p'1), 
         (p2) -- [plain] (d) -- [double,double distance=0.3ex,very thick] (p'2)
     };
     \end{feynman}
     \end{tikzpicture} 
    \caption{Typical one loop diagram that appears
    at NNLO in
    the momentum expansion, but at NLO in the standard counting. 
    \label{fig:vertpowercounting} 
     }
\end{figure}
The power counting of the pion loops for the $B^{(*)} \Bar{B}^{(*)}$ potentials (and also the $B^{(*)} B^{(*)}$ potentials that we calculate  as a byproduct)
is dependent on the three dynamical scales of the system namely, 
the pion mass, $m_\pi$, the momentum
scale $p_{\rm typ}= \sqrt{m_B \delta} \simeq  500$ MeV
and the difference of $m_B$ and $m_{B^*}$, $\delta \approx 45$ MeV.
The hard scale is  the chiral symmetry breaking
scale $\Lambda_\chi$, but it may also include the  heavy meson mass  $m_B$.
Since the
binding energies of the $Z_b$ states are generated
dynamically through the solution of the LS equation, they do not need to be considered in the power counting for the potentials.
As we aim to fit the 
available experimental data in the energy range
that covers both $Z_b$ states, we need
an effective field theory
that allows us to cover the energy range from the $B \Bar{B}$ threshold up to the
$B^{*} \Bar{B}^{*}$  threshold, which spans 90 MeV. We therefore need to treat $p_{\rm typ}$ dynamically as a soft scale. 
 It is therefore important to keep track of momentum scales, which dictate the pertinent  contributions in the loops. Thus,  the expansion parameters are
\begin{equation}
  \chi_1  = \frac{p_{\rm typ}}{\Lambda_{\chi}}
     \ , \ \chi_2 = \frac{m_{\pi}}{\Lambda_{\chi}}\ , \ \chi_3 = \frac{p_{\rm typ}}{m_B} \ , \    \chi_4  = \frac{\delta}{\Lambda_{\chi}} \ ,
\end{equation}
which numerically take values of about 1/2, 1/7 , 1/10 and 1/20
in order.
It should be noted, however, that in the charm system the
mass splitting between the pseudoscalar and vector
ground state mesons is of the order of the pion
mass, making $\chi_2$ and $\chi_4$ similar. Because of this and to keep the scheme simple, one may combine the given parameters  into
a one-parameter expansion,
\begin{equation}\label{eq:PC}
    \chi \sim \chi_1  \ , \   \chi^2 \sim \chi_1^2 \sim \chi_2 \sim \chi_3
    \sim  \chi_4 \ .
\end{equation}
In what follows we are only interested in the leading loop contributions. In particular we will use
that $m_\pi/p_{\rm typ} \sim {\mathcal O}(\chi)$.

 The implications of this  power counting scheme for the
 order assignment of one-loop diagrams is now
 illustrated on two examples, namely one specific TPE 
 contribution to the $B \Bar{B} \rightarrow B \Bar{B} $ potential seen in Fig.~\ref{fig:tripowercounting}, and a
 one-loop vertex correction shown in Fig.~\ref{fig:vertpowercounting}.
Since we are focusing on the leading loop corrections, in all cases the vertices are taken in leading order only.

The effective potential for the triangle diagram is written as, 
\begin{widetext}
 \begin{multline}
   i  V_{T}=  \sum_\lambda\int \frac{d^4l}{(2\pi)^4}
    \frac{1}{4 f_{\pi}^2}  \Big( (2l_0+q_0)  \epsilon_{cdh} (\tau_1)_h
    \Big)  \frac{i}{ (E_4 -l_0 -m_{B^*}- (\Vec{p'}+ \Vec{l})^2/(2 m_{B^*}))}  \\
     \Bigg( \frac{g}{2 f_{\pi}}  \Big( \epsilon_{j} (\lambda) (-l)_j \Big) (\tau^c_2)_d \Bigg)
     \frac{i}{(l+q)^2 - m^2_{\pi}}    \Bigg( \frac{g}{2 f_{\pi}}  \Big( \epsilon^*_{i} (\lambda) (l+q)_i \Big) (\tau^c_2)_c \Bigg)
   \frac{i}{l^2 - m^2_{\pi}} 
 \end{multline}

\end{widetext}
where 
\begin{equation}
    q = (E_3, \Vec{p'})-(E_1,\Vec{p)}= (0,\Vec{q})+
    {\cal O}(\chi) \ ,
\end{equation} 
with $\Vec{p} $ and $\Vec{p'}$ as initial and final  relative momenta of the heavy
mesons, respectively,  and $E_1$ ($E_2$) and $E_3$ ($E_4$) for the energies of the meson (anti-meson) in the initial and final states, respectively. For this discussion we assume that the external states are on their mass shell
 and the total energy is at the $B^* \Bar{B}^*$ threshold. 
Then $q \sim p_{\rm typ}$. Using

\begin{equation}
    \epsilon_{cdh}(\tau_1)_h   (\tau_2)_d  (\tau_2)_c  = - 2i(\Vec{\tau_1} \cdot  \Vec{\tau_2}) \, ,
\end{equation}
and
\begin{equation}
 \sum_\lambda   \epsilon_{i}^*(\lambda) \epsilon_{j}(\lambda) = \delta_{ij}
\end{equation}
we get,
\begin{equation}
     V_{T}= \frac{g^2}{4f^4_{\pi}} (\Vec{\tau_1} \cdot  \Vec{\tau_2}) I_{tr}
\end{equation}
where the pertinent integral is given by,
\begin{multline}\label{Itr1}
 I_{tr}= \frac{i}{2}  \int  \frac{d^4l}{(2\pi)^4}  \bigg( \frac{2l_0+q_0}{l_0+\delta + (2\Vec{p'} \Vec{l}+ \Vec{l}^2)/(2 m_{B^*})}\bigg)\\
 \times  \frac{(\vec{l}+\vec{q})\cdot\vec{l}}{\big[(l+q)^2 -m^2_{\pi} + i \epsilon \big] \big[l^2-m^2_{\pi}+ i \epsilon\big] } 
\end{multline}
One observes that the pion propagators drive $l_0 \sim l\sim p_{\rm typ}$~\cite{Baru:2013zpa,Hanhart:2003pg}.
Indeed,  keeping the leading terms in the pion propagator, one finds
\begin{equation}\label{eq:l0_pion}
(l+q)^2 -m^2_{\pi} + i \epsilon =  l_0^2 - (\vec l+\vec q)^2+ i \epsilon +{\cal O}(\chi^3), 
\end{equation}
which yields $l_0\sim q \sim p_{\rm typ}$
 as well as $|\vec l|\sim
p_{\rm typ}$. Then, one observes that  $l_0$ is the dominant term in the   expression within the parentheses in the first line 
of Eq.~\eqref{Itr1}.
Because of this,
we can drop all terms 
except $l_0$, since all other terms appear to be suppressed relative to $l_0$ either as $\delta/p_{\rm typ} \sim {\cal O}(\chi)$ or $p_{\rm typ}/m_B$ which is counted as ${\cal O}(\chi^2)$ according to Eq.~\eqref{eq:PC}. Thus the integral to evaluate is
\begin{multline}
 I_{tr}{=}\ i \! \int \! \frac{d^4l}{(2\pi)^4}  
   \frac{(\vec{l}+\vec{q})\cdot\vec{l}}{\big[(l{+}q)^2 {-}m^2_{\pi} {+} i \epsilon \big] \big[l^2{-}m^2_{\pi}{+} i \epsilon\big] } \ ,\label{eq:Itr}
\end{multline}
with the dominant contribution coming from $l \sim q \sim  p_{\rm typ}$.
As long as multiple scales enter the expansion, any given loop contributes
at various orders simultaneously~\cite{Hanhart:2003pg,Baru:2013zpa}.
The  lowest order at which the pion loops start to contribute to $B^{(*)}\bar B^{(*)}$ and  $B^{(*)}  B^{(*)}$ scattering  is ${\cal O}(\chi^2)$. 
To derive results at ${\cal O}(\chi^2)$, all scales except for $q \sim p_{\rm typ}$
  can be initially omitted in the loop expressions.
 But the same loops also contain  contributions additionally suppressed by  $(m_\pi^2/p_{\rm typ}^2)$, thus contributing at $ {\cal O}(\chi^4)$.
 To illustrate this point, we use dimensional regularisation to find for the integral  in Eq.~(\ref{eq:Itr}) the following expression
\begin{multline}
   I_{tr} =-\frac{1}{16 \pi^2} 
\Bigg\{ \bigg( \frac{5}{12} \vec{q}\,^2{+} \frac{3}{2} m^2_{\pi} \bigg) \mathcal{R} {-}
\frac{13}{36} \vec{q}\,^2{-} \frac{m^2_{\pi}}{3}\\
{+} \bigg(\frac{5}{6} \vec{q}\,^2{+} 3 m^2_{\pi} \bigg) \ln{\bigg(\frac{m_{\pi}}{\mu}\bigg)} 
{+}\bigg( \frac{5}{6} \vec{q}\,^2{+} \frac{4}{3} m^2_{\pi} \bigg) L(q) \Bigg\}\\
=-\frac{5\vec{q}\,^2}{96 \pi^2} 
\Bigg\{ \frac{\mathcal{R}}{2} {-}
\frac{13}{30}{+}\ln{\bigg(\frac{m_{\pi}}{\mu}\bigg)} 
{+} L(q) \Bigg\}{+}{\cal O}(\chi^4) \ ,\label{eq:Itreval}
\end{multline}
where $\mu$ is the renormalization scale in dimensional regularization, 
\begin{eqnarray} \label{Eq:Lq}\nonumber
L(q)&=& \frac{\sqrt{4m^2_{\pi}+ q^2}}{q} \ln{\bigg(\frac{\sqrt{4m^2_{\pi}+ q^2}+q}{2 m_{\pi}}\bigg)}  \\
&=& \ln{\bigg(\frac{q}{m_{\pi}}\bigg)} + {\mathcal O}(\chi)
\ .
\end{eqnarray}
and 
\begin{equation}
    \mathcal{R}= -\frac{2}{\xi} + \gamma_E -1-\ln{(4\pi)}
    \label{eq:Rdef}
\end{equation}
with $\xi=4-D$ and $\gamma_E\approx 0.57$ denoting the Euler-Mascheroni constant. 
 Thus, the dominant loop contribution,
 as evaluated in the last lines of Eqs.~(\ref{eq:Itreval}-\ref{Eq:Lq}), corresponds to ${\mathcal O}(\chi^2)$,  while in general, the loops (see the first lines in Eqs.~(\ref{eq:Itreval}-\ref{Eq:Lq})) also give rise to higher-order terms.  Since the leading order potential for  scattering of two heavy particles appears at ${\cal O}(\chi^0)$
and, as in the two--nucleon system, 
there are no contributions
at ${\cal O}(\chi)$, the  loops at ${\cal O}(\chi^2)$  are, by convention, associated with the next nonvanishing order, referred to as NLO.
\par
For the vertex correction (seen in Fig.~\ref{fig:vertpowercounting}) the integration variable can always be chosen such that the pion propagator in the loop does not contain any external variable. 
Therefore,  contrary to Eq.~\eqref{eq:l0_pion},
the momentum $q \sim p_{\rm typ}$ does not enter the pion propagator in the loop. 
Hence,  the energy scale $l_0$ in this case is given by either $\delta$ or $m_{\pi}$, since both are being counted at the same order. At the same time the momentum scale is also given by $m_{\pi}$.
Combining all factors, we  conclude 
that the vertex corrections are suppressed as compared to the TPE discussed above by a factor of $(m_{\pi}/p_{\rm typ})^2= \chi^2 $. Due to this observation, all the vertex correction terms can be ignored in this 
study, since they start to contribute only
at order N$^3$LO.

 \section{Effective Potentials}
 \label{sec:effpots}
In this section
 the effective potentials of  $B^{(*)} \Bar{B}^{(*)} \rightarrow B^{(*)} \Bar{B}^{(*)}$ and $B^{(*)} B^{(*)} \rightarrow B^{(*)} B^{(*)}$ are discussed 
 --- further details are provided in the appendices.

\subsection{
\texorpdfstring{Leading-order diagrams for ${B^{(*)} \bar{B}^{(*)} \to B^{(*)} \bar{B}^{(*)}}$}{}
}

\begin{figure}[H]
 \begin{subfigure}[h]{.3\linewidth}
        \centering
        \begin{tikzpicture}[scale=0.6]
     \begin{feynman}[]
     \vertex [](a) at (0,-0.75);
     \vertex [](p1) at (-2,0);
     \vertex [](p'1) at (2,0);
     \vertex [] (p2) at (-2,-1.5);
     \vertex [] (p'2) at (2,-1.5) ;

     \diagram*{
         (p1) -- [plain] (a) -- [plain] (p'1) ,
         (p2)  -- [plain](a) -- [plain] (p'2)
     };
     \end{feynman}
     \end{tikzpicture} 
        \caption*{$CT_{1}$}
        \label{subfig:CT_1}
    \end{subfigure}
    
   \begin{subfigure}[h]{.3\linewidth}
        \begin{tikzpicture}[scale=0.6]
     \begin{feynman}[]
     \vertex [](a) at (0,-0.75);
     \vertex [](p1) at (-2,0);
     \vertex [](p'1) at (2,0);
     \vertex [] (p2) at (-2,-1.5);
     \vertex [] (p'2) at (2,-1.5) ;

     \diagram*{
         (p1) -- [plain] (a) -- [double,double distance=0.1ex,very thick] (p'1) ,
         (p2)  -- [plain](a) -- [double,double distance=0.1ex,very thick] (p'2)
     };
     \end{feynman}
     \end{tikzpicture} 
        \caption*{$CT_{2}$}
        \label{subfig:CT_2}
    \end{subfigure}
    \begin{subfigure}[h]{0.4\linewidth}
        \centering
         \begin{tikzpicture}[scale=0.6]
     \begin{feynman}[]
     \vertex [](a) at (0,0);
     \vertex [] (b) at (0.0, -1.5);
     \vertex [](p1) at (-2,0);
     \vertex [](p'1) at (2,0);
     \vertex [] (p2) at (-2,-1.5);
     \vertex [] (p'2) at (2,-1.5) ;

     \diagram*{
         (p1) -- [plain] (a) -- [double,double distance=0.3ex,very thick] (p'1) ,
         (a) -- [scalar](b) ,
         (p2)  -- [plain](b) -- [double,double distance=0.3ex,very thick] (p'2)
     };
     \end{feynman}
     \end{tikzpicture} 
        \caption*{$O_{2}$}
        \label{subfig:02}
    \end{subfigure}%
    
     \hspace{-0.5em}
  \begin{subfigure}[h]{.3\linewidth}
        \begin{tikzpicture}[scale=0.6]
     \begin{feynman}[]
     \vertex [](a) at (0,-0.75);
     \vertex [](p1) at (-2,0);
     \vertex [](p'1) at (2,0);
     \vertex [] (p2) at (-2,-1.5);
     \vertex [] (p'2) at (2,-1.5) ;

     \diagram*{
         (p1) -- [double,double distance=0.1ex,very thick] (a) -- [double,double distance=0.1ex,very thick] (p'1) ,
         (p2)  -- [plain](a) -- [plain] (p'2)
     };
     \end{feynman}
     \end{tikzpicture} 
        \caption*{$CT_{3}$}
        \label{subfig:CT_3}
    \end{subfigure}
      \hspace{1em}
     \begin{subfigure}[h]{.3\linewidth}
        %
        \begin{tikzpicture}[scale=0.6]
     \begin{feynman}[]
     \vertex [](a) at (0,-0.75);
     \vertex [](p1) at (-2,0);
     \vertex [](p'1) at (2,0);
     \vertex [] (p2) at (-2,-1.5);
     \vertex [] (p'2) at (2,-1.5) ;

     \diagram*{
         (p1) -- [double,double distance=0.1ex,very thick] (a) -- [] (p'1) ,
         (p2)  -- [](a) -- [double,double distance=0.1ex,very thick] (p'2)
     };
     \end{feynman}
     \end{tikzpicture} 
        \caption*{$CT_{4}$}
        \label{subfig:CT_4}
    \end{subfigure}%
    \hspace{1em}
    \begin{subfigure}[h]{.3\linewidth}
         \begin{tikzpicture}[scale=0.6]
     \begin{feynman}[]
     \vertex [](a) at (0,0);
     \vertex [] (b) at (0.0, -1.5);
     \vertex [](p1) at (-2,0);
     \vertex [](p'1) at (2,0);
     \vertex [] (p2) at (-2,-1.5);
     \vertex [] (p'2) at (2,-1.5) ;

     \diagram*{
         (p1) -- [double,double distance=0.3ex,very thick] (a) -- [] (p'1) ,
         (a) -- [scalar](b) ,
         (p2)  -- [](b) -- [double,double distance=0.3ex,very thick] (p'2)
     };
     \end{feynman}
     \end{tikzpicture} 
        \caption*{$O_{4}$}
        \label{subfig:04}
    \end{subfigure}
    

     \hspace{-0.5em}
  \begin{subfigure}[h]{.3\linewidth}
        \begin{tikzpicture}[scale=0.6]
     \begin{feynman}[]
     \vertex [](a) at (0,-0.75);
     \vertex [](p1) at (-2,0);
     \vertex [](p'1) at (2,0);
     \vertex [] (p2) at (-2,-1.5);
     \vertex [] (p'2) at (2,-1.5) ;

     \diagram*{
         (p1) -- [plain] (a) -- [double,double distance=0.1ex,very thick] (p'1) ,
         (p2)  -- [double,double distance=0.1ex,very thick](a) -- [plain] (p'2)
     };
     \end{feynman}
     \end{tikzpicture} 
        \caption*{$CT_{5}$}
        \label{subfig:CT_5}
    \end{subfigure}
      \hspace{1em}
     \begin{subfigure}[h]{.3\linewidth}
        %
        \begin{tikzpicture}[scale=0.6]
     \begin{feynman}[]
     \vertex [](a) at (0,-0.75);
     \vertex [](p1) at (-2,0);
     \vertex [](p'1) at (2,0);
     \vertex [] (p2) at (-2,-1.5);
     \vertex [] (p'2) at (2,-1.5) ;

     \diagram*{
         (p1) -- [plain] (a) -- [] (p'1) ,
         (p2)  -- [double,double distance=0.1ex,very thick](a) -- [double,double distance=0.1ex,very thick] (p'2)
     };
     \end{feynman}
     \end{tikzpicture} 
        \caption*{$CT_{6}$}
        \label{subfig:CT_6}
    \end{subfigure}%
    \hspace{1em}
    \begin{subfigure}[h]{.3\linewidth}
         \begin{tikzpicture}[scale=0.6]
     \begin{feynman}[]
     \vertex [](a) at (0,0);
     \vertex [] (b) at (0.0, -1.5);
     \vertex [](p1) at (-2,0);
     \vertex [](p'1) at (2,0);
     \vertex [] (p2) at (-2,-1.5);
     \vertex [] (p'2) at (2,-1.5) ;

     \diagram*{
         (p1) -- [plain] (a) -- [double,double distance=0.3ex,very thick] (p'1) ,
         (a) -- [scalar](b) ,
         (p2)  -- [double,double distance=0.3ex,very thick](b) -- [] (p'2)
     };
     \end{feynman}
     \end{tikzpicture} 
        \caption*{$O_{5}$}
        \label{subfig:05}
    \end{subfigure}
     
       \begin{subfigure}[h]{.3\linewidth}
         \centering
        \begin{tikzpicture}[scale=0.6]
     \begin{feynman}[]
     \vertex [](a) at (0,-0.75);
     \vertex [](p1) at (-2,0);
     \vertex [](p'1) at (2,0);
     \vertex [] (p2) at (-2,-1.5);
     \vertex [] (p'2) at (2,-1.5) ;

     \diagram*{
         (p1) -- [double,double distance=0.1ex,very thick] (a) -- [double,double distance=0.1ex,very thick] (p'1) ,
         (p2)  -- [plain](a) -- [double,double distance=0.1ex,very thick] (p'2)
     };
     \end{feynman}
     \end{tikzpicture} 
        \caption*{$CT_{7}$}
        \label{subfig:CT_7}
    \end{subfigure}
    \hspace{1em}
    \begin{subfigure}[h]{.3\linewidth}
         \begin{tikzpicture}[scale=0.6]
     \begin{feynman}[]
     \vertex [](a) at (0,0);
     \vertex [] (b) at (0.0, -1.5);
     \vertex [](p1) at (-2,0);
     \vertex [](p'1) at (2,0);
     \vertex [] (p2) at (-2,-1.5);
     \vertex [] (p'2) at (2,-1.5) ;

     \diagram*{
         (p1) -- [double,double distance=0.3ex,very thick] (a) -- [double,double distance=0.3ex,very thick] (p'1) ,
         (a) -- [scalar](b) ,
         (p2)  -- [plain](b) -- [double,double distance=0.3ex,very thick] (p'2)
     };
     \end{feynman}
     \end{tikzpicture} 
        \caption*{$O_{7}$}
        \label{subfig:07}
    \end{subfigure}%
     
      \begin{subfigure}[h]{.3\linewidth}
        %
        \begin{tikzpicture}[scale=0.6]
     \begin{feynman}[]
     \vertex [](a) at (0,-0.75);
     \vertex [](p1) at (-2,0);
     \vertex [](p'1) at (2,0);
     \vertex [] (p2) at (-2,-1.5);
     \vertex [] (p'2) at (2,-1.5) ;

     \diagram*{
         (p1) -- [plain] (a) -- [double,double distance=0.1ex,very thick] (p'1) ,
         (p2)  -- [double,double distance=0.1ex,very thick](a) -- [double,double distance=0.1ex,very thick] (p'2)
     };
     \end{feynman}
     \end{tikzpicture} 
        \caption*{$CT_{8}$}
        \label{subfig:CT_8}
    \end{subfigure}%
    \hspace{1.4em}
    \begin{subfigure}[h]{.3\linewidth}
        %
         \begin{tikzpicture}[scale=0.6]
     \begin{feynman}[]
     \vertex [](a) at (0,0);
     \vertex [] (b) at (0.0, -1.5);
     \vertex [](p1) at (-2,0);
     \vertex [](p'1) at (2,0);
     \vertex [] (p2) at (-2,-1.5);
     \vertex [] (p'2) at (2,-1.5) ;

     \diagram*{
         (p1) -- [plain] (a) -- [double,double distance=0.3ex,very thick] (p'1) ,
         (a) -- [scalar](b) ,
         (p2)  -- [double,double distance=0.3ex,very thick](b) -- [double,double distance=0.3ex,very thick] (p'2)
     };
     \end{feynman}
     \end{tikzpicture} 
        \caption*{$O_{8}$}
        \label{subfig:08}
    \end{subfigure}
    
    \begin{subfigure}[h]{.3\linewidth}
        %
        \begin{tikzpicture}[scale=0.6]
     \begin{feynman}[]
     \vertex [](a) at (0,-0.75);
     \vertex [](p1) at (-2,0);
     \vertex [](p'1) at (2,0);
     \vertex [] (p2) at (-2,-1.5);
     \vertex [] (p'2) at (2,-1.5) ;

     \diagram*{
         (p1) -- [double,double distance=0.1ex,very thick] (a) -- [double,double distance=0.1ex,very thick] (p'1) ,
         (p2)  -- [double,double distance=0.1ex,very thick](a) -- [double,double distance=0.1ex,very thick] (p'2)
     };
     \end{feynman}
     \end{tikzpicture} 
        \caption*{$CT_{9}$}
        \label{subfig:CT_9}
    \end{subfigure}%
    \hspace{1.3em}
    \begin{subfigure}[h]{.3\linewidth}
        %
         \begin{tikzpicture}[scale=0.6]
     \begin{feynman}[]
     \vertex [](a) at (0,0);
     \vertex [] (b) at (0.0, -1.5);
     \vertex [](p1) at (-2,0);
     \vertex [](p'1) at (2,0);
     \vertex [] (p2) at (-2,-1.5);
     \vertex [] (p'2) at (2,-1.5) ;

     \diagram*{
         (p1) -- [double,double distance=0.3ex,very thick] (a) -- [double,double distance=0.3ex,very thick] (p'1) ,
         (a) -- [scalar](b) ,
         (p2)  -- [double,double distance=0.3ex,very thick](b) -- [double,double distance=0.3ex,very thick] (p'2)
     };
     \end{feynman}
     \end{tikzpicture} 
        \caption*{$O_{9}$}
        \label{subfig:09}
    \end{subfigure}
    \caption{LO contributions to the $B^{(*)} \Bar{B}^{(*)} \rightarrow B^{(*)} \Bar{B}^{(*)}$ scattering potential. Here single (double) solid lines denote $B$ ($B^*$) mesons and dashed lines represent pions. 
     \label{fig:LO diagrams for B Bbar--> B Bbar}
    }
\end{figure}
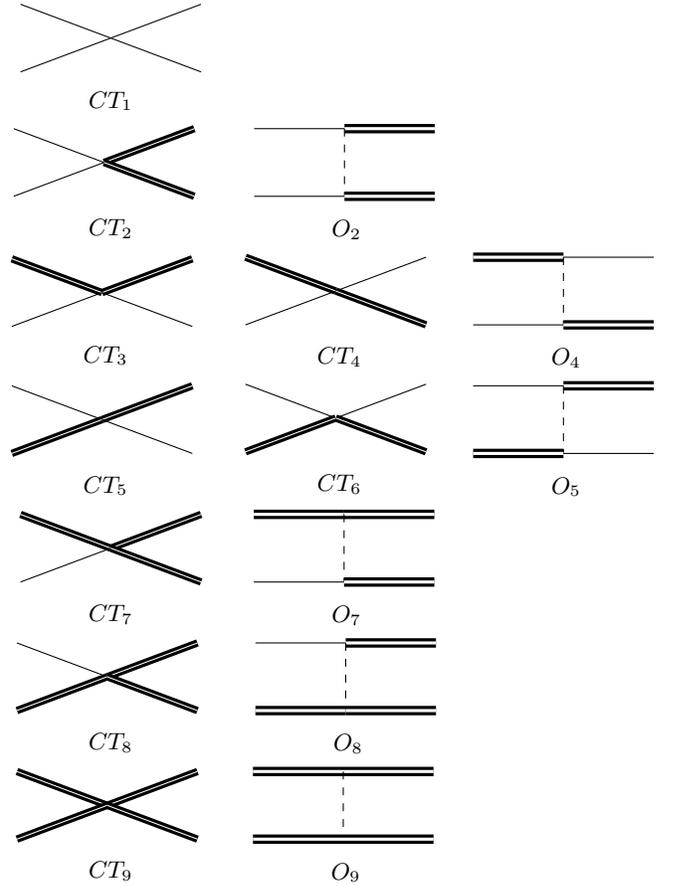

\subsubsection{LO  contact terms }
The relevant contact term (CTs) diagrams at leading order ($\mathcal{O}(\chi^0)$) are shown
in the left column of Fig.~\ref{fig:LO diagrams for B Bbar--> B Bbar}.
The CTs contain the momentum independent terms proportional to the LECs $C$'s  
from the Lagrangian in Eq.~\eqref{eq:lagrangian}
as well as the $S$-$D$ transition term, $D_{12}$,
promoted to leading order as described in the
introduction.

\subsubsection{LO one-pion exchange}

The vertices for the one-pion exchange were derived in section \ref{lag} with $q$ as the pion momentum. 
Since the kinetic energies of the heavy mesons are suppressed
by $p_{\rm typ}/m_B\sim \mathcal{O}(\chi^2)$ relative
to the momenta, we can safely set them to zero here.
At the same time, using that 
$\delta/p_{\rm typ}\sim {\cal O}(\chi)$ from Eq.~\eqref{eq:PC}, 
this implies that the
energy transfer, $q^0= E_p - E_{p'}$, can be
 dropped completely.
Taking all this together, we get
 for the $B^* \Bar{B} \rightarrow B \Bar{B}^*$
 potential
\begin{eqnarray} 
  V_{O_{4}}&=&
    -  \frac{g^2}{ 4 f^2_{\pi} } (\vec{\tau_1} \cdot \vec{\tau_2}^c) (\epsilon^*_{2',n} \epsilon_{1,i})
    \frac{q_iq_n}{\Vec{q}\,^2+m^2_{\pi}} \ ,
\end{eqnarray}
where the external polarisation vectors for incoming and outgoing $B^*$ meson ($\Bar{B}^*$) are denoted as $\epsilon_{1,i}$ and $\epsilon^*_{1',k}$ ($\epsilon_{2,l}$ and $\epsilon^*_{2',n})$, respectively.

The isospin factor for  heavy-meson heavy-antimeson scattering reads 
\begin{equation}
    \vec{\tau_1} \cdot \vec{\tau_2}^c = -\vec{\tau_1} \cdot \vec{\tau_2}=  3-2I(I+1), 
    \label{eq:isospinBBbar}
\end{equation}
which evaluates to 3 for isoscalar and -1 for isotriplet states.

\subsection{Next-to-leading order diagrams}

At next-to-leading-order, $\mathcal{O}(\chi^2)$,
there are momentum-dependent contact interactions  and the  TPE diagrams.
There are three types of TPE diagrams: triangle-diagrams, football-diagrams and box-diagrams.
In the main part of the paper we discuss  general properties of these  diagrams while
the complete expressions 
are provided in the Appendix~\ref{alleffpotentials}.

\begin{widetext}

\begin{figure}[H]
 \begin{subfigure}[h]{.3\linewidth}
        \centering
        \begin{tikzpicture}[scale=0.7]
     \begin{feynman}[]
     \vertex [](a) at (0,-0.75);
     \vertex [](p1) at (-2,0);
     \vertex [](p'1) at (2,0);
     \vertex [] (p2) at (-2,-1.5);
     \vertex [] (p'2) at (2,-1.5) ;

     \diagram*{
         (p1) -- [plain] (a) -- [plain] (p'1) ,
         (p2)  -- [plain](a) -- [plain] (p'2)
     };
     \end{feynman}
     \end{tikzpicture} 
        \caption*{$CT_{1}'$}
        \label{subfig:CT_1N}
    \end{subfigure}
     \begin{subfigure}[h]{.3\linewidth}
        \centering
         \begin{tikzpicture}[scale=0.8]
     \begin{feynman}[]
     \vertex [](a) at (0,0);
     \vertex [] (b) at (0,-1.5);
     \vertex [](p1) at (-2,0);
      \vertex [](p2) at (-2,-1.5);
     \vertex [](p1) at (-2,0);
      \vertex [](p'1) at (2,0);
     \vertex [] (p2) at (-2,-1.5);
     \vertex [] (p'2) at (2,-1.5) ;
      
     \diagram*{
         (p1) -- [plain ] 
         (a) -- [ plain,] (p'1) ,
         (a) -- [scalar,quarter right](b) ,
         (a) -- [scalar,quarter left](b) ,
         (p2)  -- [plain,]
         (b) -- [plain ] (p'2)
     };
     \end{feynman}
     \end{tikzpicture}
        \caption*{$F_{1}$}
        \label{F1}
    \end{subfigure}
     \hspace{-0.5em}
   \begin{subfigure}[h]{.3\linewidth}
        \centering
      \begin{tikzpicture}[scale=.8]
     \begin{feynman}[]
     \vertex [](a) at (0,0);
     \vertex [] (b) at (-0.8, -1.5);
     \vertex [] (c) at (0.8, -1.5);
     \vertex [](p1) at (-2,0);
      \vertex [](p2) at (-2,-1.5);
     \vertex [](p1) at (-2,0);
      \vertex [](p'1) at (2,0);
     \vertex [] (p2) at (-2,-1.5);
     \vertex [] (p'2) at (2,-1.5) ;
      
     \diagram*{
         (p1) -- [plain   ] 
         (a) -- [plain ] (p'1) ,
         (a) -- [scalar](b) ,
         (a) -- [scalar](c) ,
         (p2)  -- [plain]
         (b) -- [double,double distance=0.3ex,very thick] (c),
         (c) -- [plain](p'2)
     };
     \end{feynman}
     \end{tikzpicture}
        \caption*{$T_{1.1}$}
        \label{T1.1}
    \end{subfigure}%
      \hspace{-0.5em}
    \centering
    \begin{subfigure}[h]{.3\linewidth}
        \centering
        \begin{tikzpicture}[scale=0.8]
     \begin{feynman}[]
     \vertex [](a) at (0,0);
     \vertex [] (b) at (-0.8, 1.5);
     \vertex [] (c) at (0.8, 1.5);
     \vertex [](p1) at (-2,1.5);
      \vertex [](p2) at (-2,0);
     \vertex [](p1) at (-2,0);
      \vertex [](p'1) at (2,0);
     \vertex [] (p2) at (-2,1.5);
     \vertex [] (p'2) at (2,1.5) ;
      
     \diagram*{
         (p1) -- [plain  ] 
         (a) -- [ plain] (p'1) ,
         (a) -- [scalar](b) ,
         (a) -- [scalar](c) ,
         (p2)  -- [plain]
         (b) -- [double,double distance=0.3ex,very thick,] (c),
         (c) -- [plain](p'2)
     };
     \end{feynman}
     \end{tikzpicture}
        \caption*{$T_{1.2}$}
        \label{T1.2}
    \end{subfigure}%
    \hspace{-0.5em}
     \begin{subfigure}[h]{.3\linewidth}
        \centering
         \begin{tikzpicture}[scale=0.8]
     \begin{feynman}[]
     \vertex [](a) at (-0.8,0);
      \vertex [](d) at (0.8,0);
     \vertex [] (b) at (-0.8, -1.5);
     \vertex [] (c) at (0.8, -1.5);
     \vertex [](p1) at (-2,0);
      \vertex [](p2) at (-2,-1.5);
     \vertex [](p1) at (-2,0);
      \vertex [](p'1) at (2,0);
     \vertex [] (p2) at (-2,-1.5);
     \vertex [] (p'2) at (2,-1.5) ;

     \diagram*{
         (p1) -- [plain  ] 
         (a) -- [ double,double distance=0.3ex,very thick] (d),
         (d) -- [ plain] (p'1),
         (a) -- [scalar](b) ,
         (d) -- [scalar](c) ,
         (p2)  -- [plain]
         (b) -- [double,double distance=0.3ex,very thick] (c),
         (c) -- [plain](p'2)
     };
     \end{feynman}
     \end{tikzpicture}
        \caption*{$B_{1}$}
        \label{subfig:B1}
    \end{subfigure}
     \hspace{-0.5em}
    \centering
    \begin{subfigure}[h]{.3\linewidth}
        \centering
         \begin{tikzpicture}[scale=.8]
     \begin{feynman}[]
     \vertex [](a) at (-0.8,0);
      \vertex [](d) at (0.8,0);
     \vertex [] (b) at (-0.8, -1.5);
     \vertex [] (c) at (0.8, -1.5);
     \vertex [](p1) at (-2,0);
      \vertex [](p2) at (-2,-1.5);
     \vertex [](p1) at (-2,0);
      \vertex [](p'1) at (2,0);
     \vertex [] (p2) at (-2,-1.5);
     \vertex [] (p'2) at (2,-1.5) ;
     
     \diagram*{
         (p1) -- [plain  ] 
         (a) -- [ double,double distance=0.3ex,very thick] (d),
         (d) -- [ plain] (p'1),
         (a) -- [scalar ](c) ,
         (d) -- [scalar](b) ,
         (p2)  -- [plain]
         (b) -- [double,double distance=0.3ex,very thick] (c),
         (c) -- [plain](p'2)
     };
     \end{feynman}
     \end{tikzpicture}
        \caption*{$C_{1}$}
        \label{subfig:C1}
    \end{subfigure}%
     \hspace{-0.5em}
    \caption{ NLO diagrams for $B \Bar{B} \rightarrow B \Bar{B} $}
    \label{fig:diagrams for B Bbar--.> B Bbar}
\end{figure}
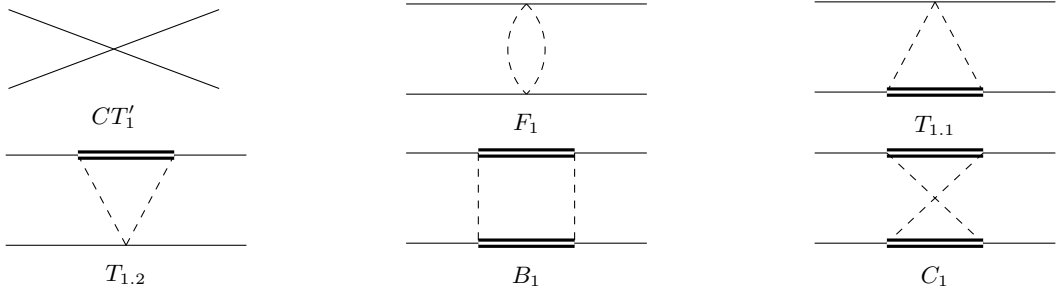
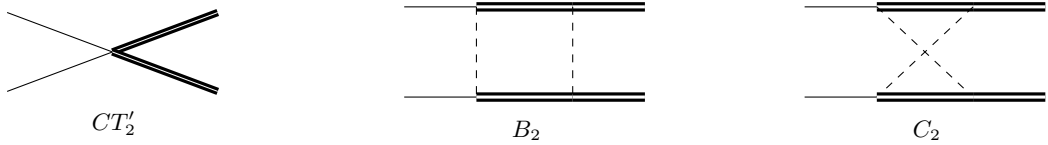
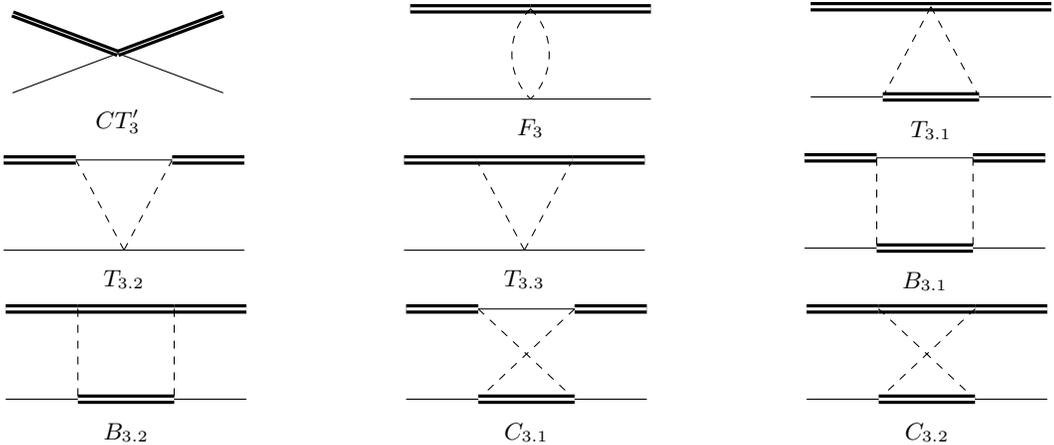

\begin{figure}[H]
\begin{subfigure}[h]{.3\linewidth}
        \centering
        \begin{tikzpicture}[scale=0.7]
     \begin{feynman}[]
     \vertex [](a) at (0,-0.75);
     \vertex [](p1) at (-2,0);
     \vertex [](p'1) at (2,0);
     \vertex [] (p2) at (-2,-1.5);
     \vertex [] (p'2) at (2,-1.5) ;

     \diagram*{
         (p1) -- [plain] (a) -- [double,double distance=0.1ex,very thick] (p'1) ,
         (p2)  -- [plain](a) -- [double,double distance=0.1ex,very thick] (p'2)
     };
     \end{feynman}
     \end{tikzpicture} 
        \caption*{$CT_{2}'$}
        \label{subfig:CT_2N}
    \end{subfigure}
  \centering
  \begin{subfigure}[h]{.3\linewidth}
    \centering
        \begin{tikzpicture}[scale=0.8]
     \begin{feynman}[]
     \vertex [](a) at (-0.8,0);
      \vertex [](d) at (0.8,0);
     \vertex [] (b) at (-0.8, -1.5);
     \vertex [] (c) at (0.8, -1.5);
     \vertex [](p1) at (-2,0);
      \vertex [](p2) at (-2,-1.5);
     \vertex [](p1) at (-2,0);
      \vertex [](p'1) at (2,0);
     \vertex [] (p2) at (-2,-1.5);
     \vertex [] (p'2) at (2,-1.5) ;
     
     \diagram*{
         (p1) -- [ plain  ] 
         (a) -- [double,double distance=0.3ex,very thick, ] (d),
         (d) -- [  double,double distance=0.3ex,very thick,] (p'1),
         (a) -- [scalar,  ](b) ,
         (d) -- [scalar, ](c) ,
         (p2)  -- [plain]
         (b) -- [double,double distance=0.3ex,very thick, ] (c),
         (c) -- [double,double distance=0.3ex,very thick](p'2)
     };
     \end{feynman}
     \end{tikzpicture}
        \caption*{$B_{2}$}
        \label{subfig:$B_{2}$}
    \end{subfigure}%
    \hspace{-0.5em}
    \begin{subfigure}[h]{.3\linewidth}
    \centering
        \begin{tikzpicture}[scale=0.8]
     \begin{feynman}[]
     \vertex [](a) at (-0.8,0);
      \vertex [](d) at (0.8,0);
     \vertex [] (b) at (-0.8, -1.5);
     \vertex [] (c) at (0.8, -1.5);
     \vertex [](p1) at (-2,0);
      \vertex [](p2) at (-2,-1.5);
     \vertex [](p1) at (-2,0);
      \vertex [](p'1) at (2,0);
     \vertex [] (p2) at (-2,-1.5);
     \vertex [] (p'2) at (2,-1.5) ;
      
     \diagram*{
         (p1) -- [plain  ] 
         (a) -- [ double,double distance=0.3ex,very thick, ] (d),
         (d) -- [ double,double distance=0.3ex,very thick,] (p'1),
         (a) -- [scalar, ](c) ,
         (d) -- [scalar,](b) ,
         (p2)  -- [plain]
         (b) -- [double,double distance=0.3ex,very thick, ] (c),
         (c) -- [double,double distance=0.3ex,very thick](p'2)
     };
     \end{feynman}
     \end{tikzpicture}
        \caption*{$C_{2}$}
        \label{subfig:C2}
    \end{subfigure}%
     \hspace{-0.5em}
    \caption{ NLO diagrams for $B \Bar{B} \rightarrow B^* \Bar{B}^* $}
    \label{fig:diagrams for B Bbar--.> B* Bbar*}
\end{figure}
\begin{figure}[H]
 \begin{subfigure}[h]{.3\linewidth}
        \centering
        \begin{tikzpicture}[scale=0.7]
     \begin{feynman}[]
     \vertex [](a) at (0,-0.75);
     \vertex [](p1) at (-2,0);
     \vertex [](p'1) at (2,0);
     \vertex [] (p2) at (-2,-1.5);
     \vertex [] (p'2) at (2,-1.5) ;

     \diagram*{
         (p1) -- [double,double distance=0.1ex,very thick] (a) -- [double,double distance=0.1ex,very thick] (p'1) ,
         (p2)  -- [plain](a) -- [plain] (p'2)
     };
     \end{feynman}
     \end{tikzpicture} 
        \caption*{$CT_{3}'$}
        \label{subfig:CT_3N}
    \end{subfigure}
\begin{subfigure}[h]{.3\linewidth}
        \centering
         \begin{tikzpicture}[scale=0.8]
     \begin{feynman}[]
     \vertex [](a) at (0,0);
     \vertex [] (b) at (0,-1.5);
     \vertex [](p1) at (-2,0);
      \vertex [](p2) at (-2,-1.5);
     \vertex [](p1) at (-2,0);
      \vertex [](p'1) at (2,0);
     \vertex [] (p2) at (-2,-1.5);
     \vertex [] (p'2) at (2,-1.5) ;
      
     \diagram*{
         (p1) -- [double,double distance=0.3ex,very thick, ] 
         (a) -- [ double,double distance=0.3ex,very thick,] (p'1) ,
         (a) -- [scalar,quarter right](b) ,
         (a) -- [scalar,quarter left](b) ,
         (p2)  -- [plain,]
         (b) -- [plain ] (p'2)
     };
     \end{feynman}
     \end{tikzpicture}
        \caption*{$F_{3}$}
        \label{subfig:$F_{3}$}
    \end{subfigure}%
    \hspace{-0.5em}
    \begin{subfigure}[h]{.3\linewidth}
        \centering
        \begin{tikzpicture}[scale=0.8]
          \begin{feynman}[]
     \vertex [](a) at (0,0);
     \vertex [] (b) at (-0.8, -1.5);
     \vertex [] (c) at (0.8, -1.5);
     \vertex [](p1) at (-2,0);
      \vertex [](p2) at (-2,-1.5);
     \vertex [](p1) at (-2,0);
      \vertex [](p'1) at (2,0);
     \vertex [] (p2) at (-2,-1.5);
     \vertex [] (p'2) at (2,-1.5) ;
     
     \diagram*{
         (p1) -- [double,double distance=0.3ex,very thick,  ] 
         (a) -- [ double,double distance=0.3ex,very thick,] (p'1) ,
         (a) -- [scalar, ](b) ,
         (a) -- [scalar, ](c) ,
         (p2)  -- [plain]
         (b) -- [double,double distance=0.3ex,very thick,] (c),
         (c) -- [plain](p'2)
     };
     \end{feynman}
     \end{tikzpicture}
        \caption*{$T_{3.1}$}
        \label{subfig:T3.1}
    \end{subfigure}%
     \hspace{-0.5em}
    \begin{subfigure}[h]{.3\linewidth}
        \centering
        \begin{tikzpicture}[scale=0.8]
     \begin{feynman}[]
     \vertex [](a) at (0,0);
     \vertex [] (b) at (-0.8, 1.5);
     \vertex [] (c) at (0.8, 1.5);
     \vertex [](p1) at (-2,1.5);
      \vertex [](p2) at (-2,0);
     \vertex [](p1) at (-2,0);
      \vertex [](p'1) at (2,0);
     \vertex [] (p2) at (-2,1.5);
     \vertex [] (p'2) at (2,1.5) ;
    
     \diagram*{
         (p1) -- [plain  ] 
         (a) -- [ plain] (p'1) ,
         (a) -- [scalar](b) ,
         (a) -- [scalar, ](c) ,
         (p2)  -- [double,double distance=0.3ex,very thick,]
         (b) -- [plain,] (c),
         (c) -- [double,double distance=0.3ex,very thick,](p'2)
     };
     \end{feynman}
     \end{tikzpicture}
        \caption*{$T_{3.2}$}
        \label{subfig:T3.2}
    \end{subfigure}%
    \hspace{-0.5em}
     \begin{subfigure}[h]{.3\linewidth}
        \centering
        \begin{tikzpicture}[scale=0.8]
     \begin{feynman}[]
     \vertex [](a) at (0,0);
     \vertex [] (b) at (-0.8, 1.5);
     \vertex [] (c) at (0.8, 1.5);
     \vertex [](p1) at (-2,1.5);
      \vertex [](p2) at (-2,0);
     \vertex [](p1) at (-2,0);
      \vertex [](p'1) at (2,0);
     \vertex [] (p2) at (-2,1.5);
     \vertex [] (p'2) at (2,1.5) ;
    
     \diagram*{
         (p1) -- [plain  ] 
         (a) -- [ plain] (p'1) ,
         (a) -- [scalar](b) ,
         (a) -- [scalar, ](c) ,
         (p2)  -- [double,double distance=0.3ex,very thick,]
         (b) -- [double,double distance=0.3ex,very thick,] (c),
         (c) -- [double,double distance=0.3ex,very thick,](p'2)
     };
     \end{feynman}
     \end{tikzpicture}
        \caption*{$T_{3.3}$}
        \label{subfig:T3.3}
    \end{subfigure}%
    \hspace{-0.5em}
     \begin{subfigure}[h]{.3\linewidth}
        \centering
        \begin{tikzpicture}[scale=0.8]
     \begin{feynman}[]
     \vertex [](a) at (-0.8,0);
      \vertex [](d) at (0.8,0);
     \vertex [] (b) at (-0.8, -1.5);
     \vertex [] (c) at (0.8, -1.5);
     \vertex [](p1) at (-2,0);
      \vertex [](p2) at (-2,-1.5);
     \vertex [](p1) at (-2,0);
      \vertex [](p'1) at (2,0);
     \vertex [] (p2) at (-2,-1.5);
     \vertex [] (p'2) at (2,-1.5) ;

     \diagram*{
         (p1) -- [ double,double distance=0.3ex,very thick  ] 
         (a) -- [plain,] (d),
         (d) -- [  double,double distance=0.3ex,very thick] (p'1),
         (a) -- [scalar,  ](b) ,
         (d) -- [scalar,](c) ,
         (p2)  -- [plain]
         (b) -- [double,double distance=0.3ex,very thick, ] (c),
         (c) -- [plain](p'2)
     };
     \end{feynman}
     \end{tikzpicture}
        \caption*{$B_{3.1}$}
        \label{subfig:B3.1}
    \end{subfigure}
     \hspace{-0.5em}
    \centering
    \begin{subfigure}[h]{.3\linewidth}
        \centering
         \begin{tikzpicture}[scale=0.8]
     \begin{feynman}[]
     \vertex [](a) at (-0.8,0);
      \vertex [](d) at (0.8,0);
     \vertex [] (b) at (-0.8, -1.5);
     \vertex [] (c) at (0.8, -1.5);
     \vertex [](p1) at (-2,0);
      \vertex [](p2) at (-2,-1.5);
     \vertex [](p1) at (-2,0);
      \vertex [](p'1) at (2,0);
     \vertex [] (p2) at (-2,-1.5);
     \vertex [] (p'2) at (2,-1.5) ;

     \diagram*{
         (p1) -- [ double,double distance=0.3ex,very thick  ] 
         (a) -- [double,double distance=0.3ex,very thick, ] (d),
         (d) -- [  double,double distance=0.3ex,very thick] (p'1),
         (a) -- [scalar, ](b) ,
         (d) -- [scalar,](c) ,
         (p2)  -- [plain]
         (b) -- [double,double distance=0.3ex,very thick, ] (c),
         (c) -- [plain](p'2)
     };
     \end{feynman}
     \end{tikzpicture}
        \caption*{$B_{3.2}$}
        \label{subfig:B3.2}
    \end{subfigure}%
     \hspace{-0.5em}
    \centering
    \begin{subfigure}[h]{.3\linewidth}
        \centering
         \begin{tikzpicture}[scale=0.8]
     \begin{feynman}[]
     \vertex [](a) at (-0.8,0);
      \vertex [](d) at (0.8,0);
     \vertex [] (b) at (-0.8, -1.5);
     \vertex [] (c) at (0.8, -1.5);
     \vertex [](p1) at (-2,0);
      \vertex [](p2) at (-2,-1.5);
     \vertex [](p1) at (-2,0);
      \vertex [](p'1) at (2,0);
     \vertex [] (p2) at (-2,-1.5);
     \vertex [] (p'2) at (2,-1.5) ;
      
     \diagram*{
         (p1) -- [double,double distance=0.3ex,very thick  ] 
         (a) -- [ plain,] (d),
         (d) -- [ double,double distance=0.3ex,very thick] (p'1),
         (a) -- [scalar,  ](c) ,
         (d) -- [scalar, ](b) ,
         (p2)  -- [plain]
         (b) -- [double,double distance=0.3ex,very thick,] (c),
         (c) -- [plain](p'2)
     };
     \end{feynman}
     \end{tikzpicture}
        \caption*{$C_{3.1}$}
        \label{subfig:C3.1}
    \end{subfigure}%
     \hspace{-0.5em}
     \begin{subfigure}[h]{.3\linewidth}
        \centering
           \begin{tikzpicture}[scale=0.8]
     \begin{feynman}[]
     \vertex [](a) at (-0.8,0);
      \vertex [](d) at (0.8,0);
     \vertex [] (b) at (-0.8, -1.5);
     \vertex [] (c) at (0.8, -1.5);
     \vertex [](p1) at (-2,0);
      \vertex [](p2) at (-2,-1.5);
     \vertex [](p1) at (-2,0);
      \vertex [](p'1) at (2,0);
     \vertex [] (p2) at (-2,-1.5);
     \vertex [] (p'2) at (2,-1.5) ;
     
     \diagram*{
         (p1) -- [double,double distance=0.3ex,very thick  ] 
         (a) -- [ double,double distance=0.3ex,very thick, ] (d),
         (d) -- [ double,double distance=0.3ex,very thick] (p'1),
         (a) -- [scalar, ](c) ,
         (d) -- [scalar, ](b) ,
         (p2)  -- [plain]
         (b) -- [double,double distance=0.3ex,very thick, ] (c),
         (c) -- [plain](p'2)
     };
     \end{feynman}
     \end{tikzpicture}
        \caption*{$C_{3.2}$}
        \label{subfig:C3.2}
    \end{subfigure}
     \hspace{-0.5em}
    \caption{NLO diagrams for $B^* \Bar{B} \rightarrow B^* \Bar{B} $}
    \label{fig:diagrams for B* Bbar--.> B* Bbar}
\end{figure}
\begin{figure}[H]
\centering
   \begin{subfigure}[h]{.3\linewidth}
    \centering
        \begin{tikzpicture}[scale=0.7]
     \begin{feynman}[]
     \vertex [](a) at (0,-0.75);
     \vertex [](p1) at (-2,0);
     \vertex [](p'1) at (2,0);
     \vertex [] (p2) at (-2,-1.5);
     \vertex [] (p'2) at (2,-1.5) ;

     \diagram*{
         (p1) -- [double,double distance=0.1ex,very thick] (a) -- [] (p'1) ,
         (p2)  -- [](a) -- [double,double distance=0.1ex,very thick] (p'2)
     };
     \end{feynman}
     \end{tikzpicture} 
        \caption*{$CT_{4}'$}
        \label{subfig:CT_4N}
    \end{subfigure}
    \begin{subfigure}[h]{.3\linewidth}
    \centering
        \begin{tikzpicture}[scale=0.8]
     \begin{feynman}[]
     \vertex [](a) at (-0.8,0);
      \vertex [](d) at (0.8,0);
     \vertex [] (b) at (-0.8, -1.5);
     \vertex [] (c) at (0.8, -1.5);
     \vertex [](p1) at (-2,0);
      \vertex [](p2) at (-2,-1.5);
     \vertex [](p1) at (-2,0);
      \vertex [](p'1) at (2,0);
     \vertex [] (p2) at (-2,-1.5);
     \vertex [] (p'2) at (2,-1.5) ;
     
     \diagram*{
         (p1) -- [ double,double distance=0.3ex,very thick  ] 
         (a) -- [double,double distance=0.3ex,very thick, ] (d),
         (d) -- [  plain] (p'1),
         (a) -- [scalar,  ](b) ,
         (d) -- [scalar, ](c) ,
         (p2)  -- [plain]
         (b) -- [double,double distance=0.3ex,very thick, ] (c),
         (c) -- [double,double distance=0.3ex,very thick](p'2)
     };
     \end{feynman}
     \end{tikzpicture}
        \caption*{$B_{4}$}
        \label{subfig:$B_{4}$}
    \end{subfigure}%
    \hspace{-0.5em}
    \begin{subfigure}[h]{.3\linewidth}
    \centering
        \begin{tikzpicture}[scale=0.8]
     \begin{feynman}[]
     \vertex [](a) at (-0.8,0);
      \vertex [](d) at (0.8,0);
     \vertex [] (b) at (-0.8, -1.5);
     \vertex [] (c) at (0.8, -1.5);
     \vertex [](p1) at (-2,0);
      \vertex [](p2) at (-2,-1.5);
     \vertex [](p1) at (-2,0);
      \vertex [](p'1) at (2,0);
     \vertex [] (p2) at (-2,-1.5);
     \vertex [] (p'2) at (2,-1.5) ;
      
     \diagram*{
         (p1) -- [double,double distance=0.3ex,very thick  ] 
         (a) -- [ double,double distance=0.3ex,very thick, ] (d),
         (d) -- [ plain] (p'1),
         (a) -- [scalar, ](c) ,
         (d) -- [scalar,](b) ,
         (p2)  -- [plain]
         (b) -- [double,double distance=0.3ex,very thick, ] (c),
         (c) -- [double,double distance=0.3ex,very thick](p'2)
     };
     \end{feynman}
     \end{tikzpicture}
        \caption*{$C_{4}$}
        \label{subfig:C4}
    \end{subfigure}
    \caption{NLO diagrams for $B^* \Bar{B} \rightarrow B \Bar{B}^* $}
    \label{fig:diagrams for B* Bbar--.> B Bbar*}
\end{figure}

\begin{figure}[H]
     \begin{subfigure}[h]{.3\linewidth}
         \centering
        \begin{tikzpicture}[scale=0.7]
     \begin{feynman}[]
     \vertex [](a) at (0,-0.75);
     \vertex [](p1) at (-2,0);
     \vertex [](p'1) at (2,0);
     \vertex [] (p2) at (-2,-1.5);
     \vertex [] (p'2) at (2,-1.5) ;

     \diagram*{
         (p1) -- [double,double distance=0.1ex,very thick] (a) -- [double,double distance=0.1ex,very thick] (p'1) ,
         (p2)  -- [plain](a) -- [double,double distance=0.1ex,very thick] (p'2)
     };
     \end{feynman}
     \end{tikzpicture} 
        \caption*{$CT_{5}'$}
        \label{subfig:CT_5N}
    \end{subfigure}
    \begin{subfigure}[h]{.3\linewidth}
        \centering
        \begin{tikzpicture}[scale=0.8]
          \begin{feynman}[]
     \vertex [](a) at (0,0);
     \vertex [] (b) at (-0.8, -1.5);
     \vertex [] (c) at (0.8, -1.5);
     \vertex [](p1) at (-2,0);
      \vertex [](p2) at (-2,-1.5);
     \vertex [](p1) at (-2,0);
      \vertex [](p'1) at (2,0);
     \vertex [] (p2) at (-2,-1.5);
     \vertex [] (p'2) at (2,-1.5) ;
     
     \diagram*{
         (p1) -- [double,double distance=0.3ex,very thick,  ] 
         (a) -- [ double,double distance=0.3ex,very thick,] (p'1) ,
         (a) -- [scalar, ](b) ,
         (a) -- [scalar, ](c) ,
         (p2)  -- [plain]
         (b) -- [double,double distance=0.3ex,very thick,] (c),
         (c) -- [double,double distance=0.3ex,very thick](p'2)
     };
     \end{feynman}
     \end{tikzpicture}
        \caption*{$T_{5}$}
        \label{subfig:T5}
    \end{subfigure}%
     \hspace{-0.5em}
     \begin{subfigure}[h]{.3\linewidth}
        \centering
        \begin{tikzpicture}[scale=0.8]
     \begin{feynman}[]
     \vertex [](a) at (-0.8,0);
      \vertex [](d) at (0.8,0);
     \vertex [] (b) at (-0.8, -1.5);
     \vertex [] (c) at (0.8, -1.5);
     \vertex [](p1) at (-2,0);
      \vertex [](p2) at (-2,-1.5);
     \vertex [](p1) at (-2,0);
      \vertex [](p'1) at (2,0);
     \vertex [] (p2) at (-2,-1.5);
     \vertex [] (p'2) at (2,-1.5) ;

     \diagram*{
         (p1) -- [ double,double distance=0.3ex,very thick  ] 
         (a) -- [plain,] (d),
         (d) -- [  double,double distance=0.3ex,very thick] (p'1),
         (a) -- [scalar,  ](b) ,
         (d) -- [scalar,](c) ,
         (p2)  -- [plain]
         (b) -- [double,double distance=0.3ex,very thick, ] (c),
         (c) -- [double,double distance=0.3ex,very thick](p'2)
     };
     \end{feynman}
     \end{tikzpicture}
        \caption*{$B_{5.1}$}
        \label{subfig:B5.1}
    \end{subfigure}
     \hspace{-0.5em}
    \centering
    \begin{subfigure}[h]{.3\linewidth}
        \centering
         \begin{tikzpicture}[scale=0.8]
     \begin{feynman}[]
     \vertex [](a) at (-0.8,0);
      \vertex [](d) at (0.8,0);
     \vertex [] (b) at (-0.8, -1.5);
     \vertex [] (c) at (0.8, -1.5);
     \vertex [](p1) at (-2,0);
      \vertex [](p2) at (-2,-1.5);
     \vertex [](p1) at (-2,0);
      \vertex [](p'1) at (2,0);
     \vertex [] (p2) at (-2,-1.5);
     \vertex [] (p'2) at (2,-1.5) ;

     \diagram*{
         (p1) -- [ double,double distance=0.3ex,very thick  ] 
         (a) -- [double,double distance=0.3ex,very thick, ] (d),
         (d) -- [  double,double distance=0.3ex,very thick] (p'1),
         (a) -- [scalar, ](b) ,
         (d) -- [scalar,](c) ,
         (p2)  -- [plain]
         (b) -- [double,double distance=0.3ex,very thick, ] (c),
         (c) -- [double,double distance=0.3ex,very thick](p'2)
     };
     \end{feynman}
     \end{tikzpicture}
        \caption*{$B_{5.2}$}
        \label{subfig:B5.2}
    \end{subfigure}%
     \hspace{-0.5em}
    \centering
    \begin{subfigure}[h]{.3\linewidth}
        \centering
         \begin{tikzpicture}[scale=0.8]
     \begin{feynman}[]
     \vertex [](a) at (-0.8,0);
      \vertex [](d) at (0.8,0);
     \vertex [] (b) at (-0.8, -1.5);
     \vertex [] (c) at (0.8, -1.5);
     \vertex [](p1) at (-2,0);
      \vertex [](p2) at (-2,-1.5);
     \vertex [](p1) at (-2,0);
      \vertex [](p'1) at (2,0);
     \vertex [] (p2) at (-2,-1.5);
     \vertex [] (p'2) at (2,-1.5) ;
      
     \diagram*{
         (p1) -- [double,double distance=0.3ex,very thick  ] 
         (a) -- [ plain,] (d),
         (d) -- [ double,double distance=0.3ex,very thick] (p'1),
         (a) -- [scalar,  ](c) ,
         (d) -- [scalar, ](b) ,
         (p2)  -- [plain]
         (b) -- [double,double distance=0.3ex,very thick,] (c),
         (c) -- [double,double distance=0.3ex,very thick](p'2)
     };
     \end{feynman}
     \end{tikzpicture}
        \caption*{$C_{5.1}$}
        \label{subfig:C5.1}
    \end{subfigure}%
     \hspace{-0.5em}
     \begin{subfigure}[h]{.3\linewidth}
        \centering
           \begin{tikzpicture}[scale=0.8]
     \begin{feynman}[]
     \vertex [](a) at (-0.8,0);
      \vertex [](d) at (0.8,0);
     \vertex [] (b) at (-0.8, -1.5);
     \vertex [] (c) at (0.8, -1.5);
     \vertex [](p1) at (-2,0);
      \vertex [](p2) at (-2,-1.5);
     \vertex [](p1) at (-2,0);
      \vertex [](p'1) at (2,0);
     \vertex [] (p2) at (-2,-1.5);
     \vertex [] (p'2) at (2,-1.5) ;
     
     \diagram*{
         (p1) -- [double,double distance=0.3ex,very thick  ] 
         (a) -- [ double,double distance=0.3ex,very thick, ] (d),
         (d) -- [ double,double distance=0.3ex,very thick] (p'1),
         (a) -- [scalar, ](c) ,
         (d) -- [scalar, ](b) ,
         (p2)  -- [plain]
         (b) -- [double,double distance=0.3ex,very thick, ] (c),
         (c) -- [double,double distance=0.3ex,very thick](p'2)
     };
     \end{feynman}
     \end{tikzpicture}
        \caption*{$C_{5.2}$}
        \label{subfig:C5.2}
    \end{subfigure}
    \caption{NLO diagrams for $B^* \Bar{B} \rightarrow B^* \Bar{B}^* $}
    \label{fig:diagrams for B* Bbar--.> B* Bbar*}
\end{figure}

\begin{figure}[H]
   \begin{subfigure}[h]{.3\linewidth}
    \centering
        \begin{tikzpicture}[scale=0.7]
     \begin{feynman}[]
     \vertex [](a) at (0,-0.75);
     \vertex [](p1) at (-2,0);
     \vertex [](p'1) at (2,0);
     \vertex [] (p2) at (-2,-1.5);
     \vertex [] (p'2) at (2,-1.5) ;

     \diagram*{
         (p1) -- [plain] (a) -- [double,double distance=0.1ex,very thick] (p'1) ,
         (p2)  -- [double,double distance=0.1ex,very thick](a) -- [double,double distance=0.1ex,very thick] (p'2)
     };
     \end{feynman}
     \end{tikzpicture} 
        \caption*{$CT_{6}'$}
        \label{subfig:CT_6N}
    \end{subfigure}
    \begin{subfigure}[h]{.3\linewidth}
        \centering
        \begin{tikzpicture}[scale=0.8]
     \begin{feynman}[]
     \vertex [](a) at (0,0);
     \vertex [] (b) at (-0.8, 1.5);
     \vertex [] (c) at (0.8, 1.5);
     \vertex [](p1) at (-2,1.5);
      \vertex [](p2) at (-2,0);
     \vertex [](p1) at (-2,0);
      \vertex [](p'1) at (2,0);
     \vertex [] (p2) at (-2,1.5);
     \vertex [] (p'2) at (2,1.5) ;
    
     \diagram*{
         (p1) -- [double,double distance=0.3ex,very thick  ] 
         (a) -- [ double,double distance=0.3ex,very thick] (p'1) ,
         (a) -- [scalar](b) ,
         (a) -- [scalar, ](c) ,
         (p2)  -- [plain]
         (b) -- [double,double distance=0.3ex,very thick,] (c),
         (c) -- [double,double distance=0.3ex,very thick,](p'2)
     };
     \end{feynman}
     \end{tikzpicture}
        \caption*{$T_{6}$}
        \label{subfig:T6}
    \end{subfigure}%
     \hspace{-0.5em}
     \begin{subfigure}[h]{.3\linewidth}
        \centering
        \begin{tikzpicture}[scale=0.8]
     \begin{feynman}[]
     \vertex [](a) at (-0.8,0);
      \vertex [](d) at (0.8,0);
     \vertex [] (b) at (-0.8, -1.5);
     \vertex [] (c) at (0.8, -1.5);
     \vertex [](p1) at (-2,0);
      \vertex [](p2) at (-2,-1.5);
     \vertex [](p1) at (-2,0);
      \vertex [](p'1) at (2,0);
     \vertex [] (p2) at (-2,-1.5);
     \vertex [] (p'2) at (2,-1.5) ;

     \diagram*{
         (p1) -- [ plain  ] 
         (a) -- [double,double distance=0.3ex,very thick,] (d),
         (d) -- [  double,double distance=0.3ex,very thick] (p'1),
         (a) -- [scalar,  ](b) ,
         (d) -- [scalar,](c) ,
         (p2)  -- [double,double distance=0.3ex,very thick]
         (b) -- [plain, ] (c),
         (c) -- [double,double distance=0.3ex,very thick](p'2)
     };
     \end{feynman}
     \end{tikzpicture}
        \caption*{$B_{6.1}$}
        \label{subfig:B6.1}
    \end{subfigure}
     \hspace{-0.5em}
    \centering
    \begin{subfigure}[h]{.3\linewidth}
        \centering
         \begin{tikzpicture}[scale=0.8]
     \begin{feynman}[]
     \vertex [](a) at (-0.8,0);
      \vertex [](d) at (0.8,0);
     \vertex [] (b) at (-0.8, -1.5);
     \vertex [] (c) at (0.8, -1.5);
     \vertex [](p1) at (-2,0);
      \vertex [](p2) at (-2,-1.5);
     \vertex [](p1) at (-2,0);
      \vertex [](p'1) at (2,0);
     \vertex [] (p2) at (-2,-1.5);
     \vertex [] (p'2) at (2,-1.5) ;

     \diagram*{
         (p1) -- [ plain  ] 
         (a) -- [double,double distance=0.3ex,very thick, ] (d),
         (d) -- [  double,double distance=0.3ex,very thick] (p'1),
         (a) -- [scalar, ](b) ,
         (d) -- [scalar,](c) ,
         (p2)  -- [double,double distance=0.3ex,very thick]
         (b) -- [double,double distance=0.3ex,very thick, ] (c),
         (c) -- [double,double distance=0.3ex,very thick](p'2)
     };
     \end{feynman}
     \end{tikzpicture}
        \caption*{$B_{6.2}$}
        \label{subfig:B6.2}
    \end{subfigure}%
     \hspace{-0.5em}
    \centering
    \begin{subfigure}[h]{.3\linewidth}
        \centering
         \begin{tikzpicture}[scale=0.8]
     \begin{feynman}[]
     \vertex [](a) at (-0.8,0);
      \vertex [](d) at (0.8,0);
     \vertex [] (b) at (-0.8, -1.5);
     \vertex [] (c) at (0.8, -1.5);
     \vertex [](p1) at (-2,0);
      \vertex [](p2) at (-2,-1.5);
     \vertex [](p1) at (-2,0);
      \vertex [](p'1) at (2,0);
     \vertex [] (p2) at (-2,-1.5);
     \vertex [] (p'2) at (2,-1.5) ;
      
     \diagram*{
         (p1) -- [plain  ] 
         (a) -- [ double,double distance=0.3ex,very thick,] (d),
         (d) -- [ double,double distance=0.3ex,very thick] (p'1),
         (a) -- [scalar,  ](c) ,
         (d) -- [scalar, ](b) ,
         (p2)  -- [double,double distance=0.3ex,very thick]
         (b) -- [plain,] (c),
         (c) -- [double,double distance=0.3ex,very thick](p'2)
     };
     \end{feynman}
     \end{tikzpicture}
        \caption*{$C_{6.1}$}
        \label{subfig:C6.1}
    \end{subfigure}%
     \begin{subfigure}[h]{.3\linewidth}
        \centering
           \begin{tikzpicture}[scale=0.8]
     \begin{feynman}[]
     \vertex [](a) at (-0.8,0);
      \vertex [](d) at (0.8,0);
     \vertex [] (b) at (-0.8, -1.5);
     \vertex [] (c) at (0.8, -1.5);
     \vertex [](p1) at (-2,0);
      \vertex [](p2) at (-2,-1.5);
     \vertex [](p1) at (-2,0);
      \vertex [](p'1) at (2,0);
     \vertex [] (p2) at (-2,-1.5);
     \vertex [] (p'2) at (2,-1.5) ;
     
     \diagram*{
         (p1) -- [plain  ] 
         (a) -- [ double,double distance=0.3ex,very thick, ] (d),
         (d) -- [ double,double distance=0.3ex,very thick] (p'1),
         (a) -- [scalar, ](c) ,
         (d) -- [scalar, ](b) ,
         (p2)  -- [double,double distance=0.3ex,very thick]
         (b) -- [double,double distance=0.3ex,very thick, ] (c),
         (c) -- [double,double distance=0.3ex,very thick](p'2)
     };
     \end{feynman}
     \end{tikzpicture}
        \caption*{$C_{6.2}$}
        \label{subfig:C6.2}
    \end{subfigure}
    \caption{NLO diagrams for $B \Bar{B}^* \rightarrow B^* \Bar{B}^* $}
    \label{fig:diagrams for B Bbar*--.> B* Bbar*}
\end{figure}

\begin{figure}[H]
\begin{subfigure}[h]{.3\linewidth}
    \centering
        \begin{tikzpicture}[scale=0.7]
     \begin{feynman}[]
     \vertex [](a) at (0,-0.75);
     \vertex [](p1) at (-2,0);
     \vertex [](p'1) at (2,0);
     \vertex [] (p2) at (-2,-1.5);
     \vertex [] (p'2) at (2,-1.5) ;

     \diagram*{
         (p1) -- [double,double distance=0.1ex,very thick] (a) -- [double,double distance=0.1ex,very thick] (p'1) ,
         (p2)  -- [double,double distance=0.1ex,very thick](a) -- [double,double distance=0.1ex,very thick] (p'2)
     };
     \end{feynman}
     \end{tikzpicture} 
        \caption*{$CT_{7}'$}
        \label{subfig:CT_7N}
    \end{subfigure}
 \begin{subfigure}[h]{.3\linewidth}
        \centering
         \begin{tikzpicture}[scale=0.8]
     \begin{feynman}[]
     \vertex [](a) at (0,0);
     \vertex [] (b) at (0,-1.5);
     \vertex [](p1) at (-2,0);
      \vertex [](p2) at (-2,-1.5);
     \vertex [](p1) at (-2,0);
      \vertex [](p'1) at (2,0);
     \vertex [] (p2) at (-2,-1.5);
     \vertex [] (p'2) at (2,-1.5) ;
      
     \diagram*{
         (p1) -- [double,double distance=0.3ex,very thick, ] 
         (a) -- [ double,double distance=0.3ex,very thick,] (p'1) ,
         (a) -- [scalar,quarter right](b) ,
         (a) -- [scalar,quarter left](b) ,
         (p2)  -- [double,double distance=0.3ex,very thick]
         (b) -- [double,double distance=0.3ex,very thick ] (p'2)
     };
     \end{feynman}
     \end{tikzpicture}
        \caption*{$F_{7}$}
        \label{subfig:$F_{7}$}
    \end{subfigure}%
    \hspace{-0.5em}
    \begin{subfigure}[h]{.3\linewidth}
        \centering
        \begin{tikzpicture}[scale=0.8]
          \begin{feynman}[]
     \vertex [](a) at (0,0);
     \vertex [] (b) at (-0.8, -1.5);
     \vertex [] (c) at (0.8, -1.5);
     \vertex [](p1) at (-2,0);
      \vertex [](p2) at (-2,-1.5);
     \vertex [](p1) at (-2,0);
      \vertex [](p'1) at (2,0);
     \vertex [] (p2) at (-2,-1.5);
     \vertex [] (p'2) at (2,-1.5) ;
     
     \diagram*{
         (p1) -- [double,double distance=0.3ex,very thick,  ] 
         (a) -- [ double,double distance=0.3ex,very thick,] (p'1) ,
         (a) -- [scalar, ](b) ,
         (a) -- [scalar, ](c) ,
         (p2)  -- [double,double distance=0.3ex,very thick,]
         (b) -- [plain] (c),
         (c) -- [double,double distance=0.3ex,very thick,](p'2)
     };
     \end{feynman}
     \end{tikzpicture}
        \caption*{$T_{7.1}$}
        \label{subfig:T7.1}
    \end{subfigure}%
     \hspace{-0.5em}
    \begin{subfigure}[h]{.3\linewidth}
        \centering
        \begin{tikzpicture}[scale=0.8]
     \begin{feynman}[]
     \vertex [](a) at (0,0);
     \vertex [] (b) at (-0.8, 1.5);
     \vertex [] (c) at (0.8, 1.5);
     \vertex [](p1) at (-2,1.5);
      \vertex [](p2) at (-2,0);
     \vertex [](p1) at (-2,0);
      \vertex [](p'1) at (2,0);
     \vertex [] (p2) at (-2,1.5);
     \vertex [] (p'2) at (2,1.5) ;
    
     \diagram*{
         (p1) -- [double,double distance=0.3ex,very thick  ] 
         (a) -- [ double,double distance=0.3ex,very thick] (p'1) ,
         (a) -- [scalar](b) ,
         (a) -- [scalar, ](c) ,
         (p2)  -- [double,double distance=0.3ex,very thick,]
         (b) -- [plain,] (c),
         (c) -- [double,double distance=0.3ex,very thick,](p'2)
     };
     \end{feynman}
     \end{tikzpicture}
        \caption*{$T_{7.2}$}
        \label{subfig:T7.2}
    \end{subfigure}%
    \hspace{-0.5em}
     \begin{subfigure}[h]{.3\linewidth}
        \centering
        \begin{tikzpicture}[scale=0.8]
          \begin{feynman}[]
     \vertex [](a) at (0,0);
     \vertex [] (b) at (-0.8, -1.5);
     \vertex [] (c) at (0.8, -1.5);
     \vertex [](p1) at (-2,0);
      \vertex [](p2) at (-2,-1.5);
     \vertex [](p1) at (-2,0);
      \vertex [](p'1) at (2,0);
     \vertex [] (p2) at (-2,-1.5);
     \vertex [] (p'2) at (2,-1.5) ;
     
     \diagram*{
         (p1) -- [double,double distance=0.3ex,very thick,  ] 
         (a) -- [ double,double distance=0.3ex,very thick,] (p'1) ,
         (a) -- [scalar, ](b) ,
         (a) -- [scalar, ](c) ,
         (p2)  -- [double,double distance=0.3ex,very thick,]
         (b) -- [double,double distance=0.3ex,very thick] (c),
         (c) -- [double,double distance=0.3ex,very thick,](p'2)
     };
     \end{feynman}
     \end{tikzpicture}
        \caption*{$T_{7.3}$}
        \label{subfig:T7.3}
    \end{subfigure}%
     \hspace{-0.5em}
      \begin{subfigure}[h]{.3\linewidth}
        \centering
        \begin{tikzpicture}[scale=0.8]
     \begin{feynman}[]
     \vertex [](a) at (0,0);
     \vertex [] (b) at (-0.8, 1.5);
     \vertex [] (c) at (0.8, 1.5);
     \vertex [](p1) at (-2,1.5);
      \vertex [](p2) at (-2,0);
     \vertex [](p1) at (-2,0);
      \vertex [](p'1) at (2,0);
     \vertex [] (p2) at (-2,1.5);
     \vertex [] (p'2) at (2,1.5) ;
    
     \diagram*{
         (p1) -- [double,double distance=0.3ex,very thick  ] 
         (a) -- [ double,double distance=0.3ex,very thick] (p'1) ,
         (a) -- [scalar](b) ,
         (a) -- [scalar, ](c) ,
         (p2)  -- [double,double distance=0.3ex,very thick,]
         (b) -- [double,double distance=0.3ex,very thick,] (c),
         (c) -- [double,double distance=0.3ex,very thick,](p'2)
     };
     \end{feynman}
     \end{tikzpicture}
        \caption*{$T_{7.4}$}
        \label{subfig:T7.4}
    \end{subfigure}%
    \hspace{-0.5em}
     \begin{subfigure}[h]{.3\linewidth}
        \centering
        \begin{tikzpicture}[scale=0.8]
     \begin{feynman}[]
     \vertex [](a) at (-0.8,0);
      \vertex [](d) at (0.8,0);
     \vertex [] (b) at (-0.8, -1.5);
     \vertex [] (c) at (0.8, -1.5);
     \vertex [](p1) at (-2,0);
      \vertex [](p2) at (-2,-1.5);
     \vertex [](p1) at (-2,0);
      \vertex [](p'1) at (2,0);
     \vertex [] (p2) at (-2,-1.5);
     \vertex [] (p'2) at (2,-1.5) ;

     \diagram*{
         (p1) -- [ double,double distance=0.3ex,very thick  ] 
         (a) -- [plain,] (d),
         (d) -- [  double,double distance=0.3ex,very thick] (p'1),
         (a) -- [scalar,  ](b) ,
         (d) -- [scalar,](c) ,
         (p2)  -- [double,double distance=0.3ex,very thick,]
         (b) -- [plain ] (c),
         (c) -- [double,double distance=0.3ex,very thick,](p'2)
     };
     \end{feynman}
     \end{tikzpicture}
        \caption*{$B_{7.1}$}
        \label{subfig:B7.1}
    \end{subfigure}
     \hspace{-0.5em}
    \centering
    \begin{subfigure}[h]{.3\linewidth}
        \centering
         \begin{tikzpicture}[scale=0.8]
     \begin{feynman}[]
     \vertex [](a) at (-0.8,0);
      \vertex [](d) at (0.8,0);
     \vertex [] (b) at (-0.8, -1.5);
     \vertex [] (c) at (0.8, -1.5);
     \vertex [](p1) at (-2,0);
      \vertex [](p2) at (-2,-1.5);
     \vertex [](p1) at (-2,0);
      \vertex [](p'1) at (2,0);
     \vertex [] (p2) at (-2,-1.5);
     \vertex [] (p'2) at (2,-1.5) ;

     \diagram*{
         (p1) -- [ double,double distance=0.3ex,very thick  ] 
         (a) -- [double,double distance=0.3ex,very thick, ] (d),
         (d) -- [  double,double distance=0.3ex,very thick] (p'1),
         (a) -- [scalar, ](b) ,
         (d) -- [scalar,](c) ,
         (p2)  -- [double,double distance=0.3ex,very thick]
         (b) -- [plain ] (c),
         (c) -- [double,double distance=0.3ex,very thick](p'2)
     };
     \end{feynman}
     \end{tikzpicture}
        \caption*{$B_{7.2}$}
        \label{subfig:B7.2}
    \end{subfigure}%
     \hspace{-0.5em}
    \centering
     \begin{subfigure}[h]{.3\linewidth}
        \centering
        \begin{tikzpicture}[scale=0.8]
     \begin{feynman}[]
     \vertex [](a) at (-0.8,0);
      \vertex [](d) at (0.8,0);
     \vertex [] (b) at (-0.8, -1.5);
     \vertex [] (c) at (0.8, -1.5);
     \vertex [](p1) at (-2,0);
      \vertex [](p2) at (-2,-1.5);
     \vertex [](p1) at (-2,0);
      \vertex [](p'1) at (2,0);
     \vertex [] (p2) at (-2,-1.5);
     \vertex [] (p'2) at (2,-1.5) ;

     \diagram*{
         (p1) -- [ double,double distance=0.3ex,very thick  ] 
         (a) -- [plain,] (d),
         (d) -- [  double,double distance=0.3ex,very thick] (p'1),
         (a) -- [scalar,  ](b) ,
         (d) -- [scalar,](c) ,
         (p2)  -- [double,double distance=0.3ex,very thick,]
         (b) -- [double,double distance=0.3ex,very thick ] (c),
         (c) -- [double,double distance=0.3ex,very thick,](p'2)
     };
     \end{feynman}
     \end{tikzpicture}
        \caption*{$B_{7.3}$}
        \label{subfig:B7.3}
    \end{subfigure}
     \hspace{-0.5em}
     \begin{subfigure}[h]{.3\linewidth}
        \centering
         \begin{tikzpicture}[scale=0.8]
     \begin{feynman}[]
     \vertex [](a) at (-0.8,0);
      \vertex [](d) at (0.8,0);
     \vertex [] (b) at (-0.8, -1.5);
     \vertex [] (c) at (0.8, -1.5);
     \vertex [](p1) at (-2,0);
      \vertex [](p2) at (-2,-1.5);
     \vertex [](p1) at (-2,0);
      \vertex [](p'1) at (2,0);
     \vertex [] (p2) at (-2,-1.5);
     \vertex [] (p'2) at (2,-1.5) ;

     \diagram*{
         (p1) -- [ double,double distance=0.3ex,very thick  ] 
         (a) -- [double,double distance=0.3ex,very thick, ] (d),
         (d) -- [  double,double distance=0.3ex,very thick] (p'1),
         (a) -- [scalar, ](b) ,
         (d) -- [scalar,](c) ,
         (p2)  -- [double,double distance=0.3ex,very thick]
         (b) -- [double,double distance=0.3ex,very thick ] (c),
         (c) -- [double,double distance=0.3ex,very thick](p'2)
     };
     \end{feynman}
     \end{tikzpicture}
        \caption*{$B_{7.4}$}
        \label{subfig:B7.4}
    \end{subfigure}%
     \hspace{-0.5em}
    \begin{subfigure}[h]{.3\linewidth}
        \centering
         \begin{tikzpicture}[scale=0.8]
     \begin{feynman}[]
     \vertex [](a) at (-0.8,0);
      \vertex [](d) at (0.8,0);
     \vertex [] (b) at (-0.8, -1.5);
     \vertex [] (c) at (0.8, -1.5);
     \vertex [](p1) at (-2,0);
      \vertex [](p2) at (-2,-1.5);
     \vertex [](p1) at (-2,0);
      \vertex [](p'1) at (2,0);
     \vertex [] (p2) at (-2,-1.5);
     \vertex [] (p'2) at (2,-1.5) ;
      
     \diagram*{
         (p1) -- [double,double distance=0.3ex,very thick  ] 
         (a) -- [ plain,] (d),
         (d) -- [ double,double distance=0.3ex,very thick] (p'1),
         (a) -- [scalar,  ](c) ,
         (d) -- [scalar, ](b) ,
         (p2)  -- [double,double distance=0.3ex,very thick]
         (b) -- [plain,] (c),
         (c) -- [double,double distance=0.3ex,very thick](p'2)
     };
     \end{feynman}
     \end{tikzpicture}
        \caption*{$C_{7.1}$}
        \label{subfig:C7.1}
    \end{subfigure}%
     \hspace{-0.5em}
     \begin{subfigure}[h]{.3\linewidth}
        \centering
           \begin{tikzpicture}[scale=0.8]
     \begin{feynman}[]
     \vertex [](a) at (-0.8,0);
      \vertex [](d) at (0.8,0);
     \vertex [] (b) at (-0.8, -1.5);
     \vertex [] (c) at (0.8, -1.5);
     \vertex [](p1) at (-2,0);
      \vertex [](p2) at (-2,-1.5);
     \vertex [](p1) at (-2,0);
      \vertex [](p'1) at (2,0);
     \vertex [] (p2) at (-2,-1.5);
     \vertex [] (p'2) at (2,-1.5) ;
     
     \diagram*{
         (p1) -- [double,double distance=0.3ex,very thick  ] 
         (a) -- [ double,double distance=0.3ex,very thick, ] (d),
         (d) -- [ double,double distance=0.3ex,very thick] (p'1),
         (a) -- [scalar, ](c) ,
         (d) -- [scalar](b) ,
         (p2)  -- [double,double distance=0.3ex,very thick]
         (b) -- [plain ] (c),
         (c) -- [double,double distance=0.3ex,very thick](p'2)
     };
     \end{feynman}
     \end{tikzpicture}
        \caption*{$C_{7.2}$}
        \label{subfig:C7.2}
    \end{subfigure}%
     \hspace{-0.5em}
        \begin{subfigure}[h]{.3\linewidth}
        \centering
         \begin{tikzpicture}[scale=0.8]
     \begin{feynman}[]
     \vertex [](a) at (-0.8,0);
      \vertex [](d) at (0.8,0);
     \vertex [] (b) at (-0.8, -1.5);
     \vertex [] (c) at (0.8, -1.5);
     \vertex [](p1) at (-2,0);
      \vertex [](p2) at (-2,-1.5);
     \vertex [](p1) at (-2,0);
      \vertex [](p'1) at (2,0);
     \vertex [] (p2) at (-2,-1.5);
     \vertex [] (p'2) at (2,-1.5) ;
      
     \diagram*{
         (p1) -- [double,double distance=0.3ex,very thick  ] 
         (a) -- [ plain,] (d),
         (d) -- [ double,double distance=0.3ex,very thick] (p'1),
         (a) -- [scalar,  ](c) ,
         (d) -- [scalar, ](b) ,
         (p2)  -- [double,double distance=0.3ex,very thick]
         (b) -- [double,double distance=0.3ex,very thick,] (c),
         (c) -- [double,double distance=0.3ex,very thick](p'2)
     };
     \end{feynman}
     \end{tikzpicture}
        \caption*{$C_{7.3}$}
        \label{subfig:C7.3}
    \end{subfigure}%
     \hspace{-0.5em}
     \begin{subfigure}[h]{.3\linewidth}
        \centering
           \begin{tikzpicture}[scale=0.8]
     \begin{feynman}[]
     \vertex [](a) at (-0.8,0);
      \vertex [](d) at (0.8,0);
     \vertex [] (b) at (-0.8, -1.5);
     \vertex [] (c) at (0.8, -1.5);
     \vertex [](p1) at (-2,0);
      \vertex [](p2) at (-2,-1.5);
     \vertex [](p1) at (-2,0);
      \vertex [](p'1) at (2,0);
     \vertex [] (p2) at (-2,-1.5);
     \vertex [] (p'2) at (2,-1.5) ;
     
     \diagram*{
         (p1) -- [double,double distance=0.3ex,very thick  ] 
         (a) -- [ double,double distance=0.3ex,very thick, ] (d),
         (d) -- [ double,double distance=0.3ex,very thick] (p'1),
         (a) -- [scalar, ](c) ,
         (d) -- [scalar](b) ,
         (p2)  -- [double,double distance=0.3ex,very thick]
         (b) -- [ double,double distance=0.3ex,very thick ] (c),
         (c) -- [double,double distance=0.3ex,very thick](p'2)
     };
     \end{feynman}
     \end{tikzpicture}
        \caption*{$C_{7.4}$}
        \label{subfig:C7.4}
    \end{subfigure}%
     \hspace{-0.5em}
    \caption{NLO diagrams for $B^* \Bar{B}^* \rightarrow B^* \Bar{B}^* $}
    \label{fig:diagrams for B* Bbar*--.> B* Bbar*}
\end{figure}

\begin{figure}[H]
   \begin{subfigure}[h]{.3\linewidth}
        \centering
      \begin{tikzpicture}[scale=.8]
     \begin{feynman}[]
     \vertex [](a) at (0,0);
     \vertex [] (b) at (-0.8, -1.5);
     \vertex [] (c) at (0.8, -1.5);
     \vertex [](p1) at (-2,0);
      \vertex [](p2) at (-2,-1.5);
     \vertex [](p1) at (-2,0);
      \vertex [](p'1) at (2,0);
     \vertex [] (p2) at (-2,-1.5);
     \vertex [] (p'2) at (2,-1.5) ;
      
     \diagram*{
         (p1) -- [plain   ] 
         (a) -- [plain ] (p'1) ,
         (a) -- [scalar](b) ,
         (a) -- [scalar](c) ,
         (p2)  -- [double,double distance=0.3ex,very thick]
         (b) -- [double,double distance=0.3ex,very thick] (c),
         (c) -- [plain](p'2)
     };
     \end{feynman}
     \end{tikzpicture}
        \caption*{$T_{8}$}
        \label{T8}
    \end{subfigure}%
      \hspace{-0.5em}
     \begin{subfigure}[h]{.3\linewidth}
        \centering
         \begin{tikzpicture}[scale=0.8]
     \begin{feynman}[]
     \vertex [](a) at (-0.8,0);
      \vertex [](d) at (0.8,0);
     \vertex [] (b) at (-0.8, -1.5);
     \vertex [] (c) at (0.8, -1.5);
     \vertex [](p1) at (-2,0);
      \vertex [](p2) at (-2,-1.5);
     \vertex [](p1) at (-2,0);
      \vertex [](p'1) at (2,0);
     \vertex [] (p2) at (-2,-1.5);
     \vertex [] (p'2) at (2,-1.5) ;

     \diagram*{
         (p1) -- [plain  ] 
         (a) -- [ double,double distance=0.3ex,very thick] (d),
         (d) -- [ plain] (p'1),
         (a) -- [scalar](b) ,
         (d) -- [scalar](c) ,
         (p2)  -- [double,double distance=0.3ex,very thick]
         (b) -- [double,double distance=0.3ex,very thick] (c),
         (c) -- [plain](p'2)
     };
     \end{feynman}
     \end{tikzpicture}
        \caption*{$B_{8}$}
        \label{subfig:B8}
    \end{subfigure}
     \hspace{-0.5em}
    \centering
    \begin{subfigure}[h]{.3\linewidth}
        \centering
         \begin{tikzpicture}[scale=.8]
     \begin{feynman}[]
     \vertex [](a) at (-0.8,0);
      \vertex [](d) at (0.8,0);
     \vertex [] (b) at (-0.8, -1.5);
     \vertex [] (c) at (0.8, -1.5);
     \vertex [](p1) at (-2,0);
      \vertex [](p2) at (-2,-1.5);
     \vertex [](p1) at (-2,0);
      \vertex [](p'1) at (2,0);
     \vertex [] (p2) at (-2,-1.5);
     \vertex [] (p'2) at (2,-1.5) ;
     
     \diagram*{
         (p1) -- [plain  ] 
         (a) -- [ double,double distance=0.3ex,very thick] (d),
         (d) -- [ plain] (p'1),
         (a) -- [scalar ](c) ,
         (d) -- [scalar](b) ,
         (p2)  -- [double,double distance=0.3ex,very thick]
         (b) -- [double,double distance=0.3ex,very thick] (c),
         (c) -- [plain](p'2)
     };
     \end{feynman}
     \end{tikzpicture}
        \caption*{$C_{8}$}
        \label{subfig:C8}
    \end{subfigure}%
     \hspace{-0.5em}
    \caption{NLO diagrams for $B \Bar{B}^* \rightarrow B \Bar{B} $}
    \label{fig:diagrams for B Bbar*--.> B Bbar}
\end{figure}

\end{widetext}

\subsubsection{Contact interactions at NLO}
The relevant CTs at NLO,  $\mathcal{O}(\chi^2)$,  are the two momentum-dependent terms proportional to $D_{10}$ and $D_{11}$, as seen in the Lagrangian in Eq.~\eqref{eq:lagrangian}. 
The chiral expansion formally also generates momentum independent subleading contact terms proportional to $m_\pi^2$, which in the standard 
power counting would appear at the same order. In the momentum counting scheme imposed here,
however, those are suppressed by $(m_\pi/p_{\rm typ})^2{\sim}\chi^2$ and thus start to contribute only  at $\mathcal{O}(\chi^4)$. 
 In addition, for a fixed pion mass, 
the $m_\pi^2$-dependent CT's only lead to a redefinition of the LO CT's.

\subsubsection{Triangle and Football diagrams}
\label{triandfb}
For the $B^{(*)} \Bar{B}^{(*)}$ case, the sum of all  triangle diagrams vanishes.
In this section we demonstrate this explicitly for the
$B \Bar{B} \rightarrow B \Bar{B}$ channel, however, the same pattern applies
to all the other potentials analogously. 
For the $B \Bar{B} \rightarrow B \Bar{B}$ potential, as shown in   Fig.~\ref{fig:diagrams for B Bbar--.> B Bbar}, we have two triangle diagrams denoted as $T_{1.1}$ and $T_{1.2}$. 
The potential from the first diagram is (The labels on the potentials given below refer to those in the  related figures), 
 \begin{widetext}
     \begin{multline}
    i  V_{T_{1.1}}=  \sum_\lambda\int \frac{d^4l}{(2\pi)^4}
    \frac{1}{4 f_{\pi}^2}  \Big( 2l_0  \epsilon_{cdh} (\tau_1)_h
    \Big)  \frac{i}{l^2 - m^2_{\pi}}  \,
    \frac{i}{(-l_0 +i \epsilon )}  \\   \times 
      \Bigg( \frac{g}{2 f_{\pi}}  \Big( \epsilon_{j} (\lambda) (-l)_j \Big) (\tau^c_2)_d \Bigg)
      \Bigg( \frac{g}{2 f_{\pi}}  \Big( \epsilon^*_{i} (\lambda) (l+q)_i \Big) (\tau^c_2)_c \Bigg)
   \frac{i}{(l+q)^2 - m^2_{\pi}} \ ,
 \end{multline}
 \end{widetext}
The potential can thus be written as
\begin{equation}
     V_{T_{1.1}}= \frac{g^2}{4 f^4_{\pi}} 
     (\Vec{\tau_1} \cdot  \Vec{\tau_2}) I_{tr} \ ,
\end{equation}
where we used $ \sum_\lambda   \epsilon_{i}^*(\lambda) \epsilon_{j}(\lambda) = \delta_{ij}$. The analytic expression for the integral $I_{tr}$ is
provided in Eq.~(\ref{eq:Itreval}).
The second diagram
simplifies to
\begin{equation}
    V_{T_{1.2}}= -\frac{g^2}{4 f^4_{\pi}} (\Vec{\tau_1} \cdot  \Vec{\tau_2}) I_{tr} \ ,
\end{equation}
where the change in sign resulted from the appearance of the
charge conjugate isospin matrix at the lower, $\pi \bar B\to \pi\bar B$ vertex.
 The total contribution therefore cancels. 
 If, instead, we  calculated the $BB$ potential,
 $V_{T_{1.2}}$ would appear with a positive sign,
and  the two triangle contributions would be summed.

The potential for the football diagram reads
\begin{equation}
     V_{F_1}= \frac{1 }{2 f^4_{\pi}}(\Vec{\tau_1} \cdot  \Vec{\tau_2}) I_{fb}
\end{equation}
where $I_{fb}$ is given by,
\begin{equation}
    I_{fb}= i \int  \frac{d^4l}{(2\pi)^4}  \frac{(l^0)^2}{\big[(l{+}q)^2 {-}m^2_{\pi} {+}i \epsilon\big] \big[l^2{-}m^2_{\pi} {+}i \epsilon\big]} \ .
\end{equation}
The evaluation of this integral  is provided  in the Appendix \ref{sec:football}.
A closed form expression for this integral at the order $\chi^2$ reads
\begin{equation}
   I_{fb} 
=\frac{\vec{q}\,^2}{96 \pi^2} 
\Bigg\{ \frac{\mathcal{R}}{2}{-}
\frac{5}{6}{+} \ln{\bigg(\frac{m_{\pi}}{\mu}\bigg)} 
{+} L(q) \Bigg\}+{\cal O}(\chi^4).
\end{equation}

\subsubsection{Box diagrams}
\label{box}
The box diagrams are made from four pion--heavy-meson vertices contracted
with two pion propagators. These form either a planar box or a crossed box. 
To avoid double counting,  only
the irreducible two-heavy meson part of the planar box is kept in the potential,
as the reducible parts are generated through iterations in the Lippmann-Schwinger equation.
Again, for illustration,  the $B \Bar{B} \rightarrow B \Bar{B}$ case
is discussed explicitly.
For the planar box $B_1$ in Fig.~\ref{fig:diagrams for B Bbar--.> B Bbar} one finds
\begin{widetext}
    \begin{multline}
  i  V_{B_{1}}= \sum_{\lambda_1, \lambda_2 }\int \frac{d^4l}{(2\pi)^4} 
   \Bigg[ \frac{g}{2 f_{\pi}}  \Big( \epsilon_{i} (\lambda_1) (-q_2)_{i} \Big) (\tau_1)_d \Bigg]
   \frac{i}{-l_0 +i \epsilon }
   \Bigg[ \frac{g}{2 f_{\pi}} \Big( \epsilon^*_{j} (\lambda_1) (q_1)_{j} \Big) (\tau_1)_c \Bigg]
  \frac{i}{q_1^2 - m^2_{\pi} } \\
   \Bigg[ \frac{g}{2 f_{\pi}}  \Big( \epsilon_{n} (\lambda_2) (q_2)_n \Big) (\tau^c_2)_d \Bigg]  \frac{i}{q_2^2 - m^2_{\pi} } 
  \Bigg[ \frac{g}{2 f_{\pi}}  \Big( \epsilon^*_{m} (\lambda_2) (-q_1)_m \Big) (\tau^c_2)_c \Bigg]
  \frac{i}{l_0 +i \epsilon }
\end{multline}
\end{widetext}
Here, $q_1$ and $q_2$ are the momenta of the two pions and, to the order we are working,  one finds $q_1 {=} (l_0, \vec{p}{-}\Vec{l})$, $q_2{=} (l_0, \vec{p'}{-}\Vec{l})$ and $q{=}(0,\Vec{p'}{-}\Vec{p}){=}  (0, \Vec{q}) $,  where $l$ is the loop momentum.
This can be written as
 \begin{equation}
 V_{B_{1}} = \frac{g^4}{16 f^4_{\pi}}   \Big( 3 - 2(\Vec{\tau_1} \cdot  \Vec{\tau_2}) \Big)   \,  I^{(1)}_{box}
 \end{equation}
where we used $(\tau_1)_d   (\tau_1)_c  (\tau_2)_d (\tau_2)_c = \Big( 3 - 2(\Vec{\tau_1} \cdot  \Vec{\tau_2}) \Big)$  and 
\begin{equation}\label{Eq:Ibox1}
    I^{(1)}_{box} = i \int \frac{d^4l}{(2\pi)^4}\! \frac{(\vec{q_2} \cdot \vec{q_1})^2}{\big[l^0{+}i\epsilon\big] \big[l^0{-}i\epsilon\big] \big[q_2^2 {-}m^2_{\pi}\big] \big[q_1^2{-}m^2_{\pi}\big]  } \ .
\end{equation}
For the crossed box ($C_1$ in Fig.~\ref{fig:diagrams for B Bbar--.> B Bbar}) the simplified expression reads
\begin{equation}
  V_{C_{1}}= 
  \frac{g^4}{16 f^4_{\pi}}   \Big(-3 - 2(\Vec{\tau_1} \cdot  \Vec{\tau_2}) \Big)  \, I^{(2)}_{box}
\end{equation}
where, we used $(\tau_1)_d   (\tau_1)_c  (\tau_2)_c (\tau_2)_d =  3 + 2(\Vec{\tau_1} \cdot  \Vec{\tau_2}) $, 
and $I^{(2)}_{box}$ is given by,
\begin{equation}\label{Eq:Ibox2}
    I^{(2)}_{box} = i \int \frac{d^4l}{(2\pi)^4} \frac{(\vec{q_2} \cdot \vec{q_1})^2}{[l_0{-}i\epsilon]^2 \big[q_2^2 {-}m^2_{\pi}\big] \big[q_1^2{-}m^2_{\pi}\big]  } \ . 
\end{equation}
The only difference between $I^{(1)}_{box}$ and $I^{(2)}_{box}$ is the sign
in the $i\epsilon$ term in one of the heavy meson propagators.
Just because of this sign change the former integral contains
a reducible contribution, while the latter does not.
In Ref.~\cite{Machleidt:2011} and Appendix~\ref{sec:planarbox} it is shown that the irreducible contribution of the integral $ I^{(1)}_{box}$ is identical to $I^{(2)}_{box}$.
Accordingly, 
the resulting irreducible contribution of the box diagrams to the effective potential for $B \Bar{B} \rightarrow B \Bar{B}$ reads
\begin{equation}
 V^{box}_{B \Bar{B} \rightarrow B \Bar{B}}= V_{B_{1}} + V_{C_{1}}   =  -\frac{g^4}{4 f^4_{\pi}}  (\Vec{\tau_1} \cdot  \Vec{\tau_2})  \,  I^{(2)}_{box}
\end{equation}
The derivation of this integral is provided in Appendix~\ref{Sec:I2box}, the final result at ${\cal O}(\chi^2)$ reads 
\begin{equation}
I^{(2)}_{box} 
=   \frac{-23\vec{q}\,^2}{96 \pi^2} \Bigg\{ \frac{\mathcal{R}}{2} {+} \frac{5}{138}{+} \ln{\bigg(\frac{m_{\pi}}{\mu}\bigg)} {+} L(q)\Bigg\}
{+}{\cal O}(\chi^4).
\label{eq:Ibox}
\end{equation}

The effective potentials for the OPE and TPE in the case of all possible transitions for  $B^{(*)} \Bar{B}^{(*)} \rightarrow B^{(*)} \Bar{B}^{(*)}$ scatterings are listed in the Appendix \ref{alleffpotentials}.

\subsection{ $ {B^{(*)} B^{(*)} \to B^{(*)} B^{(*)}}$ scattering}
\label{BBpots}
The same topologies as discussed above also contribute to $B^{(*)} {B}^{(*)}$ scattering. 
However, in contrast to $\bar B$, for $B$ mesons there is no  charge-conjugated Pauli matrix ($\boldsymbol{\tau^c}= \boldsymbol{-\tau}$). It is evident that this change does not affect the box diagrams as there is an even number of pion-emission vertices on each heavy meson line. However, this sign change is important for the one-pion exchange, the football diagrams and also some triangle diagrams, with an odd number of vertices in the antimeson line. For example, for elastic $BB^*$ scattering  these changes will affect  diagrams $O_4$ and $O_5$ in Fig.~\ref{fig:LO diagrams for B Bbar--> B Bbar} and diagrams $F_3$,  $T_{3.2}$ and   $T_{3.3}$  in Fig.~\ref{fig:diagrams for B* Bbar--.> B* Bbar}.  As a consequence of this, for $B$-mesons  the triangle diagrams do not cancel.  
Additionally, we note that the isospin coefficients for scattering of two heavy mesons are generally different from those of one heavy meson one heavy antimeson, because of the differences in their isospin wave functions. In particular,   the $\vec \tau_1\cdot\vec \tau_2$  matrix element  for a heavy meson-antimeson system is given by Eq.~\eqref{eq:isospinBBbar},  resulting in  -3  for $I=0$ and +1 for $I=1$.  The same values are obtained for isospin matrix elements in elastic transitions between two identical bosons, namely $PP\to PP$ and $VV\to VV$, where $P$ and $V$ represent pseudoscalar and vector mesons, respectively. However, for the $PV$ system, the matrix element depends on the topology, namely on whether the initial $PV$ transform to the final $VP$ (u-channel) or $PV$ state (t-channel)
\bea\label{Eq:iso_u}  
\langle (PV)_I|\vec \tau_1\cdot \vec \tau_2 |(VP)_I\rangle_{\mathrm{u-channel}}&=& \begin{cases} 3\ ;\ {I\!=\!0} \\ 1\ ; \ {I\!=\!1}\end{cases}\\\label{Eq:iso_t}
\langle (PV)_I|\vec \tau_1\cdot \vec \tau_2|(PV)_I\rangle_{\mathrm{t-channel}}&=& \begin{cases} -3\ ;\ {I\!=\!0} \\ 1\ ; \ {I\!=\!1,}\end{cases} 
\eea 
Specifically, the isoscalar transition in the u-channel has a different sign, as compared to all other isoscalar transitions. 

The explicit expressions for the $B^{(*)} B^{(*)} \rightarrow B^{(*)} B^{(*)}$ potentials are provided in Appendix \ref{alleffpotentials}.

\section{Partial wave Decomposition}
\label{sec:pwd}
In this section, the effective potentials from the earlier section are decomposed into four channels
$J^{PC}= 0^{++},1^{++}, 1^{+-}, 2^{++}$. For those we chose the
following bases~\cite{Baru:2019xnh}:

\begin{eqnarray}\nonumber
    0^{++} &:&  \alpha,\beta=  
\Big\{ B \Bar{B}(^1S_0), B^* \Bar{B}^*(^1S_0), B^* \Bar{B}^*(^5D_0)     \Big\}\\
\nonumber
1^{++} &:&  \alpha,\beta= \Big\{ B \Bar{B}^*(^3 S_1,+), B \Bar{B}^*(^3 D_1,+), B^* \Bar{B}^*(^5D_1)     \Big\}\\
\nonumber
1^{+-} &:& \alpha,\beta= \Big\{ B \Bar{B}^*(^3 S_1,-), B \Bar{B}^*(^3 D_1,-), \\ \nonumber
& &\qquad \qquad \qquad \qquad \qquad \ \ B^* \Bar{B}^*(^3 S_1), B^* \Bar{B}^*(^3 D_1)    \Big\}\\
\nonumber
2^{++} &:& \alpha,\beta= \Big\{ B \Bar{B}(^1 D_2), B \Bar{B}^*(^3 D_2), B^* \Bar{B}^*(^5 S_2), \\ & &  \qquad B^* \Bar{B}^*(^1 D_2), B^* \Bar{B}^*(^5 D_2), B^* \Bar{B}^*(^5 G_2)    \Big\} \ .
\label{eq:basisstates}
\end{eqnarray}
The individual partial waves are labeled as $^{2S+1}L_{J}$ with $S$, $L$ and $J$ denoting the total spin, the angular momentum and the total momentum, respectively.
The sign in parentheses for $B\Bar{B}^*$ states corresponds to their $C$-parity which is given by
\begin{equation}
 |{B \Bar{B}^*, \pm}\rangle =  \frac{1}{\sqrt{2}} \big( |B \Bar{B}^*\rangle \pm |B^* \Bar{B}\rangle \big) \ .
\end{equation}
The partial wave decomposition (PWD) of the potentials is done using the formalism of Ref.~\cite{Baru:2019xnh}, which gives

 \begin{multline}
    \begin{split}
         V_{\alpha \beta}(J^{PC})  &= \frac{1}{2J+1} \int \frac{d \Omega_n}{4 \pi} \frac{d \Omega_{n'}}{4 \pi} \, \\
         &\hspace{2cm}\times \mathrm{Tr} \big[ P^{\dagger} (\alpha, \vec{n})V  P(\beta, \vec{n'} ) \big]
    \end{split}
 \end{multline}
where  $\vec{n}= \vec{p}/|\vec{p}|$ ,  $\vec{n'}= \vec{p'}/|\vec{p'}|$, $P^{\dagger} (\alpha, \vec{n})$ and $ P(\beta, \vec{n'} )$ are outgoing and incoming  normalised projectors respectively with $\alpha $ and $\beta$ being the bases states mentioned in Eq.~(\ref{eq:basisstates}) and finally,  $V$ are the potentials calculated earlier. Due to the spacial symmetry of this $2\rightarrow2$ reaction, only the angle
$\theta$ between the incoming and outgoing momentum needs to be considered and we use the notation
$x=\cos{(\theta)}$. 
Then the above expression simplifies to
\begin{equation}
     V_{\alpha \beta} \big(J^{PC}\big) {=} \frac{1}{2J{+}1} \int_{{-}1}^{{+}1} \frac{dx}{2} \, \mathrm{Tr} \big[ P^{\dagger} (\alpha, \vec{n'}) \,V \, P(\beta, \vec{n} ) \big]\ ,
 \end{equation}
 with the trace taken over the spin indices. The partial wave projectors used to calculate the potentials are 
 provided in Ref.~\cite{Baru:2019xnh} and repeated in Appendix \ref{PWP}
 to keep the presentation self-contained. 
Using the effective potentials derived here, the  scattering amplitudes can be obtained as a solution of the partial-wave decomposed coupled-channel Lippmann-Schwinger equation 
\bea
\label{eq:lse}
T_{\alpha\beta}(E,p,p') &=& V_{\alpha\beta}(E,p,p')\\
& & \hspace{-2cm} - \int_0^\Lambda\frac{\text{d}{ q}\, q^2}{2\pi^2}
V_{\alpha\gamma}(E,p,q)G(E,q)T_{\gamma\beta}(E,q,p'),\nonumber
\eea
where the sharp cutoff $\Lambda$ is introduced to render the integral equation well defined. Alternatively, the potentials in the Lippmann-Schwinger equation can be regularized using the method proposed in the NN sector in Ref.~\cite{Reinert:2017usi}, which explicitly preserves long-range interactions (see also Ref.~\cite{Dong:2021lkh} for related work in the double-quarkonium sector). We expect both methods to yield similar results and defer explicit checks to future work. 

Further, the two-body propagator in the channel $\gamma$ from the list given in Eq.~\eqref{eq:basisstates} is
\bea
G_\gamma=(q^2/(2\mu_\gamma)+m_{1,\gamma}+m_{2,\gamma}-\sqrt{s}-i \epsilon )^{-1},
\eea
where $\sqrt{s}$ is the  total energy and the reduced mass of two heavy mesons with the masses $m_{1,\gamma}$ and $m_{2,\gamma}$ in 
the channel $\gamma$ reads
\bea
\mu_\gamma = \frac{m_{1,\gamma} m_{2,\gamma}}{m_{1,\gamma}+m_{2,\gamma}}.
\eea
 
 The partial wave projected potentials for $B^{(*)} \Bar{B}^{(*)} \rightarrow B^{(*)} \Bar{B}^{(*)}$ in the $J^{PC}=0^{++}$ channel are presented here and the partial wave projected potentials for the rest of the channels ($J^{PC}= 1^{++}, 1^{+-}, 2^{++}$) are listed in Appendix \ref{PWDforrest}.
The pertinent integrals used  in PWD are  abbreviated as $Q(p',p)$, $R(p',p)$, $S(p',p)$. Their explicit expressions are given in the Appendix \ref{PWDint}.
We denote $|\vec{p'}|= p'$ and $|\vec{p}|= p$.

\newpage
\begin{widetext}

\subsection{ \texorpdfstring{$\bm{B^{(*)} \bar{B}^{(*)} \rightarrow B^{(*)} \bar{B}^{(*)}}$}{}}
\subsubsection{$\mathbf{J^{PC}=0^{++}}$}
\label{Sec:0++}
\begin{equation}\label{Eq:VCT0++}
   V^{0^{++},I}_{\rm CT} =  
    \begin{pmatrix} 
	   \mathcal{C}_d^I+ \frac{1}{2} \mathcal{C}_f^I +( \mathcal{D}_d^I + \frac{1}{2} \mathcal{D}_f^I) (p^2+p'^2)   & \frac{1}{2}\sqrt{3} ( \mathcal{C}_f^I +  \mathcal{D}_f^I (p^2+p'^2)) & -\sqrt{3} \mathcal{D}_{SD}^I p'^2 \\
		\frac{1}{2}\sqrt{3} ( \mathcal{C}_f^I +  \mathcal{D}_f^I (p^2+p'^2)) &  \mathcal{C}_d^I - \frac{1}{2} \mathcal{C}_f^I +( \mathcal{D}_d^I - \frac{1}{2} \mathcal{D}_f^I) (p^2+p'^2)  & -\mathcal{D}_{SD}^I p'^2  \\
		-\sqrt{3} \mathcal{D}_{SD}^I p^2 & -\mathcal{D}_{SD}^I p^2 & 0
	\end{pmatrix},
\end{equation}
where the index $I$ stands for isospin and the parameters are convenient linear combinations of the Lagrangian parameters of Eq.~(\ref{eq:lagrangian}).

\begin{equation}
 V_{\rm OPE}^{0^{++}}=   - \, \frac{g^2}{4 f^2_{\pi}} (\vec{\tau_1}\cdot \vec{\tau_2})
    \begin{pmatrix} 
	 0 & \frac{1}{\sqrt{3}} Q_2 & \frac{1}{\sqrt{6}} (Q_2-3Q_{n'})  \\
		\frac{1}{\sqrt{3}} Q_2 & -\frac{2}{3} Q_2 & \frac{1}{\sqrt{2}}(\frac{Q_2}{3}-Q_{n'})  \\
		\frac{1}{\sqrt{6}} (Q_2-3Q_{n}) & \frac{1}{\sqrt{2}}(\frac{Q_2}{3}-Q_{n}) & \frac{3}{2} (5Q_2-6Q_n-6Q_{n'}+18Q_x-9Q_{x^2})
	\end{pmatrix}
\end{equation}
\begin{equation}
 V_{\rm TPE}^{0^{++}} =   \frac{ 1}{16 \pi^2 f^4_{\pi}} 
    \begin{pmatrix} 
	 (V_{\rm TPE}^{0^{++}})_{11} &  \frac{\sqrt{3}}{4} \, g^4 (V_{\rm TPE}^{0^{++}})_{12}  &  \frac{3 \sqrt{3}}{8 \sqrt{2}}\, g^4 (V_{\rm TPE}^{0^{++}})_{13}  \\
    \frac{\sqrt{3}}{4} \, g^4 (V_{\rm TPE}^{0^{++}})_{21} & (V_{\rm TPE}^{0^{++}})_{22}  &  \frac{-1}{16\sqrt{2}} \,g^4(V_{\rm TPE}^{0^{++}})_{23}  \\
		\frac{3 \sqrt{3}}{8 \sqrt{2}} \, g^4(V_{\rm TPE}^{0^{++}})_{31}  &   \frac{- 1}{16\sqrt{2}}\,g^4 (V_{\rm TPE}^{0^{++}})_{32} & (V_{\rm TPE}^{0^{++}})_{33}
	\end{pmatrix}
\end{equation}
where,
\begin{equation}
    (V_{\rm TPE}^{0^{++}})_{11}= \Bar{S}_0(p',p)
\end{equation}

\begin{equation}\label{Eq:V0pp12}
    (V_{\rm TPE}^{0^{++}})_{12}= (V_{\rm TPE}^{0^{++}})_{21}= (p'\,^2+ p^2) \bigg[-\mathcal{R}+1- 2 \ln{\bigg(\frac{m_{\pi}}{\mu}\bigg)} \bigg] -  2 R_2(p',p)
\end{equation}
\begin{equation}
     (V_{\rm TPE}^{0^{++}})_{13}=\frac{2}{3}(p'\,^2) \bigg[-\mathcal{R}+1- 2 \ln{\bigg(\frac{m_{\pi}}{\mu}\bigg)} \bigg] -2 R_{n'}(p',p)+\frac{2}{3} R_2(p',p)
\end{equation}

\begin{equation}
     (V_{\rm TPE}^{0^{++}})_{22}   = \Bar{S}_0 
     +  \frac{g^4}{4} \Bigg\{ 
     2 \mathcal{R}(p'\,^2+ p^2)  -2 (p'\,^2+ p^2) 
     +\big[ 4 (p'\,^2+ p^2) \big] \ln{\bigg(\frac{m_{\pi}}{\mu}\bigg)} +4 R_2(p',p) \Bigg\}      
\end{equation}

\begin{equation}
     (V_{\rm TPE}^{0^{++}})_{23} = 2p'\,^2  \bigg[2 \mathcal{R} -2 + 4 \ln{\bigg(\frac{m_{\pi}}{\mu}\bigg)} \bigg]  + 4 \bigg\{ 3 R_{n'}(p',p)- R_2(p',p)  \bigg\}  
\end{equation}
\begin{equation}
     (V_{\rm TPE}^{0^{++}})_{31}=\frac{2}{3}(p^2) \bigg[-\mathcal{R}+1- 2 \ln{\bigg(\frac{m_{\pi}}{\mu}\bigg)} \bigg] -2 R_{n}(p',p)+\frac{2}{3} R_2(p',p)
\end{equation}

\begin{equation}
     (V_{\rm TPE}^{0^{++}})_{32} = 2p^2  \bigg[2 \mathcal{R}-2 + 4\ln{\bigg(\frac{m_{\pi}}{\mu}\bigg)} \bigg]  +4 \bigg\{ 3 R_{n}(p',p)- R_2(p',p)  \bigg\}  
\end{equation}
\begin{multline}
    (V_{\rm TPE}^{0^{++}})_{33}= \Bar{S}_2+ \frac{g^4}{4} \Bigg\{  \frac{-15}{64p'p} \bigg[ \frac{\rho^4}{4}  -\rho^4 L(\rho) \bigg]_{p'-p}^{p'+p}  - \frac{45}{8}R_{x^2}(p',p) 
   - \frac{21}{8} \bigg( R_{n} (p',p)+ R_{n'}(p',p) 
   \\  -\frac{1}{2} R_2(p',p)
    - 3 R_{x}(p',p) \bigg)  \Bigg\}
\end{multline}
and,
\begin{multline}
     \Bar{S}_0 (p',p)
     = \int^{1}_{-1} \frac{dx}{2} \big(\vec{\tau_1}\cdot \vec{\tau_2} \big) \vec{q}\,^2 \Bigg\{  
     \mathcal{R} \bigg[\frac{23}{48}   g^4   
     + \frac{1}{24}  \bigg] 
    +  \bigg(\frac{5}{144}   g^4 - \frac{5}{72}  \bigg)  
     + \ln{\bigg(\frac{m_{\pi}}{\mu}\bigg)}  \bigg( \frac{23}{24} g^4 + \frac{1}{12} \bigg)
      + L(q)   \bigg( \frac{23}{24} g^4 + \frac{1}{12} \bigg) \Bigg\} \\
     = \big(\vec{\tau_1}\cdot \vec{\tau_2} \big)\Bigg\{ \mathcal{R}  ({p'}^2 +p^2) \bigg( \frac{23}{48}g^4  +\frac{1}{24}\bigg)  
  +({p'}^2 +p^2) \bigg( \frac{5}{144}g^4  -\frac{5}{72}\bigg)    
+({p'}^2 +p^2) \bigg( \frac{23}{24}g^4 +\frac{1}{12}\bigg) 
\ln{\bigg(\frac{m_{\pi}}{\mu}\bigg)} \\
+R_2(p',p)  \bigg(\frac{23}{24}g^4 +\frac{1}{12} \bigg)      \Bigg\}
\end{multline}
\begin{multline}
      \Bar{S}_2 (p',p) =\!\! \int^{1}_{-1}\!\!  \frac{dx}{2} \frac{(3x^2-1)}{2} (\vec{\tau_1}\cdot \vec{\tau_2} )\vec{q}^2
     \Bigg\{ \! 
     \mathcal{R} \bigg[\frac{23}{48}   g^4   
     + \frac{1}{24}  \bigg] \!
    + \! \bigg(\frac{5}{144}   g^4 - \frac{5}{72}  \bigg)  
     \! + \ln{\bigg(\frac{m_{\pi}}{\mu}\bigg)}  \bigg( \frac{23}{24} g^4 + \frac{1}{12} \bigg)
      \! + \! L(q)   \bigg( \frac{23}{24} g^4 + \frac{1}{12} \bigg) \!\! \Bigg\}
      \\= \frac{\big(\vec{\tau_1}\cdot \vec{\tau_2} \big)}{8{p'}^2p^2} \Bigg\{
\Bigg[ 3 R_6 (p',p) -6({p'}^2+p^2)R_4 (p',p) + \big(3 ({p'}^2+p^2)^2 -4{p'}^2p^2 \big) R_2(p',p)  \Bigg] \bigg( \frac{23}{24}  g^4 + \frac{1}{12} \bigg)
    \Bigg\} 
\end{multline}
\end{widetext}
\subsection{ \texorpdfstring{$\bm{B^{(*)} B^{(*)} \rightarrow B^{(*)} B^{(*)}}$}{}}
The PWD potentials can be obtained in exactly the same way as for $B^{(*)} \Bar{B}^{(*)} {\rightarrow} B^{(*)} \Bar{B}^{(*)}$. 
However , in this case  due to Bose symmetry for identical heavy mesons,  certain partial wave transitions are forbidden by quantum number constraints. In particular, 
\begin{eqnarray*}
    & &V^{I=0}(B^{(*)} B^{(*)}{\to} B^{(*)} B^{(*)}, 0^{+})=\\
    & &\hspace{2cm}V^{I=0}(B^{*} B^{*}{\to} B^{*} B^{*}, 2^{+})=0
    \end{eqnarray*}
and 
\begin{eqnarray*}
 V^{I=1}(B B^{*}{\to} B^* B^*, 1^{+})=
V^{I=1}(B^* B^*{\to} B^* B^*, 1^{+})=0 \ .
\end{eqnarray*}
Also, the additional contribution from the triangle diagrams  and the signs, as discussed  
in Sec.~\ref{BBpots}, have to be incorporated. 
The additional triangle contribution can be included in the PWD potentials of $B^{(*)} B^{(*)}$ by changing the $\Bar{S}_k(p',p)$ functions (which includes $\Bar{S}_0(p',p)$, $\Bar{S}_2(p',p)$ and $\Bar{S}_4(p',p)$) to $S_k(p',p)$ functions ($S_0(p',p)$, $S_2(p',p)$ and $S_4(p',p)$), where $S_k(p',p)$ is given by, 

\begin{widetext}
    \begin{multline}
  S_k(p',p)=  \int_{-1}^{1} \frac{dx}{2} P_k(x)  \big(\vec{\tau_1}\cdot \vec{\tau_2} \big) \vec{q}\,^2  \Bigg\{ 
     \mathcal{R} \bigg[ \frac{23}{48}  g^4 -\frac{5}{24}   g^2    - \frac{1}{24}  \bigg] 
     +  \bigg(\frac{5}{144}   g^4 +\frac{13}{72}g^2 + \frac{5}{72}  \bigg) \\
     + \ln{\bigg(\frac{m_{\pi}}{\mu}\bigg)}   \bigg( \frac{23}{24} g^4 -\frac{5}{12} g^2 - \frac{1}{12} \bigg) 
     + L(q)   \bigg( \frac{23}{24} g^4 - \frac{5}{12} g^2 - \frac{1}{12} \bigg)   \Bigg\} \ ,
     \label{eq:BBpot}
\end{multline}
\end{widetext}

 where $P_k(x)$ denotes the $k$-th Legendre polynomial ($k=0,2,4$ ; in our case). 
 
\section{Checks of the results}
\label{sec:check}
A non-trivial cross check of our results is provided by the  renormalisation of the formally divergent loops, which has
to work at each order in the power counting separately.
Specifically, the divergent terms ($\cal R$-terms---see Eq.~(\ref{eq:Rdef})), contained in
the loop contributions of $B^{(*)}\bar B^{(*)}$ (and similarly $B^{(*)} B^{(*)}$) TPE potentials must consistently match to the LECs of the contact interactions. This matching has to work for each type of diagram separately: the footballs, scaling with $g^0$, the triangles $\sim g^2$ and the boxes $\sim g^4$.

We take the case of $(V_{\rm TPE}^{0^{++}})_{12}$ from Eq.~\eqref{Eq:V0pp12} in Sec.~\ref{Sec:0++} and relate it to $(V_{\rm CT}^{0^{++}})_{12}$ in the $0^{++}$ channel as an example.
By equating the divergence of the TPE loop integral with the LECs ($ \mathcal{C}_f^I,  \mathcal{D}_f^I$) for the $g^4$ contribution, one finds for the divergent parts of the counter terms 
\begin{equation}
     \mathcal{C}_f^I=0
\end{equation}
\begin{equation}
     \mathcal{D}_f^I= \frac{  -\mathcal{R} g^4 }{ 32 \pi^2 f_{\pi}^4}. 
\end{equation}
Similarly, equating the divergence of $(V_{\rm TPE}^{0^{++}})_{11}$ and $(V_{\rm TPE}^{0^{++}})_{22}$ with  $(V_{\rm CT}^{0^{++}})_{11}$ and  $(V_{\rm CT}^{0^{++}})_{22}$, respectively for $g^4$ contribution, results in
\begin{equation}
     \mathcal{C}_d^I= 0
\end{equation}
\begin{equation}
     \mathcal{D}_d^I= \frac{  \mathcal{R}g^4 }{64 \pi^2 f_{\pi}^4} \bigg( \frac{23}{12}(\vec{\tau_1}\cdot \vec{\tau_2} )+1 \bigg)
\end{equation}
Following the same procedure for   $(V_{\rm CT}^{0^{++}})_{13}$ and $(V_{\rm TPE}^{0^{++}})_{13}$ with its prefactors, one finds the
divergent part of $\mathcal{D}_{SD}^I$ as
\begin{equation}
     \mathcal{D}_{SD}^I= \frac{ \sqrt{2}\mathcal{R} g^4 }{ 128  \pi^2 f_{\pi}^4} \ .
     \label{D_sd}
\end{equation}
Similarly, when performing this exercise for all $g^0$ terms in $B^{(*)}\bar B^{(*)}$ case, one finds 
\begin{equation}
    \mathcal{C}_d^I =  \mathcal{C}_f^I= \mathcal{D}_f^I =\mathcal{D}_{SD}^I=0
\end{equation}
\begin{equation}
   \mathcal{D}_d^I= \frac{\mathcal{R}}{384 \pi^2 f_{\pi}^4} (\vec{\tau_1}\cdot \vec{\tau_2} )
\end{equation}
and in the $g^2$ case 
 \begin{equation}
    \mathcal{C}_d^I =  \mathcal{C}_f^I= \mathcal{D}_f^I =\mathcal{D}_{SD}^I=0
\end{equation}
\begin{equation}
   \mathcal{D}_d^I= \frac{-5\mathcal{R} g^2}{384 \pi^2 f_{\pi}^4} (\vec{\tau_1}\cdot \vec{\tau_2} )
\end{equation}
The renormalisation program provides a non-trivial cross
check of the calculations. 
For example, 
the divergent terms of the PWD potentials 
$(V_{\rm CT}^{0^{++}})_{13}$ and $(V_{\rm CT}^{0^{++}})_{23}$ corresponding to 
 $B \Bar{B}(^1S_0) {\rightarrow} B^* \Bar{B}^*(^5D_0)$ and $B^* \Bar{B}^*(^1S_0) {\rightarrow} B^* \Bar{B}^*(^5D_0)$, respectively,  must be related such that their infinities can be 
absorbed into one single $ \mathcal{D}_{SD}^I$ term in Eq.~\eqref{D_sd}.
Performing the same exercise for the remaining transitions, we confirm that $\mathcal{C}_d^I$, $\mathcal{C}_f^I $, $\mathcal{D}_d^I $ and $\mathcal{D}_f^I $ absorb all the divergences of $B \Bar{B}(^1S_0) {\rightarrow} B \Bar{B}(^1S_0)$, 
$B \Bar{B}(^1S_0) {\rightarrow} B^* \Bar{B}^*(^1S_0)$ and 
$B^* \Bar{B}^*(^1S_0) {\rightarrow} B^* \Bar{B}^*(^1S_0)$ PWD potentials consistently.
Moreover, we verified that with the counter terms determined above, all 
$B^{(*)}\bar B^{(*)}$ transitions with the 
quantum numbers $0^{++}, 1^{++}, 1^{+-}$ and $2^{++}$ are finite. 

As a further non-trivial cross check
we verified that the triangle contributions
add up in the $B^{(*)}B^{(*)}$ case and cancel for $B^{(*)} \Bar{B}^{(*)}$, by computing the diagrams
in particle basis for both cases.

\section{ Plots of potentials}
\label{sec:plots}

In this section we present some representative plots for the TPE potentials in momentum space. 
 Please note that, although in what follows we refer to the results for $B$-mesons, the heavy meson mass does not explicitly enter the TPE contributions at the given order. Therefore, these results are equally applicable to $D^{(*)}D^{(*)}$ and $D^{(*)}\bar D^{(*)}$ scattering at the order we are working. The only difference to keep in mind is that for the physical pion mass, the three-body cuts in the TPE diagrams will contribute to the imaginary part of the scattering amplitude. These contributions are not captured by our power counting but may become important if one aims for a high-accuracy calculation of this quantity. This is particularly important in the context of accurately extracting the width of the $T_{cc}$ state from data \cite{Du:2021zzh}.

In  what follows, for the sake of illustration, we present the heavy meson heavy (anti)meson TPE potentials for various partial waves and typical momentum regimes. We also compare the momentum dependence of these potentials with that from the pure LO+NLO contact interactions and demonstrate in this way that, to the accuracy asigned to this chiral order, the full potentials
are represented well by the contact terms only.
This suggests that the nonanalytic contributions are suppressed. In this section, we  also discuss the  contributions of the individual TPE diagrams to the full results. To generate the plots, we set $\mu = 1.156$ GeV and neglect the $R$ term, as it contains divergences that are absorbed by the contact terms, as discussed in the previous section.

\begin{figure*}[t]
 \begin{subfigure}[h]{0.32\linewidth}
        \centering
        \includegraphics[width=\linewidth]{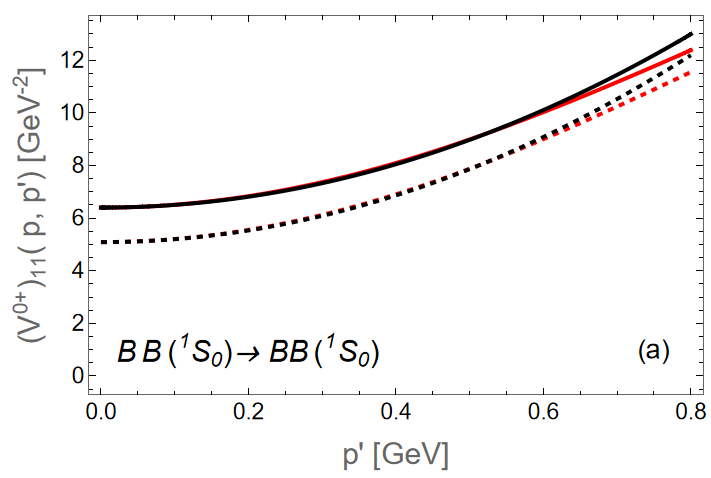}
        \caption*{}
    \end{subfigure}
     \begin{subfigure}[h]{0.32\linewidth}
        \centering
         \includegraphics[width=\linewidth]{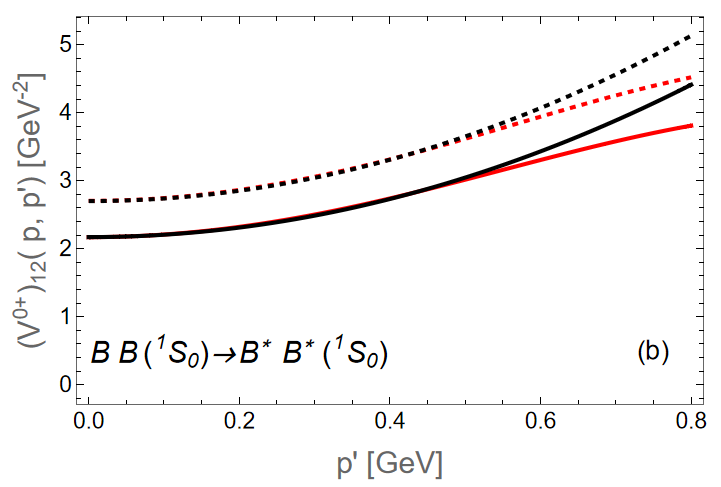}
        \caption*{}
    \end{subfigure}
  \begin{subfigure}[h]{0.32\linewidth}
        \centering
         \includegraphics[width=\linewidth]{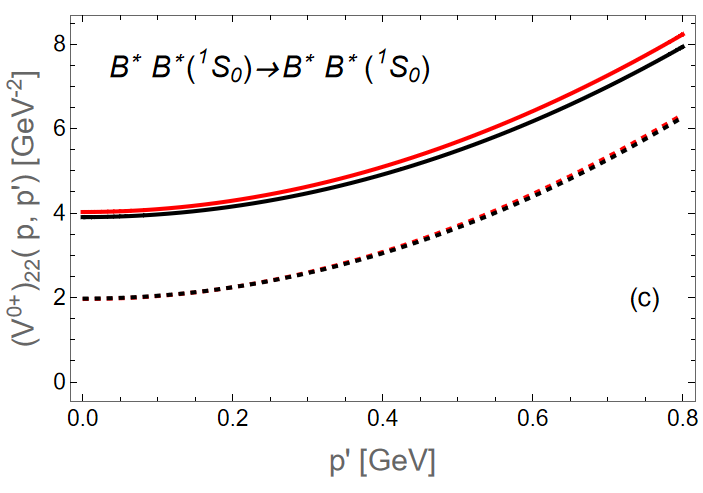}
        \caption*{}
    \end{subfigure}
      \begin{subfigure}[h]{0.32\linewidth}
        \centering
         \includegraphics[width=\linewidth]{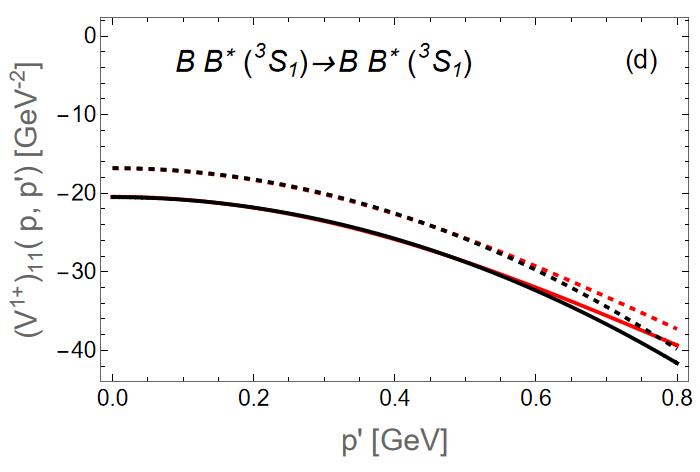}
        \caption*{}
    \end{subfigure}
    \begin{subfigure}[h]{0.32\linewidth}
        \centering
         \includegraphics[width=\linewidth]{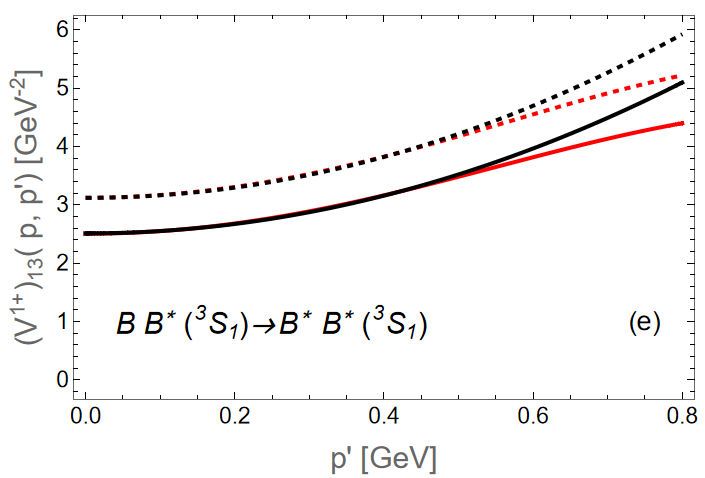}
        \caption*{}
    \end{subfigure}
      \begin{subfigure}[h]{0.32\linewidth}
        \centering
         \includegraphics[width=1.\linewidth]{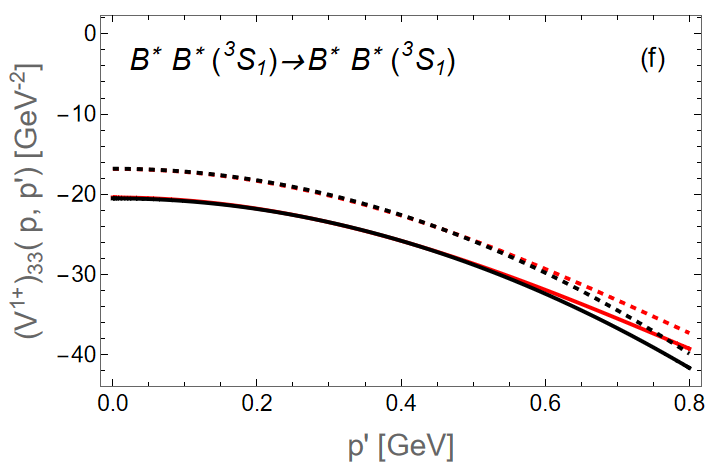}
        \caption*{}
    \end{subfigure}
    \caption{ \label{fig:plots for B B system ptyp}
    The TPE  and the contact potentials of the $B^{(*)} B^{(*)}$ system in  the momentum space, allowed by Bose symmetry, as a function 
    of the final momentum $p'$ for the initial momentum $p=500$ MeV. 
    The solid (dotted) red lines represent the full (expanded to ${\mathcal O}(Q^2)$) TPE potential, while the solid (dotted) black lines show the results of the fit of the contact potential at $p=500$ MeV  to the full (expanded) TPE potential, as described in the main text.   Figures (a)-(c) depict transitions in the $0^{+}$ channels for $I=1$, namely $ (V_{\rm TPE}^{0^{+}})_{11}$, $ (V_{\rm TPE}^{0^{+}})_{12}$ and $ (V_{\rm TPE}^{0^{+}})_{22}$, respectively. Figures (d)-(f) show transitions in the $1^{+}$ channels for $I=0$, namely $ (V_{\rm TPE}^{1^{+}})_{11}$, $ (V_{\rm TPE}^{1^{+}})_{13}$ and $ (V_{\rm TPE}^{1^{+}})_{33}$ respectively. 
 }
\end{figure*}

\begin{figure*}[t]
    \begin{subfigure}[h]{0.32\linewidth}
        \centering
        \includegraphics[width=\linewidth]{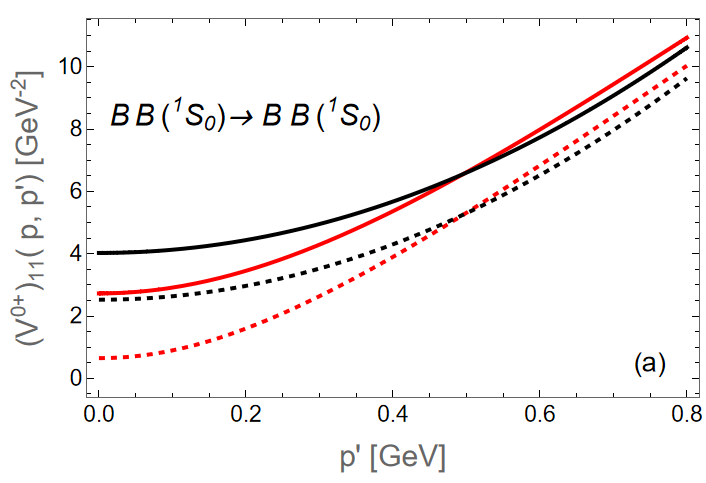}
        \caption*{}
        \label{barBV11for0}
    \end{subfigure}
    \begin{subfigure}[h]{0.32\linewidth}
       \centering
         \includegraphics[width=\linewidth]{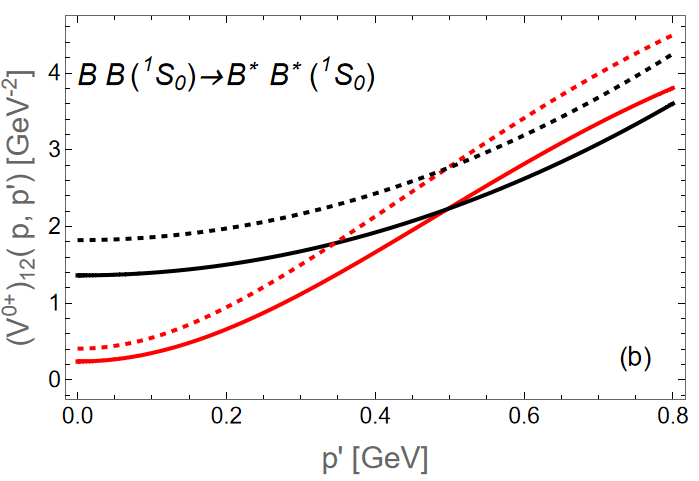}
        
      \caption*{}
        \label{barBV12for0}
    \end{subfigure}
     \hspace{-0.5em}
  \begin{subfigure}[h]{0.32\linewidth}
        \centering
       \includegraphics[width=\linewidth]{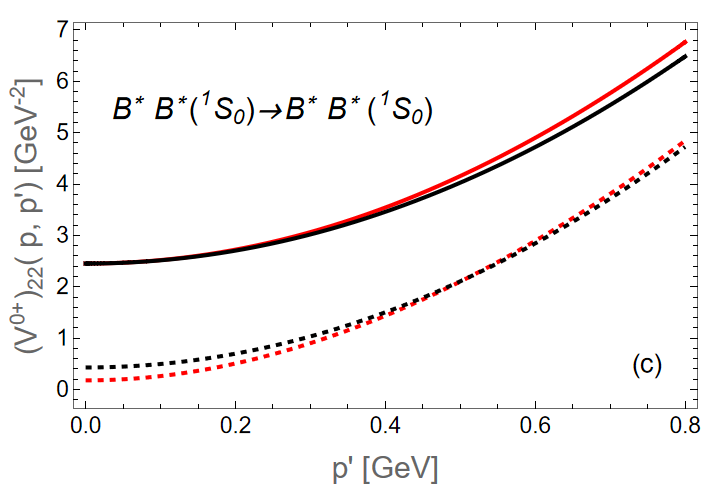}
        
        \caption*{}
        \label{barBV22for0}
   \end{subfigure}
      \begin{subfigure}[h]{0.32\linewidth}
        \centering
         \includegraphics[width=\linewidth]{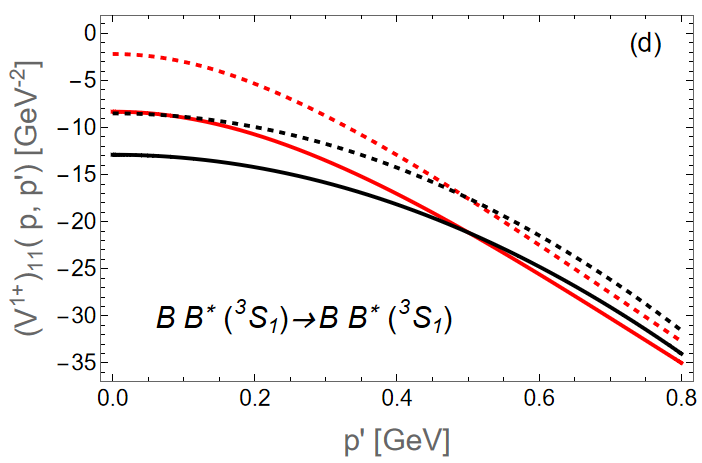}
        \caption*{}
    \end{subfigure}
   \hspace{0.5em}
    \begin{subfigure}[h]{0.32\linewidth}
        \centering
         \includegraphics[width=\linewidth]{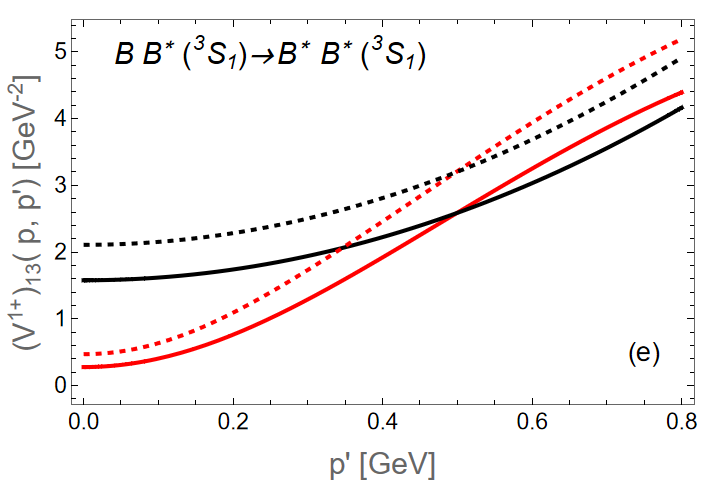}
        \caption*{}
    \end{subfigure}
      \begin{subfigure}[h]{0.33\linewidth}
        \centering
         \includegraphics[width=1.\linewidth]{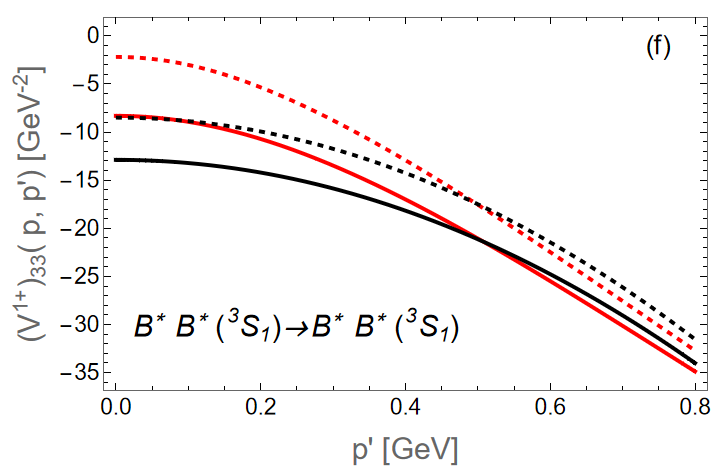}
        \caption*{}
    \end{subfigure}
    \caption{Same as in Fig.~\ref{fig:plots for B B system ptyp} but for the initial momentum $p=m_\pi$.
 }
    \label{fig:plots for B B system mpi}
\end{figure*}

Figs.~\ref{fig:plots for B Bbar system ptyp} and~\ref{fig:plots for B Bbar system mpi} show the results  for the $I=1$ $B^{(*)}\bar B^{(*)}$ TPE potentials with initial momenta of $p= 500$ MeV and $p=m_{\pi}$ respectively, plotted as a function of the final momentum\footnote{Note that the characteristic momentum $p_{\rm typ}\approx 500$ MeV for both BB- and DD-meson systems}.
Similarly,   Figs.~\ref{fig:plots for B B system ptyp} and Fig.~\ref{fig:plots for B B system mpi} show the results  for the  $B^{(*)}B^{(*)}$ system under the same  conditions for the initial momenta. In the $B^{(*)}B^{(*)}$ system, however, only those transitions  are shown that are allowed by Bose symmetry. 
In all these plots,  the red solid lines represent the full potentials, which include terms additionally suppressed by  $(m_\pi^2/p_{\rm typ}^2)$, while 
the red dashed lines correspond to the potentials expanded to ${\mathcal O}(Q^2)$ -- see, e.g., Eqs.~\eqref{eq:Itreval} and \eqref{Eq:Lq} for details. As one can see for all cases
shown, the deviation between these curves is well below the expectations from the power counting. As expected, the discrepancies for $p = m_\pi$ are slightly larger than those for $p\sim p_{\rm typ}$. 
Additionally, it is observed that the TPE potentials in the kinematic region where $ p \sim p' \sim p_{\rm typ} $ are larger compared to those at lower momenta. Specifically, $V_{\rm TPE}(m_\pi, m_\pi) $  is significantly smaller than $V_{\rm TPE}(p_{\rm typ}, p_{\rm typ}) $ in line with the power counting.

\begin{widetext}

\begin{figure}[H]
 \begin{subfigure}[h]{0.32\linewidth}
        \centering
        \includegraphics[width=\linewidth]{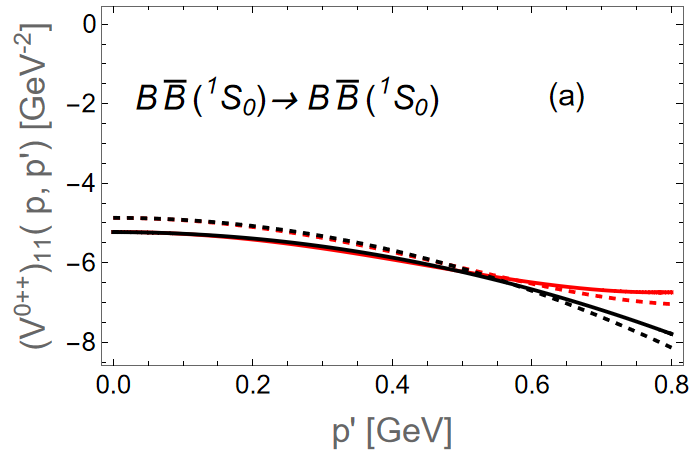}
        \caption*{}
    \end{subfigure}
     \begin{subfigure}[h]{0.32\linewidth}
        \centering
         \includegraphics[width=\linewidth]{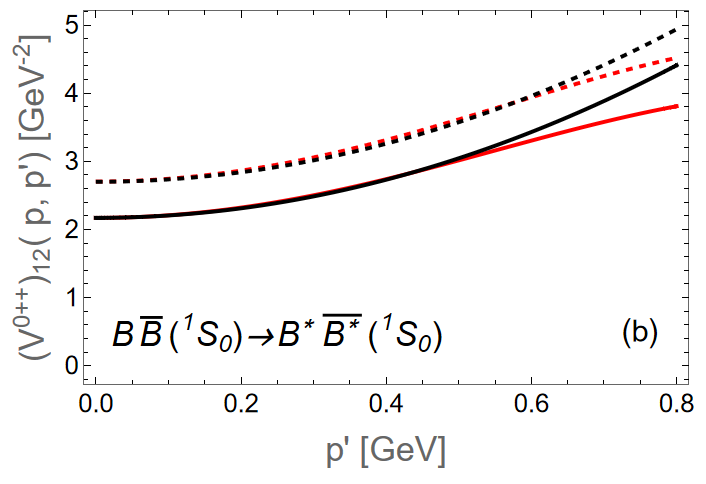}
        \caption*{}
    \end{subfigure}
  \begin{subfigure}[h]{0.32\linewidth}
        \centering
         \includegraphics[width=\linewidth]{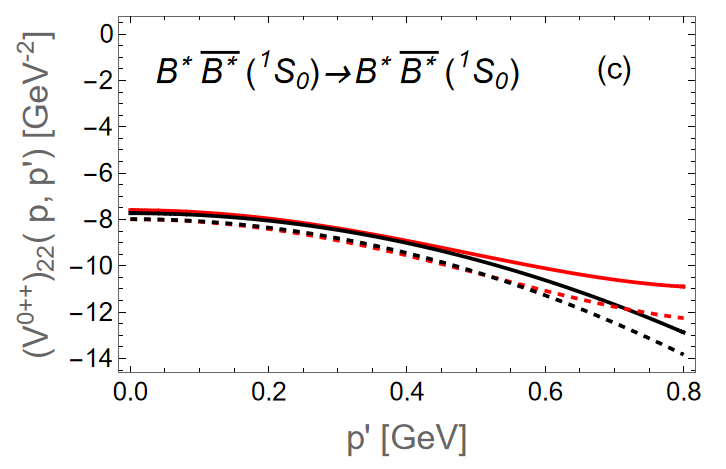}
        \caption*{}
    \end{subfigure}
      \begin{subfigure}[h]{0.32\linewidth}
        \centering
         \includegraphics[width=\linewidth]{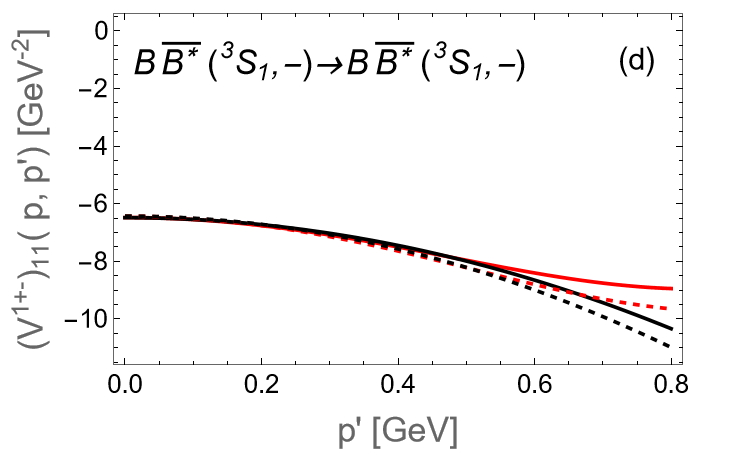}
        \caption*{}
    \end{subfigure}
    \hspace{0.5em}
    \begin{subfigure}[h]{0.32\linewidth}
        \centering
         \includegraphics[width=\linewidth]{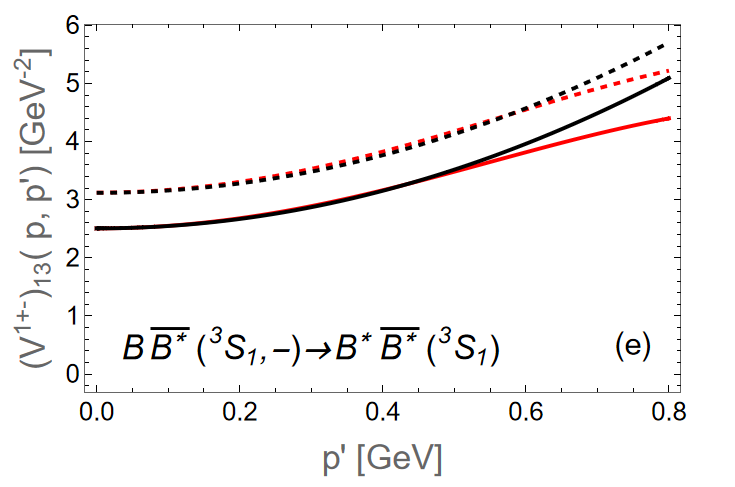}
        \caption*{}
    \end{subfigure}
      \hspace{2em}
      \begin{subfigure}[h]{0.32\linewidth}
        \centering
         \includegraphics[width=1.\linewidth]{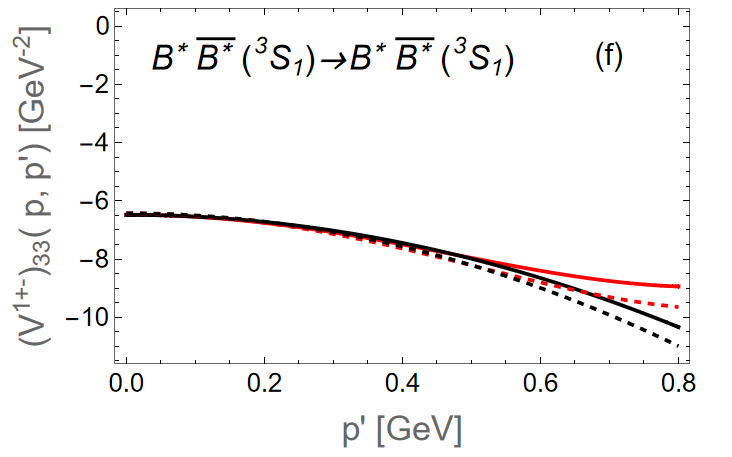}
        \caption*{}
    \end{subfigure}
    \caption{   \label{fig:plots for B Bbar system ptyp}
    The TPE and contact potentials of the  $B^{(*)} \Bar{B}^{(*)}$ system in the momentum space as a function 
    of the final momentum $p'$ for the initial momentum $p=500$ MeV. The solid (dotted) red lines represent the full (expanded to ${\mathcal O}(Q^2)$) TPE potential, while the solid (dotted) black lines illustrate the results of the fit of the contact  potential
    at $p=500$ MeV   to the full (expanded) TPE potential, as described in the main text.
        Figures (a)-(c) depict transitions in the $0^{++}$ channels, namely $ (V_{\rm TPE}^{0^{++}})_{11}$, $ (V_{\rm TPE}^{0^{++}})_{12}$ and $ (V_{\rm TPE}^{0^{++}})_{22}$, respectively. Figures (d)-(f) show transitions in the $1^{+-}$ channels, namely $ (V_{\rm TPE}^{1^{+-}})_{11}$, $ (V_{\rm TPE}^{1^{+-}})_{13}$ and $ (V_{\rm TPE}^{1^{+-}})_{33}$, respectively. All   figures correspond to $I=1$.}
\end{figure}
\begin{figure}[H]
    \begin{subfigure}[h]{0.32\linewidth}
        \centering
        \includegraphics[width=\linewidth]{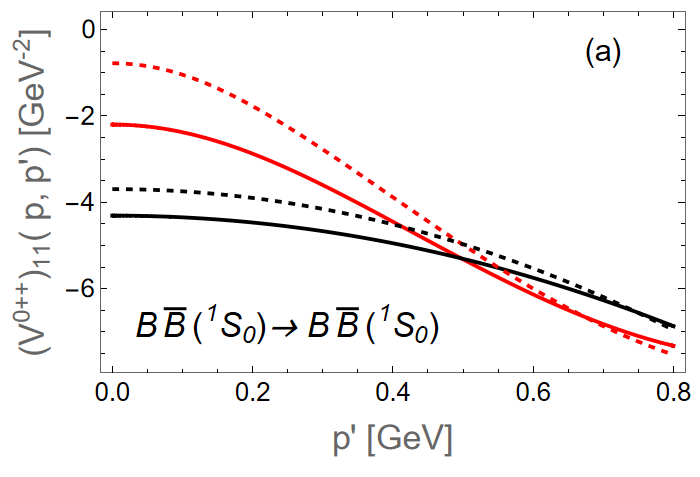}
        \caption*{}
    \end{subfigure}
     \begin{subfigure}[h]{0.32\linewidth}
        \centering
         \includegraphics[width=\linewidth]{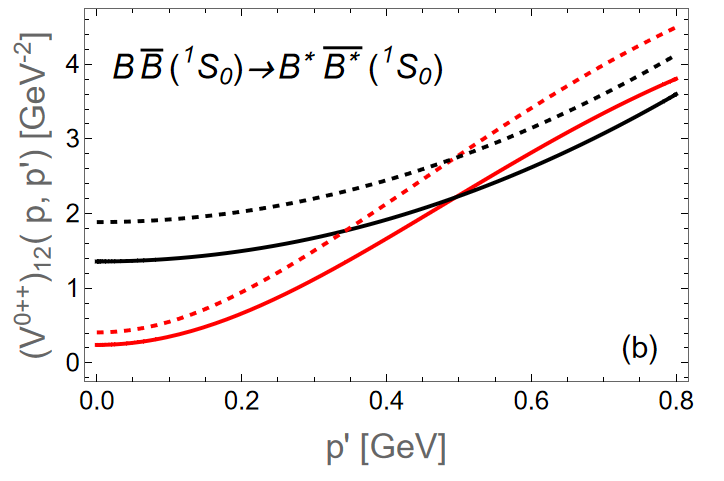}
        \caption*{}
    \end{subfigure}
  \begin{subfigure}[h]{0.32\linewidth}
        \centering
         \includegraphics[width=\linewidth]{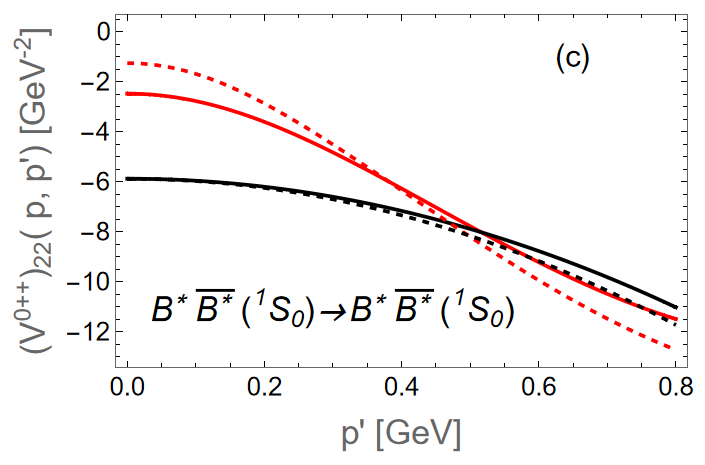}
        \caption*{}
    \end{subfigure}
      \begin{subfigure}[h]{0.32\linewidth}
        \centering
         \includegraphics[width=\linewidth]{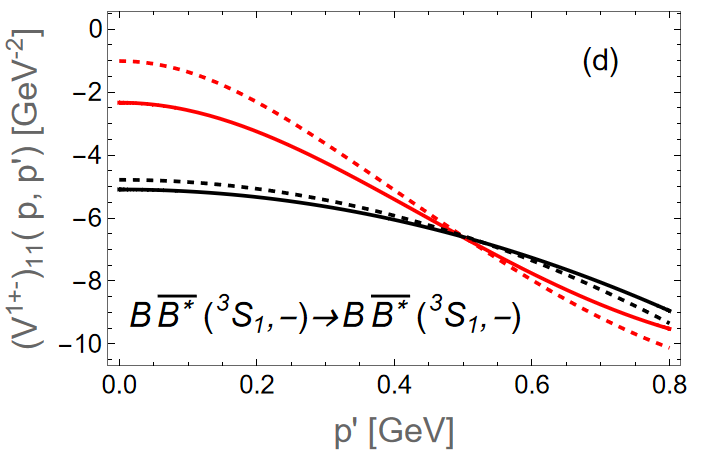}
        \caption*{}
    \end{subfigure}
   \hspace{0.5em}
    \begin{subfigure}[h]{0.32\linewidth}
        \centering
         \includegraphics[width=\linewidth]{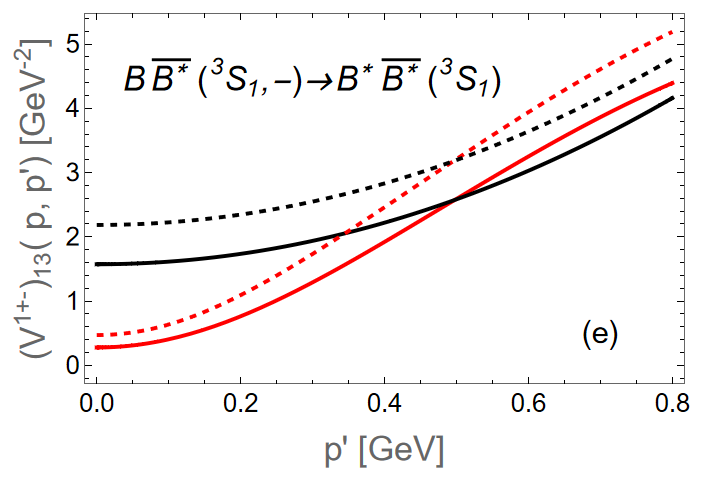}
        \caption*{}
    \end{subfigure}
      \begin{subfigure}[h]{0.33\linewidth}
        \centering
         \includegraphics[width=1.\linewidth]{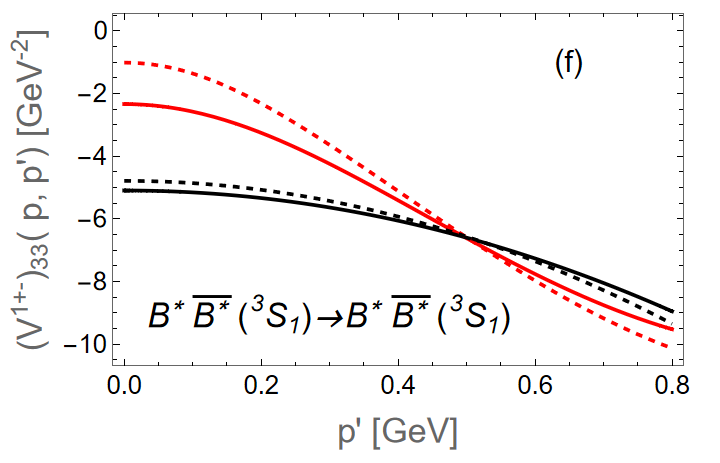}
        \caption*{}
    \end{subfigure}
    \caption{\label{fig:plots for B Bbar system mpi}
    Same as in Fig.~\ref{fig:plots for B Bbar system ptyp} but for the initial momentum $p=m_\pi$. 
    }
\end{figure}

\end{widetext}
    
\begin{figure*}[t]
 \begin{subfigure}[h]{0.32\linewidth}
        \centering
        \includegraphics[width=\linewidth]{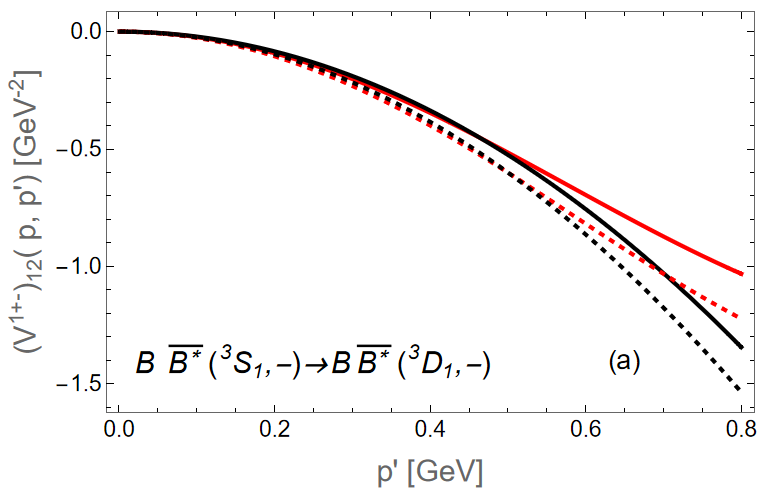}
        \caption*{}
    \end{subfigure}
     \begin{subfigure}[h]{0.32\linewidth}
        \centering
         \includegraphics[width=\linewidth]{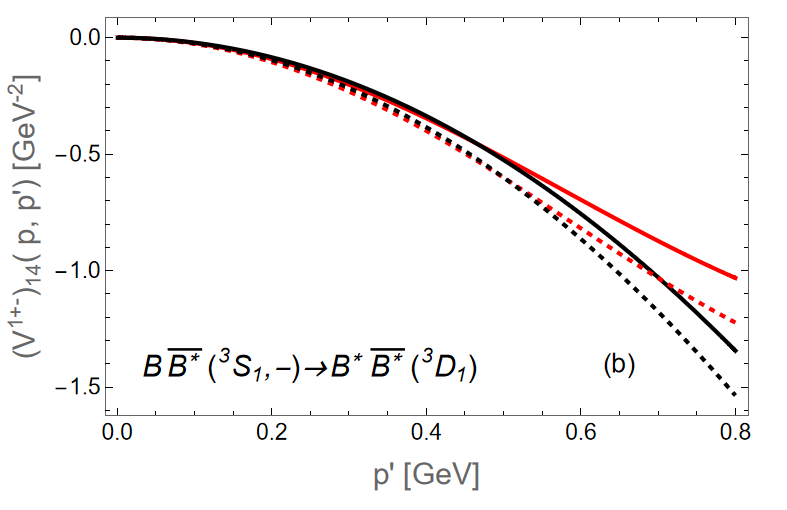}
        \caption*{}
    \end{subfigure}
  \begin{subfigure}[h]{0.32\linewidth}
        \centering
         \includegraphics[width=\linewidth]{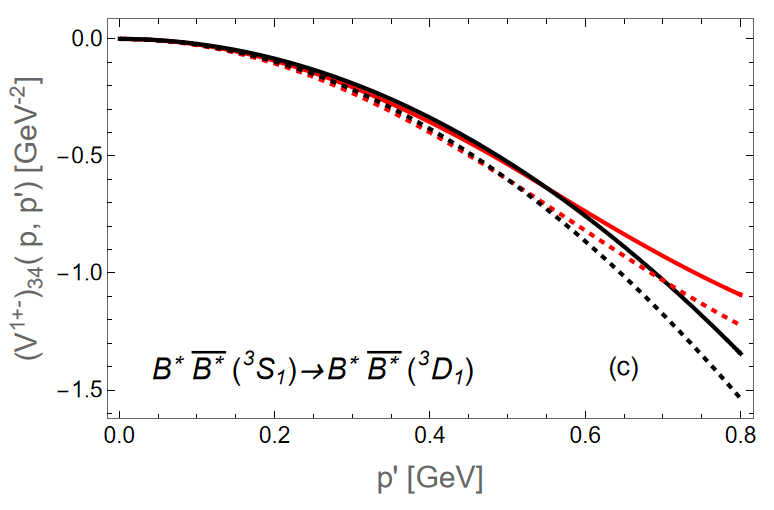}
        \caption*{}
    \end{subfigure}
    \caption{The TPE and contact $S$ to $D$ wave  transitions of the  $B^{(*)} \Bar{B}^{(*)}$  
    system in the momentum space are presented as a function 
    of the final momentum for the initial momentum $p=500$ MeV. Since the $S$ to $D$ transitions arise solely from the box diagrams, which are the same for $B^{(*)} \Bar{B}^{(*)}$ and $B^{(*)} 
    {B}^{(*)}$ before the partial wave decomposition, only the $B^{(*)} \Bar{B}^{(*)}$ contributions are shown. 
   The solid (dotted) red lines represent the full (expanded to ${\mathcal O}(Q^2)$) TPE potential, while the solid (dotted) black lines show the results of the fit of the contact potential
    at $p=500$ MeV   to the full (expanded) potential, as described in the main text.
    Figures (a), (b) and (c) depict  $ (V_{\rm TPE}^{1^{+-}})_{12}$, $ (V_{\rm TPE}^{1^{+-}})_{14}$ and $ (V_{\rm TPE}^{1^{+-}})_{34}$ respectively.
    \label{fig:plots for SD transition ptyp}}
\end{figure*}

\begin{figure*}[t]
    \begin{subfigure}[h]{0.32\linewidth}
        \centering
        \includegraphics[width=\linewidth]{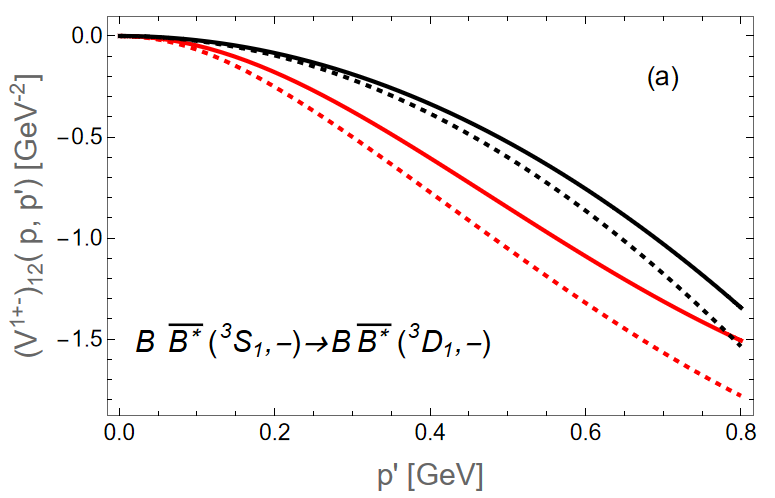}
        \caption*{}
    \end{subfigure}
     \begin{subfigure}[h]{0.32\linewidth}
        \centering
         \includegraphics[width=\linewidth]{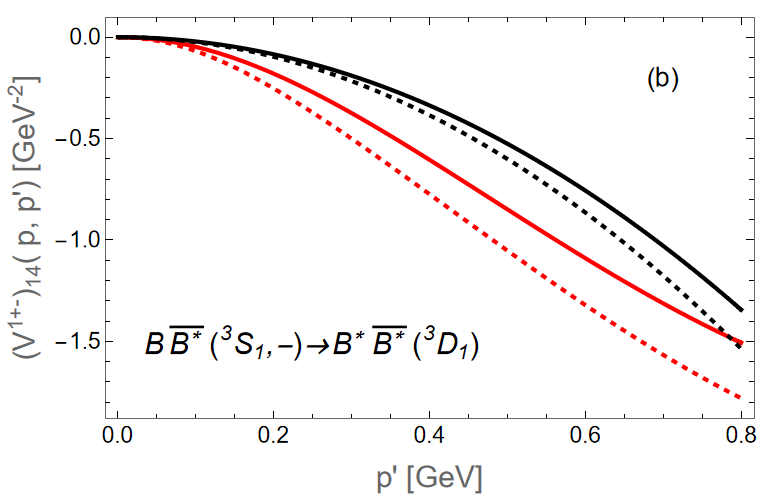}
        \caption*{}
    \end{subfigure}
  \begin{subfigure}[h]{0.32\linewidth}
        \centering
         \includegraphics[width=\linewidth]{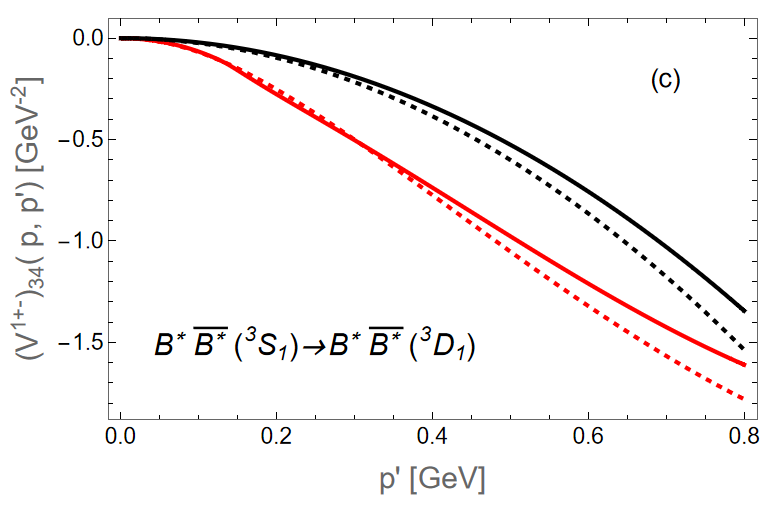}
        \caption*{}
    \end{subfigure}
    \caption{Same as in Fig.~\ref{fig:plots for SD transition ptyp} but for the initial momentum $p=m_\pi$.
    \label{fig:plots for SD transition mpi}}
\end{figure*}

\begin{figure*}[t]
    \begin{subfigure}[h]{0.32\linewidth}
        \centering
        \includegraphics[width=\linewidth]{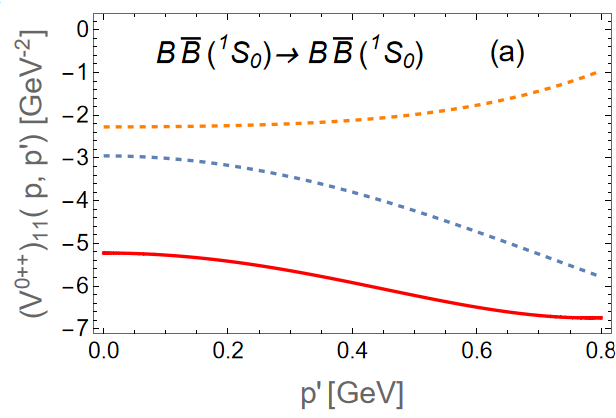}
        \caption*{}
        \label{barBV11for0_contr}
    \end{subfigure}
    \begin{subfigure}[h]{0.32\linewidth}
       \centering
         \includegraphics[width=\linewidth]{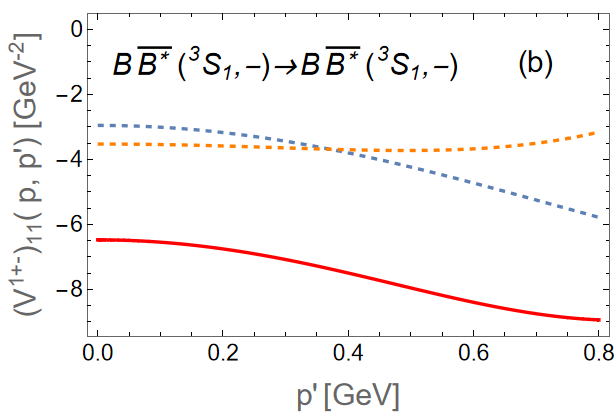}
        
      \caption*{}
        \label{barBV12for0_contr}
    \end{subfigure}
     \hspace{-0.5em}
  \begin{subfigure}[h]{0.32\linewidth}
        \centering
       \includegraphics[width=\linewidth]{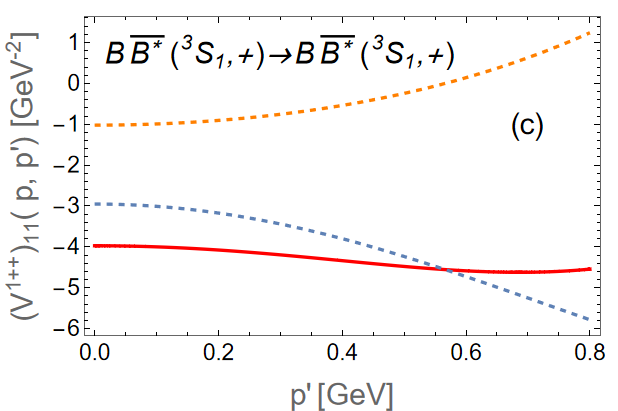}
        
        \caption*{}
        \label{barBV22for0_contr}
   \end{subfigure}
      \begin{subfigure}[h]{0.32\linewidth}
        \centering
         \includegraphics[width=\linewidth]{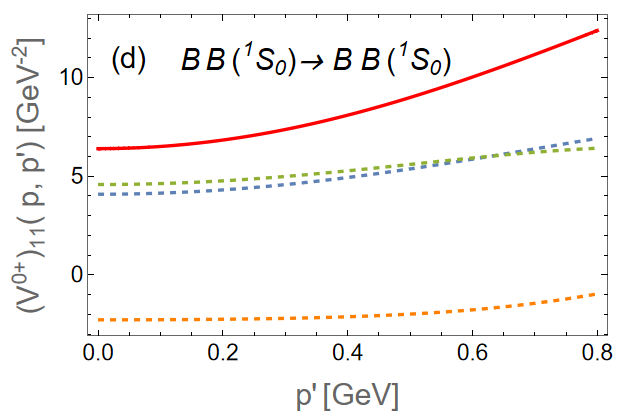}
        \caption*{}
    \end{subfigure}
   \hspace{0.5em}
    \begin{subfigure}[h]{0.32\linewidth}
        \centering
         \includegraphics[width=\linewidth]{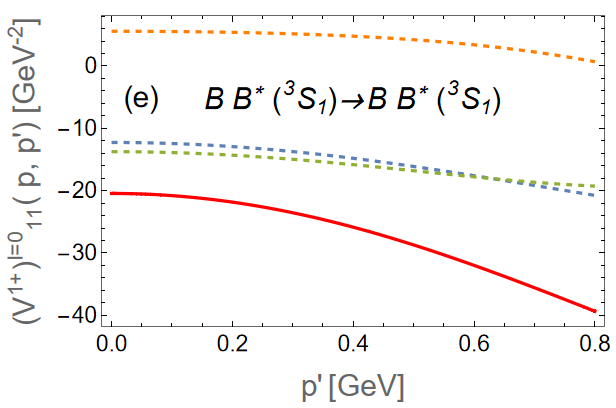}
        \caption*{}
    \end{subfigure}
      \begin{subfigure}[h]{0.33\linewidth}
        \centering
         \includegraphics[width=1.\linewidth]{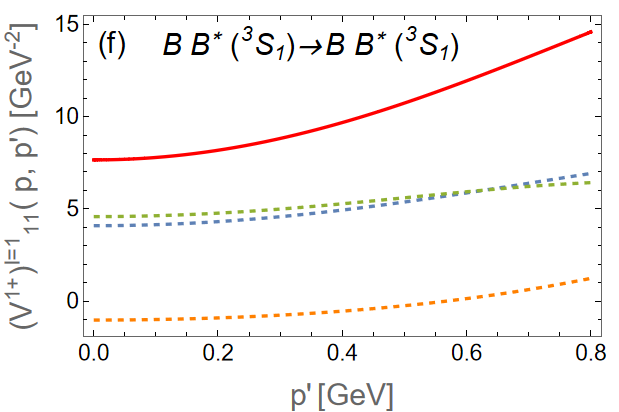}
        \caption*{}
    \end{subfigure}
    \caption{The individual contributions to the TPE potentials 
    for the  $B^{(*)} \Bar{B}^{(*)}$ (first line) and the  $B^{(*)} {B}^{(*)}$ (second line) system in momentum
    space for $p = 500$ MeV.
    Here, the red solid line represents the full TPE potential and the blue dashed, green dashed and orange dashed lines represent the contributions proportional to $g^0$ (footballs), $g^2$ (triangles), $g^4$ (boxes), respectively.
     Figures (a), (b) and (c) depict  $ (V_{\rm TPE}^{0^{++}})_{11}$, $ (V_{\rm TPE}^{1^{+-}})_{11}$ and $ (V_{\rm TPE}^{1^{++}})_{11}$ for $I=1$ in the  $B^{(*)} \Bar{B}^{(*)}$ system, respectively. Figures (d), (e) and (f)  depict  $ (V_{\rm TPE}^{0^{+}})_{11}$ for $I=1$, $ (V_{\rm TPE}^{1^{+}})_{11}$ for $I=0$ and $ (V_{\rm TPE}^{1^{+}})_{11}$ for $I=1$ in the  $B^{(*)} {B}^{(*)}$ system, respectively.
    }
    \label{fig:plots for gn diagrams}
\end{figure*}

As the corresponding black lines we also show the results of fits of the pertinent contact terms from Eq.~\eqref{Eq:VCT0++} (see also Eqs.~\eqref{Eq:VCT1++}, \eqref{Eq:VCT1+-} and \eqref{Eq:VCT2++})
to the full potentials. The fits were done for $p=500$ MeV, with the LO contact terms fixed by the values of the potentials at $p'=0$, and the  NLO contact terms adjusted by fitting the slope.
 By fitting   two different TPE potentials (e.g $B \Bar{B}(^1S_0) {\rightarrow} B\Bar{B}(^1S_0)$ and $B \Bar{B}(^1S_0) {\rightarrow} B^* \Bar{B}^*(^1S_0)$) 
we were able to individually extract the values of the low-energy constants
$\mathcal{C}_d$, $\mathcal{C}_f $, $\mathcal{D}_d $ and $\mathcal{D}_f $,
that represent the TPE contributions best. 
All  other coupled-channel transitions with different $J^{\rm PC}$ are therefore predictions, with all  LEC's fixed. 
Remarkably, the curves representing the contact interactions for $ p \simeq p_{\rm typ} $ show excellent agreement with the TPE contributions across the entire momentum range of $p' $, including $p' \simeq p_{\rm typ}$.
The deviations are more
sizeable for $p=m_\pi$, however, not exceeding expectations from the power counting. 
Moreover, the contact terms fixed from the $^1S_0$ transition also represent well the
TPE potentials for the other quantum numbers.  
For illustration, we show the $1^{+-}$
potentials, which are shown  in the lower panels in Figs.~\ref{fig:plots for B Bbar system ptyp} and~\ref{fig:plots for B Bbar system mpi}.
Similar to $S$-$S$ transitions, we   extracted the value of $\mathcal{D}_{SD}$ for a
particular $S$-$D$ transition, which was chosen to be $B \Bar{B}^*(^3S_1) {\rightarrow} B\Bar{B}^*(^3D_1)$. With this value fixed, the $S$-$D$ contact interaction  also provides reasonable estimates for all other 
$S$-$D$ transitions as illustrated in Fig.~\ref{fig:plots for SD transition ptyp} for $p=500$ MeV and Fig.~\ref{fig:plots for SD transition mpi} for $p=m_\pi$.
 In these plots, we present results for various $ B^{(*)}\bar{B}^{(*)}$ transitions in the $ 1^{+-}$ channel as an example; however, the results for all other transitions with different quantum numbers exhibit a very similar pattern.
The results shown in Figs.~\ref{fig:plots for B Bbar system ptyp}-\ref{fig:plots for SD transition mpi} for both $B^{(*)}\bar B^{(*)}$  and $B^{(*)}B^{(*)}$ demonstrate that the leading contribution of the TPE interactions is predominantly polynomial in momenta, while the non-analytic logarithmic contributions are suppressed.

Clearly, the values of the LECs extracted above are not their physical values, which can only
be determined from a fit to the data. What we aimed at here was more the demonstration that
the TPE contributions  can be largely represented by contact terms, suggesting, in particular,  that 
the results of Refs.~\cite{Wang:2018jlv,Baru:2019xnh} for a coupled-channel analysis of the Belle data in the context of $Z_b(10610)$
and $Z_b(10650)$ and their possible spin partners, can be interpreted as a reasonable 
representation of an NLO calculation. However, to obtain accurate results, e.g., for the pole locations of the  $Z_b$  states with fully controlled uncertainty estimates, a data analysis that includes the TPE contributions is needed.

 In Fig.~\ref{fig:plots for gn diagrams}, we  show  contributions from individual diagram types---football, triangle, and box---to the TPE scattering potential for three representative partial waves. All contributions for $B\bar B^{(*)}$  are evaluated for $I=1$. For $B B^{(*)}$, we display only the transitions allowed by Bose statistics: the elastic $BB (^1S_0)$ transition for $I=1$ (d), and the $BB^* (^3S_1)$ transitions for $I=0$ (e) and $I=1$ (f). Additionally, the elastic $B^*\bar B^* (^3S_1)$ contributions are identical to those for $B\bar{B}^* (^3S_1)$, as are $B^* B^* (^3S_1) (I=0)$ and $BB^* (^3S_1) (I=0)$, so these are not shown.

It can be seen that the box contributions,  proportional to  $g^4$ (indicated by orange dashed lines), are generally the smallest among the different types. For a given isospin,
these contributions are identical for both meson-meson and meson-antimeson scattering (see panels (a) and (d) in Fig.~\ref{fig:plots for gn diagrams} for $(^1S_0)$, and (c) and (f) in Fig.~\ref{fig:plots for gn diagrams} for $(^3S_1)$ partial waves, all with $I=1$).  The triangle contributions (terms $\sim g^2$,  shown by green dashed lines),  cancel out 
for $B ^{(*)}\bar B^{(*)}$ scattering, but do not vanish in the $B ^{(*)} B^{(*)}$ sector. They are positive for $BB(^1S_0)$ and $B{B^*}(^3S_1) (I=1)$ 
 and negative for $B{B^*}(^3S_1) (I=0)$, with these differences arising from the different isospin coefficients in Eq.~\eqref{Eq:iso_t}. 
Finally, the football diagrams, which change sign when switching from  $B ^{(*)}\bar B^{(*)}$ to  $B ^{(*)} B^{(*)}$, show a behaviour very similar 
to that of  the triangles in the  $B ^{(*)} B^{(*)}$ sector. 

All in all, the resulting  $B\bar B^{(*)}$ TPE contributions are always negative.   In contrast,  the    TPE contributions for $B B^{(*)}$ depend on isospin:  they are notably  large and negative for $I=0$,  indicating strong attraction,  while they are positive for $I=1$,  corresponding to repulsion.  It is important to note that while the results presented here are for $B$-mesons, the derived TPE contributions up to order
  ${\cal O}(Q^2)$ are independent of the heavy-meson mass. Thus, they also apply to other  heavy-meson heavy-(anti)meson systems, 
  such as $D^{(*)}D^* - D^{(*)}D^*$ scattering, particularly in the context of the $T_{cc}$ $(I=0)$ and its possible partner states.  
Recently, $DD^* \to DD^*$ scattering with $J^P=1 (^3S_1)$ was investigated on the lattice at $m_{\pi} =280$~MeV  for  both $I=0$ \cite{Padmanath:2022cvl,Collins:2024sfi} and  $I=1$ \cite{Meng:2024kkp}, see also Refs.~\cite{Chen:2016qju,Whyte:2024ihh} for related lattice investigations.  An analysis of the individual correlators contributing to both isospins revealed  that the difference between the two isospin channels is associated with isovector-vector exchanges, such as the TPE contributions in the $\rho$-meson channel,  yielding attraction for $I=0$  and repulsion for $I=1$ \cite{Meng:2024kkp}.  
We note also that Ref.~\cite{Lyu:2023xro} pointed out the role of the TPE in the $I=0$ $DD^*$ potential at separations $r \simeq 1 - 2 $ fm and nearly physical $m_\pi$, though the isospin and total spin of the $\pi\pi$ system were not specified---see also Ref.~\cite{Wang:2022jop} for a related discussion.
The pattern discussed above is in line with our calculations at the physical pion mass (see panels (e) and (f)). Additionally, repeating the calculations at $m_{\pi} =280$~MeV shows qualitatively similar results also at this pion mass, with a significant attraction for $I=0$ and a milder repulsion for $I=1$.

\newpage

\section{Comparison to earlier works}
\label{sec:comparison}

When comparing our OPE results with the ones calculated by Wang et al.~\cite{Wang:2018jlv} (given in  equations (22)-(28)), one finds that their results calculated for the isovector case
agree with ours. 
Also the PWD contact interactions in~\cite{Baru:2019xnh} (see equations (12)-(15) in this reference) agree with ours.

For the TPE contribution in $B^{(*)} \Bar{B}^{(*)} \rightarrow B^{(*)} \Bar{B}^{(*)}$ scattering, however, our results 
disagree with those of Wang et al.~\cite{Wang:2020} (given in equation (10)).
In this work, there is a net triangle contribution (terms proportional to $g^2$)
in the total TPE potential, which in our case is absent.
At the same time, their total TPE potential does not have
any box contributions (terms proportional to $g^4$), which are present in our case. 
Only in case of the football diagrams (terms proportional to $g^0$) we
agree to the results of Ref.~\cite{Wang:2020}.

Furthermore, we compared our results to the  TPE potentials of
Wang et al.~\cite{Wang:2018atz}
for $\bar B^{(*)} \bar B^{(*)}$ scattering.
Contrary to the meson--anti-meson scattering amplitudes of Ref.~\cite{Wang:2020} referred to above,
the TPE potentials for anti-meson--anti-meson scattering in \cite{Wang:2018atz}
contain all types, namely football, triangle and box contributions.
However, a direct comparison with our potentials is difficult,
since the results are given in quite different form compared to ours.

\section{Summary and Outlook}
\label{summary}

Chiral effective field theory has recently become a precision tool for analyzing low-energy few-nucleon reactions, nuclear structure, and form factors. Combined with heavy-quark spin symmetry, this model-independent approach also allows one to probe the properties and internal structure of exotic states in the quarkonium sector, provided they are located near certain hadronic thresholds. This work is part of a series dedicated to a systematic understanding of the nature and properties of the $Z_b(10610)$ and $Z_b(10650)$ states in the bottomonium spectrum   within chiral EFT.

In Refs.~\cite{Wang:2018jlv,Baru:2019xnh,Baru:2020ywb}, analyses of the experimental data for the   $Z_b$ states
in the elastic and inelastic channels were conducted, from which the pole positions of these states were extracted. Given that the spin-partner states
$Z_b(10610)$ and $Z_b(10650)$ are located near the $B\bar{B}^*$ and $B^*\bar{B}^*$ thresholds, respectively, 
and that these hadronic channels can couple to each other, 
a hierarchy of operators was developed using a power counting, where the typical momenta  $Q$  
 are associated with the coupled-channel momentum scale between $B\bar{B}^{(*)}$ and $B^*\bar{B}^{(*)}$, $Q \sim p_{\rm typ}=\sqrt{m_B \delta}$, with $\delta = m_{B^*} - m_B$ representing the $B^*-B$ mass difference. 
A very similar counting was employed in pion production in nucleon-nucleon collisions, specifically the reaction $NN\to NN\pi$, leading to significant progress in understanding the data \cite{Hanhart:2003pg,Baru:2013zpa}.

In the analyses of Refs.~\cite{Wang:2018jlv,Baru:2019xnh}, all diagrams up to and including $\mathcal{O}(Q^2)$ 
 were incorporated in the effective potential except for two-pion exchanges. These one-loop contributions contain information about intermediate-range forces, and need to be
 considered  for systematic uncertainty estimates of the theoretical results and reliable extraction of the pole positions of the $Z_b(10610)$ and $Z_b(10650)$ states from data. In this work, we complete the calculation of diagrams at order 
$\mathcal{O}(Q^2)$  by deriving the missing two-pion exchange operators.

The TPE operators at order $\mathcal{O}(Q^2)$  consist of four topologies:  triangle,  football, box and crossed box  (see Figs.~\ref{fig:diagrams for B Bbar--.> B Bbar}--\ref{fig:diagrams for B Bbar*--.> B Bbar} for details).  
We provide closed-form expressions for all of them, as   detailed    in Appendix C.  
We also present the results for the effective potentials up to $\mathcal{O}(Q^2)$  for $B^{(*)} B^{(*)}$
  scattering, a doubly bottom analogue of the doubly charm case investigated by the LHCb collaboration. 
Notably, we find that  the sum of all triangle diagrams vanishes for  all $B^{(*)}\bar B^{(*)}$ transitions,  while it yields a finite result for $B^{(*)} B^{(*)}$ case. 
An important self-consistency check of the results is provided by renormalization: The loop integrals are divergent and were regularized using dimensional regularization. These divergent parts must be absorbed by the counterterms in the course of renormalization. However, because of heavy-quark spin symmetry, at order  $\mathcal{O}(Q^2)$ 
there are only three contact terms, while the number of allowed transitions is significantly larger. 
Therefore, after absorbing three divergent contributions into the redefined contact terms, no additional divergences should occur. We verified that this is indeed the case. To ease the implementation of the results for calculating observables, we present the effective potential in the form of partial wave decomposition.

Moreover, we show that the TPE potentials can be largely represented by the contact term contributions,
suggesting that the incomplete NLO calculations presented in Refs.~\cite{Wang:2018jlv,Baru:2019xnh}
already provided a fair representation of the physics to this order, although a calculation including
the full NLO potential would be desirable to find results for the pole parameters of the $Z_b$ states
and their spin partners with controlled uncertainties.

Finally, it is important to note that, although the results discussed here are presented for $B$-mesons, the heavy meson mass does not explicitly enter  the TPE operators at the order we are working. As a result, these findings are equally valid for coupled-channel $D^{(*)}D^{(*)}$ and  $D^{(*)}\bar D^{(*)}$ scattering at the same order. The main difference to consider is that, with a physical pion mass, the three-body cuts in the TPE diagrams will contribute to the imaginary part of the $D$-meson scattering amplitudes. 
While these contributions are beyond the order we are working here,
they may become significant for high-precision calculations, particularly in the context of accurately extracting the width of the $T_{cc}$ state from data \cite{Du:2021zzh}. Apart from that, 
  we emphasize that the differences between isovector and isoscalar $J^P=1^+~ DD^*$ potentials, 
 attributed to isovector-vector exchanges in Ref.~\cite{Meng:2024kkp},
  can be naturally explained by the TPE contributions.
We stress that the TPE contributions for  $D^{(*)}D^{(*)}$ scattering can be largely absorbed into the $O(Q^2)$ contact interactions,  with  minor residual non-analytic contributions. 
This conclusion should also hold for unphysically large pion masses, at least as long as the corrections scaling as $m_\pi^2/p_{\rm typ}^2$ remain suppressed.

\begin{acknowledgments}

We would like to thank Evgeny Epelbaum, Arseniy Filin and Jambul Gegelia for valuable discussions.  This work is supported in part by the NSFC and the Deutsche Forschungsgemeinschaft (DFG) through the funds provided to the Sino-German Collaborative Research Center ``Symmetries and the Emergence of Structure in QCD'' (NSFC Grant No. 11621131001, DFG Grant No. CRC110).
\end{acknowledgments}

%

\newpage
\appendix
\section{Calculation of Pertinent Integrals}
\label{sec:pertintegrals}
\subsection{Calculation of Triangle Integral}
\label{sec:triangle}
\begin{equation}
 I_{tr}= i\int  \frac{d^4l}{(2\pi)^4} \,  \frac{l_0}{(l_0 - i \epsilon)}\, \frac{(\vec{l}+\vec{q})\cdot\vec{l}}{\big[(l+q)^2 -m^2_{\pi} + i \epsilon\big] \big[l^2-m^2_{\pi} + i \epsilon\big] }
\end{equation}
Introducing Feynman parameters, shifting $l \rightarrow l-qx$, dropping all odd powers of $l$ due to symmetry and using $q^0=0$, one finds
\begin{multline}
  I_{tr}=i\int_{0}^{1}dx   \int \frac{d^3\vec{l}}{(2\pi)^3}
 \int \frac{dl_0}{2 \pi}    \frac{l_0}{(l_0 - i \epsilon)}\\
 \frac{\vec{l}^2 - \vec{q}^2 x(1-x)
 }{\big[(l_0)^2 -\vec{l}^2 - \vec{q}\,^2 x (1-x)- m^2_{\pi} +i\epsilon\big]^2}.
\end{multline}

Executing the $l^0$- integration  with the residue theorem  and setting $\epsilon \rightarrow 0$,
\begin{equation}
  I_{tr} = \frac{i^2}{4} \int_{0}^{1}dx
    \int \frac{d^3\vec{l}}{(2\pi)^3}  
    \frac{\vec{l}^2 - \vec{q}\,^2 x(1-x)
    }{\big[\vec{l}^2 + \vec{q}\,^2 x (1-x) + m^2_{\pi} \big]^{3/2}}.
\end{equation}

Going to $(D-1)$ - dimensional spherical coordinates and inserting $\mu$,
\begin{equation}
   I_{tr}  = \frac{-\sqrt{\pi}}{(4\pi)^{D/2}} \frac{\mu ^{4-D}}{\Gamma(\frac{D-1}{2})}
    \int_{0}^{1}dx \int_{0}^{\infty} dl \frac{ l^D-l^{D-2} \vec{q}\,^2 x (1-x)
    }{\big[\vec{l}^2 + \vec{q}\,^2 x (1-x) + m^2_{\pi} \big]^{3/2}}.
\end{equation}

Executing the $l$-integration and inserting $D=4-\xi$, 
\begin{multline}
    I_{tr} = \frac{-1}{16 \pi^2} \int_{0}^{1}dx
\Bigg\{ \bigg( \frac{5}{2} \vec{q}\,^2 x (1-x)+ \frac{3}{2} m^2_{\pi} \bigg) \bigg( -\frac{2}{\epsilon} + \gamma_E -1 \\-\ln{(4\pi)}+ \ln{\bigg(\frac{\vec{q}\,^2x(1-x)+m^2_{\pi}}{\mu^2}\bigg)} \bigg)
+ 2\vec{q}\,^2x(1-x)+m^2_{\pi}
\Bigg\}
\end{multline}

Doing the $x$-integration, one finally obtains
\begin{multline}
   I_{tr} =\frac{-1}{16 \pi^2} 
\Bigg\{ \bigg( \frac{5}{12} \vec{q}\,^2+ \frac{3}{2} m^2_{\pi} \bigg) \mathcal{R} -
\frac{13}{36} \vec{q}\,^2- \frac{m^2_{\pi}}{3}\\
+ \bigg(\frac{5}{6} \vec{q}\,^2+ 3 m^2_{\pi} \bigg) \ln{\bigg(\frac{m_{\pi}}{\mu}\bigg)} 
+\bigg( \frac{5}{6} \vec{q}\,^2+ \frac{4}{3} m^2_{\pi} \bigg) L(q) \Bigg\}\\
=\frac{-\vec{q}\,^2}{16 \pi^2} 
\Bigg\{  \frac{5}{12}  \mathcal{R} -
\frac{13}{36} 
+ \frac{5}{6} \ln{\bigg(\frac{m_{\pi}}{\mu}\bigg)} 
+\frac{5}{6}  L(q) \Bigg\}+{\cal O}(\chi^4),
\label{eq:Itrfull}
\end{multline}
where $\mathcal{R}$ and $L(q)$ are given by
\begin{equation}
    \mathcal{R}= -\frac{2}{\xi} + \gamma_E -1-\ln{(4\pi)},
\end{equation}
\begin{equation}
L(q)= \frac{\sqrt{4m^2_{\pi}+ q^2}}{q} \ln{\bigg(\frac{\sqrt{4m^2_{\pi}+ q^2}+q}{2 m_{\pi}}\bigg)},
\end{equation}
and $\gamma_E$ is the Euler-Mascheroni constant.
\subsection{ Calculation of football integral}
\label{sec:football}
\begin{equation}
    I_{fb}= i\int  \frac{d^4l}{(2\pi)^4}  \frac{(l^0)^2}{\big[(l+q)^2 -m^2_{\pi} +i \epsilon\big] \big[l^2-m^2_{\pi} +i \epsilon\big]}
\end{equation}

Introducing Feynman parameters, shifting $l \rightarrow l-qx$, dropping all odd powers of $l$ due to symmetry and  executing the $l^0$-integration,
\begin{equation}
 I_{fb} = \frac{1}{4} \int_{0}^{1}dx
    \int \frac{d^3\vec{l}}{(2\pi)^3}  
    \frac{1
    }{\big[\vec{l}^2 + \vec{q}\,^2 x (1-x) + m^2_{\pi} \big] ^{1/2} }.
\end{equation}

Going to $(D-1)$ - dimensional spherical coordinates and inserting $\mu$,
\begin{multline}
   I_{fb} = \frac{\sqrt{\pi}}{(4\pi)^{D/2}} \frac{\mu ^{4-D}}{\Gamma(\frac{D-1}{2})}
    \int_{0}^{1}dx \\
    \int_{0}^{\infty} dl \frac{ l^{D-2} 
    }{\big[\vec{l}^2 + \vec{q}\,^2 x (1-x) + m^2_{\pi} + i\epsilon\big]^{1/2}}.
\end{multline}

Executing the $l$-integration and inserting $D=4-\xi$
\begin{multline}
   I_{fb} = \frac{1}{32 \pi^2} \int_{0}^{1}dx
 \big(  \vec{q}\,^2 x (1-x)+  m^2_{\pi} \big) \bigg( -\frac{2}{\xi} + \gamma_E -1-\ln{(4\pi)}\\+ \ln{\bigg(\frac{\vec{q}\,^2x(1-x)+m^2_{\pi}}{\mu^2}\bigg)} \bigg)
\end{multline}

Doing the $x$-integration, we obtain
\begin{multline}
   I_{fb} =\frac{-1}{16 \pi^2} 
\Bigg\{ -\bigg( \frac{\vec{q}\,^2}{12} + \frac{m^2_{\pi}}{2}  \bigg) \mathcal{R} +
\frac{5}{36} \vec{q}\,^2+ \frac{2 m^2_{\pi}}{3}
\\- \bigg(\frac{\vec{q}\,^2}{6} +  m^2_{\pi} \bigg) \ln{\bigg(\frac{m_{\pi}}{\mu}\bigg)} 
-\bigg( \frac{\vec{q}\,^2}{6} + \frac{4}{6} m^2_{\pi} \bigg) L(q) \Bigg\}
\\
=\frac{\vec{q}\,^2}{96 \pi^2} 
\Bigg\{ \frac{\mathcal{R}}{2}{-}
\frac{5}{6}{+} \ln{\bigg(\frac{m_{\pi}}{\mu}\bigg)} 
{+} L(q) \Bigg\}+{\cal O}(\chi^4).
\label{eq:Ifb}
\end{multline}

\subsection{Calculation of Crossed Box Integrals}
 \subsubsection{$ I^{(2)}_{box}$}
 \label{Sec:I2box}
We encounter the $I^{(2)}_{box}$ integral in the crossed box diagrams contributing to $B \Bar{B} \rightarrow B \Bar{B}$, $B^* \Bar{B} \rightarrow B^* \Bar{B}$ and  $B^* \Bar{B}^* \rightarrow B^* \Bar{B}^*$ and accordingly to the $B^{(*)} B^{(*)}$ counterparts. The $I^{(2)}_{box}$ integral is given by
\begin{equation}
   I^{(2)}_{box} = i \int \frac{d^4l}{(2\pi)^4} \frac{(\Vec{q_1}\cdot\Vec{q_2})^2}{(l^0-i\epsilon)^2 \big[q_2^2 -m^2_{\pi}+i\epsilon\big] \big[q_1^2-m^2_{\pi}+i\epsilon\big] }
\end{equation}

Expanding $q_1^2 = (l^0)^2 -\vec{q_1}^2$ and $q_2^2 = (l^0)^2 -\Vec{q_2}^2$ and introducing the Feynman parameter $x$, one finds 
   \begin{multline}
 I^{(2)}_{box} = 
  i \int_{0}^{1} dx  \int \frac{d^3\vec{l}}{(2\pi)^3} (\Vec{q_1}\cdot\Vec{q_2})^2
\\ 
\times\int \frac{dl^0}{2 \pi}   \frac{1}{ \big[l^0-i\epsilon\big]^2} \frac{1}{\big[(l_0)^2 - a^2 +i\epsilon\big]^2},
\end{multline}
where $a^2= (\Vec{q_2}^2-\Vec{q_1}^2)x + \Vec{q_1}^2 + m^2_{\pi}$.
Executing $l^0$ integration gives

\begin{equation}
   I^{(2)}_{box} = \frac{-3}{4} \int_{0}^{1}dx
    \int \frac{d^3\vec{l}}{(2\pi)^3} \frac{(\Vec{q_1}\cdot\Vec{q_2})^2}{\big[ (\vec{q_2}^2- \Vec{q_1}^2)x+ \vec{q_1}^2+m^2_{\pi}\big]^{5/2} }.
\end{equation}

Shifting $\vec{l} \rightarrow \vec{l}+ \vec{p}$ such that $\vec{q_1} = \vec{p}- \vec{l} \rightarrow -\vec{l}$ and  $\vec{q_2} = \vec{p'}- \vec{l} \rightarrow -\vec{l} +\vec{q}$ with $\vec{q}= \vec{p'}-\vec{p}$,
\begin{multline}
    I^{(2)}_{box} =  \frac{-3}{4} \int_{0}^{1}dx
  \\  \int \frac{d^3\vec{l}}{(2\pi)^3} \frac{\big(\Vec{l}\cdot (\Vec{l} -\vec{q}) \big)^2}{\big[\vec{l}^2 +(-2 \vec{l} \cdot \vec{q} + \vec{q}\,^2)x + m^2_{\pi}\big]^{5/2} }.
\end{multline}
Following the same steps as for the above integrals,
\begin{multline}
I^{(2)}_{box}= \frac{-1}{16 \pi^2} \int_{0}^{1}dx
\Bigg\{ \bigg( \bigg[-\frac{35}{2}x^2 + \frac{35}{2}x -1\bigg] \vec{q}\,^2 +\frac{15}{2} m^2_{\pi}\bigg) \mathcal{R} \\
+(-22x^2+ 22x-1)\vec{q}\,^2 + 8 m^2_{\pi} + \frac{2\vec{q}\,^4 x^2 (1-x)^2}{m^2_{\pi}+ \vec{q}\,^2 x(1-x)}\\
+\bigg( \bigg[ -\frac{35}{2}x^2 + \frac{35}{2}x -1\bigg]\vec{q}\,^2 +\frac{15}{2} m^2_{\pi}\bigg) \\ \ln{\bigg(\frac{\vec{q}\,^2 x(1-x)+m^2_{\pi}}{\mu^2}\bigg)} \Bigg\}
\end{multline}

Performing the $x$-integration, one finally obtains
\begin{multline}
I^{(2)}_{box} =   \frac{-1}{16 \pi^2} \Bigg\{ \bigg( \frac{23}{12} \vec{q}\,^2 + \frac{15}{2}m^2_{\pi} \bigg)\mathcal{R} + \frac{5}{36}\vec{q}\,^2 + \frac{8}{3}m^2_{\pi} 
+\bigg( \frac{23}{6}\vec{q}^2 \\+ 15 m^2_{\pi}\bigg) \ln{\bigg(\frac{m_{\pi}}{\mu}\bigg)} +\bigg( \frac{23}{6}\vec{q}\,^2 + \frac{10}{3}m^2_{\pi} + \frac{8 m^4_{\pi}}{4m^2_{\pi}+\vec{q}\,^2 }\bigg) L(q)\Bigg\}\\
{=}   \frac{-23\vec{q}\,^2}{96 \pi^2} \Bigg\{ \frac{\mathcal{R}}{2} {+} \frac{5}{138}{+} \ln{\bigg(\frac{m_{\pi}}{\mu}\bigg)} {+} L(q)\Bigg\}
{+}{\cal O}(\chi^4).
\label{eq:I2box}
\end{multline}

\subsection{Calculation of Planar Box integral}
\label{sec:planarbox}
 As mentioned earlier, the $I^{(1)}_{box}$ integral splits into a reducible and an irreducible TPE contribution. Because the iterations of the OPE within the Lippmann-Schwinger equation take care of the reducible part, it is omitted here and one gets  $I^{(1)}_{box} \rightarrow I^{(2)}_{box}$ \cite{Machleidt:2011}, with the box integrals  defined in Eqs.~\eqref{Eq:Ibox1} and \eqref{Eq:Ibox2}.  For the sake of completeness, 
 below we provide a derivation of this result. 

The starting point is the planar box integral
 \begin{multline}
    I^{(1)}_{box} =i \int \frac{d^4l}{(2\pi)^4} \frac{1}{\big[l^0+i\epsilon\big] \big[l^0-i\epsilon\big]} \\\frac{(\vec{q_2} \cdot \vec{q_1})^2}{ \big[q_2^2 -m^2_{\pi} +i\epsilon \big] \big[q_1^2-m^2_{\pi} +i\epsilon \big]  }.
\end{multline}
 Expanding $q_1^2= (l^0)^2- \Vec{q_1}^2$ and $q_2^2= (l^0)^2- \Vec{q_2}^2$ and introducing the Feynman parameter $x$,
    \begin{multline}
 I^{(1)}_{box} = 
  i \int_{0}^{1} dx  \int \frac{d^3\vec{l}}{(2\pi)^3} (\Vec{q_1}\cdot\Vec{q_2})^2
\\ 
\times\int \frac{dl^0}{2 \pi}   \frac{1}{\big[l^0+i\epsilon\big] \big[l^0-i\epsilon\big]} \frac{1}{\big[(l_0)^2 - a^2 +i\epsilon\big]^2},
\end{multline}
where $a^2= (\Vec{q_2}^2-\Vec{q_1}^2)x + \Vec{q_1}^2 + m^2_{\pi}$. In contrast to the crossed box integral $I^{(2)}_{box} $, the $l^0$-integration diverges as $\epsilon \rightarrow 0$, and our LO approximation of the heavy-meson propagator produces nonphysical poles.
We can avoid this pinch singularity (the singularity is squeezed between $+i \epsilon$ and $- i\epsilon$), by including higher-order corrections to the heavy meson propagator \cite{Weinberg:1991}. 
Specifically, including   ${\cal O}(Q^2)$ terms, namely the kinetic energies of the heavy mesons, 
shifts the poles in the heavy mesons propagators, making them distinct  and, consequently, avoiding the singularity. 
This can be achieved by replacing $i \epsilon \rightarrow i\zeta$, with
 \begin{equation}
     i \zeta= \frac{\Vec{k}\,^2}{2 m_B} - \frac{\Vec{l}\,^2}{2 m_B} + i\epsilon,
 \end{equation}
where $k$ is the on shell momentum of the heavy mesons, $k^2= m_B E$. 
One finds
\begin{multline}
    I^{(1)}_{box} =i\int_{0}^{1} dx  \int \frac{d^3\vec{l}}{(2\pi)^3} (\Vec{q_1}\cdot\Vec{q_2})^2
 \int \frac{dl^0}{2 \pi}   \frac{1}{\big[l^0+i\zeta \big] \big[l^0-i\zeta \big]  } \\ \frac{1}{\big[(l_0)^2 - a^2 +i\epsilon\big]^2}.
\end{multline} 
Executing $l^0$-integration, 
\begin{equation}
     I^{(1)}_{box} =i \int_{0}^{1} dx  \int \frac{d^3\vec{l}}{(2\pi)^3} (\Vec{q_1}\cdot\Vec{q_2})^2
\frac{i}{4} \, \frac{2\,a- i \zeta}{ \big[ i \zeta a^3 (a-i\zeta)^2 \big] }.
\end{equation}
Expanding the fraction for $a \gg i\zeta$ and dropping terms of order $\mathcal{O}(\zeta)$,
\begin{equation}
     I^{(1)}_{box} =i\int_{0}^{1} dx  \int \frac{d^3\vec{l}}{(2\pi)^3} (\Vec{q_1}\cdot\Vec{q_2})^2
\frac{i}{4} \,\bigg( \frac{3}{a^5 }+ \frac{2}{i\zeta \,a^4}+ \mathcal{O}(\zeta) \bigg).
\end{equation}
The first term of the expansion is the integral $I^{(2)}_{box}$ evaluated in the previous section. The second term is the iterated OPE, and the dropped term scales as $(i\zeta)/a^6$. Since $i\zeta \sim p^2_{\rm typ}/m_B$ and $a^2 \sim p^2_{\rm typ}$, the neglected term is suppressed as $\mathcal{O}( p_{\rm typ}/m_B)\sim \mathcal{O}(\chi^2)$ relative to the leading one.

Inserting $i\zeta= (\Vec{k}\,^2-\Vec{l}\,^2 + i\epsilon)/2 m_B$, one finds 
 \be 
 I^{(1)}_{box} = I^{(2)}_{box}+ \Delta I,
\ee 
 where 
  \begin{multline}\label{Eq:Delta}
 \Delta I= i^2\, m_B 
 \int_{0}^{1} dx  \\ 
\times \int \frac{d^3\vec{l}}{(2\pi)^3}\frac{(\vec{q_2} \cdot \vec{q_1})^2}{\big[ \Vec{k}^2-\Vec{l}^2 + i\epsilon \big]  \big[ (\Vec{q_2}^2-\Vec{q_1}^2)x + \Vec{q_1}^2 + m^2_{\pi} \big]^2  }   \\ 
 = i^2 \, m_B  
 \int \frac{d^3\vec{l}}{(2\pi)^3} 
 \frac{(\vec{q_2} \cdot \vec{q_1})^2}{\big[ \Vec{k}^2-\Vec{l}^2\! +\! i\epsilon\big]  \big[ \Vec{q_2}^2 + m^2_{\pi} \big] \big[\Vec{q_1}^2  + m^2_{\pi} \big]  },    
\end{multline}
which is the non-relativistic version of the twice iterated OPE. 
Indeed, the iterated potential can be written as \cite{Kaiser,Krug},
\begin{multline}
    V^{\rm eff}_{\rm TPE,it}(\Vec{p'},\Vec{p})= - m_B 
 \int \frac{d^3\vec{l}}{(2\pi)^3} 
 \frac{V^{\rm eff}_{\rm OPE}(\Vec{p'},\Vec{l}) \, V^{\rm eff}_{\rm OPE}(\Vec{l},\Vec{p}) }
 {\big[ \Vec{k}\,^2-\Vec{l}\,^2 +i\epsilon \big],    }    
\end{multline}
which is in agreement with \eqref{Eq:Delta}.

Since, we only consider irreducible contributions, we drop this reducible term and keep track of the irreducible integral  $I^2_{box}$ instead. In summary, the treatment of the planar box comes down to sign flip $+i\epsilon \rightarrow -i\epsilon$ for one of the heavy meson propagators.

\subsubsection{ Calculation of $B^* \Bar{B}^* \rightarrow B^* \Bar{B}^*$ and $B^* B^* \rightarrow B^* B^*$ pertinent loop integrals }
\label{B*B*loop}

The box contributions are identical for the $B^* \Bar{B}^* \rightarrow B^* \Bar{B}^*$ and $B^* B^* \rightarrow B^* B^*$ scattering processes. In total, one has eight box diagrams and the resulting contribution is, 
 \begin{widetext}
     \begin{equation}
    \begin{split}
        V^{box}_{B^* \Bar{B}^* \rightarrow B^* \Bar{B}^*} &= V_{B_{7.1}} +V_{B_{7.2}}+ V_{B_{7.3}}+ V_{B_{7.4}} +V_{C_{7.1}} +V_{C_{7.2}}+ V_{C_{7.3}}+ V_{C_{7.4}}
        \\ &=
         \frac{g^4}{16 f^4_{\pi}}   ( \epsilon_{1,i} \epsilon^*_{1',k} \epsilon_{2,l}  \epsilon^*_{2',n})  i \int \frac{d^4l}{(2\pi)^4} \frac{ [3A -2(\Vec{\tau_1} \cdot  \Vec{\tau_2})B]}
        {[l_0-i\epsilon]^2 \big[q_2^2 -m^2_{\pi}\big] \big[q_1^2-m^2_{\pi}\big]  }
        \end{split}
\end{equation} 
where,
\begin{multline}
    A= 2\big[ (q_2)_{k}  (q_1)_{i} (q_2)_n (q_1)_l - (q_2)_i (q_1)_k (q_2)_n (q_1)_l - (q_2)_k (q_1)_i (q_2)_l (q_1)_n 
    + (q_2)_i (q_1)_k (q_2)_l (q_1)_n \big] \\
    + 2 \delta_{ik} (\vec{q_2} \cdot \vec{q_1}) \big[ (q_2)_n (q_1)_l- (q_2)_l (q_1)_n) \big]
\end{multline}
and
\begin{equation}
    B = 2 \delta_{ik} \delta_{ln} (\vec{q_2} \cdot \vec{q_1})^2 + 2 \delta_{ln} (\vec{q_2} \cdot \vec{q_1})^2 \big[ 
    (q_2)_{k}  (q_1)_{i}- (q_2)_{i}  (q_1)_{k} \big]
\end{equation}
The last term in $A$ and the second term in   $B$ will vanish, due to tensor decomposition being symmetric. Solving the remaining terms using tensor decomposition,
\begin{multline}
V^{box}_{B^* \Bar{B}^* \rightarrow B^* \Bar{B}^*} =   \frac{g^4}{4 f^4_{\pi}} ( \epsilon_{1,i} \epsilon^*_{1',k} \epsilon_{2,l}  \epsilon^*_{2',n}) \Bigg\{ -(\Vec{\tau_1} \cdot  \Vec{\tau_2})  \delta_{ik}\delta_{ln } I^{(2)}_{box} + \frac{1}{2} \big[\delta_{il}\delta_{kn}- \delta_{in}\delta_{kl} \big] \big[I^{(3)}_{box}-I^{(2)}_{box}+\frac{1}{\vec{q}\,^2}(I^{(4)}_{box}-I^{(5)}_{box})\big] \\
+ \frac{3}{4 \vec{q}\,^4} \big[ q_k q_n\delta_{il}-q_k q_l\delta_{in}+q_i q_l\delta_{kn}-q_i q_n\delta_{kl} \big] \big[ I^{(5)}_{box}-I^{(4)}_{box} \big] \Bigg\}
\end{multline}
where,

\begin{align}
I^{(3)}_{box} &= i \int \frac{d^4l}{(2\pi)^4} \frac{\vec{q_1}^2 \vec{q_2}^2}{(l^0-i\epsilon)^2 \big[q_2^2 -m^2_{\pi}+i\epsilon\big] \big[q_1^2-m^2_{\pi}+i\epsilon\big] } \ , 
&  I^{(4)}_{box} &= i \int \frac{d^4l}{(2\pi)^4} \frac{\Vec{q_1}^2(\Vec{q_2}\cdot \Vec{q} )^2}{(l^0-i\epsilon)^2 \big[q_2^2 -m^2_{\pi}+i\epsilon\big] \big[q_1^2-m^2_{\pi}+i\epsilon\big] } \,  \nonumber \\
     I^{(5)}_{box} &= i \int \frac{d^4l}{(2\pi)^4} \frac{\Vec{q_2}^2(\Vec{q_1}\cdot \Vec{q} )^2}{(l^0-i\epsilon)^2 \big[q_2^2 -m^2_{\pi}+i\epsilon\big] \big[q_1^2-m^2_{\pi}+i\epsilon\big] } 
\end{align}
\end{widetext}

We will solve each of the three integral, starting with the  $I^{(3)}_{box}$ integral. 
Expanding $q_1^2 = (l^0)^2 -\vec{q_1}^2$ and $q_2^2 = (l^0)^2 -\Vec{q_2}^2$, using Feynman parameters and executing the $l^0$-integration  (and setting $\epsilon \rightarrow 0$).
\begin{equation}
     I^{(3)}_{box}  =  \frac{-3}{4} \int_{0}^{1}dx
    \int \frac{d^3\vec{l}}{(2\pi)^3} \frac{\vec{q_1}^2 \vec{q_2}^2}{\big[ (\vec{q_2}^2- \Vec{q_1}^2)x+ \vec{q_1}^2+m^2_{\pi}\big]^{5/2} }
\end{equation}

Shifting $\vec{l} \rightarrow \vec{l}+ \vec{p}$ such that $\vec{q_1} = \vec{p}- \vec{l} \rightarrow -\vec{l}$ and  $\vec{q_2} = \vec{p'}- \vec{l} \rightarrow -\vec{l} +\vec{q}$ with $\vec{q}= \vec{p'}-\vec{p}$,
\begin{multline}
     I^{(3)}_{box}  =  \frac{-3}{4} \int_{0}^{1}dx
    \\\int \frac{d^3\vec{l}}{(2\pi)^3} \frac{\Vec{l}\,^2 (\Vec{l} -\vec{q})^2}{\big[\vec{l}^2 +(-2 \vec{l} \cdot \vec{q} + \vec{q}\,^2)x + m^2_{\pi}\big]^{5/2} }.
\end{multline}
Following the same steps as for the earlier integrals, one finds
\begin{multline}
 I^{(3)}_{box}  =   \frac{-1}{16 \pi^2} \Bigg\{ \bigg(   \frac{15}{2}m^2_{\pi} - \frac{1}{12} \vec{q}\,^2\bigg)\mathcal{R} + \frac{1}{4}\vec{q}\,^2 +  \frac{8}{3}m^2_{\pi} 
\\+\bigg(  15 m^2_{\pi}  -\frac{1}{6}\vec{q}\,^2 \bigg) \ln{\bigg(\frac{m_{\pi}}{\mu}\bigg)} +\bigg( \frac{10}{3}m^2_{\pi} -\frac{1}{6}\vec{q}\,^2 \\ \hfill + \frac{8 m^4_{\pi}}{4m^2_{\pi}+\vec{q}\,^2 }\bigg) L(q)\Bigg\}
\\
= \frac{-\vec{q}\,^2}{16 \pi^2} \Bigg\{   \frac{-1}{12}  \mathcal{R} + \frac{1}{4} 
 -\frac{1}{6} \ln{\bigg(\frac{m_{\pi}}{\mu}\bigg)} -\frac{1}{6} L(q)\Bigg\} {+}{\cal O}(\chi^4)
\end{multline}

One proceeds in a similar manner for the integral $I^{4}_{box} $: 
\begin{multline} 
    I^{4}_{box} = i\int \frac{d^4l}{(2\pi)^4} \\
   \times \int_0^1 dx \frac{\Vec{q_1}^2(\Vec{q_2}\cdot \Vec{q} )^2}{(l^0-i\epsilon)^2 \big[(l^0)^2 - (\vec{q_2}^2- \Vec{q_1}^2)x- \vec{q_1}^2-m^2_{\pi}+i\epsilon\big]^2 }.
\end{multline} 
Executing the $l^0$-integration, setting $\epsilon \rightarrow 0$ and shifting $\vec{l} \rightarrow \vec{l}+ \vec{p}$ such that $\vec{q_1} = \vec{p}- \vec{l} \rightarrow -\vec{l}$ and  $\vec{q_2} = \vec{p'}- \vec{l} \rightarrow -\vec{l} +\vec{q}$,
\begin{equation}
     I^{4}_{box}  = i^2 \frac{3}{4} \int_{0}^{1}dx
    \int \frac{d^3\vec{l}}{(2\pi)^3} \frac{\Vec{l}\,^2 [(\Vec{l} -\vec{q}) \cdot \vec{q}]^2 }{\big[\vec{l}^2 +(-2 \vec{l} \cdot \vec{q} + \vec{q}\,^2)x + m^2_{\pi}\big]^{5/2} }
\end{equation}

Solving the $l$-integration, inserting $D=4-\epsilon$ and doing the $x$-integration, one finds
\begin{multline}
 I^{(4)}_{box}  =   \frac{-1}{16 \pi^2} \vec{q}\,^2 \Bigg\{ \bigg(   \frac{5}{2}m^2_{\pi} - \frac{7}{12} \vec{q}\,^2\bigg)\mathcal{R} - \frac{7}{36}\vec{q}\,^2 -  \frac{7}{3}m^2_{\pi} 
\\+\bigg(  5 m^2_{\pi} -\frac{7}{6}\vec{q}\,^2 \bigg) \ln{\bigg(\frac{m_{\pi}}{\mu}\bigg)} +\bigg( \frac{10}{3}m^2_{\pi} -\frac{7}{6}\vec{q}\,^2 \\ \hfill + \frac{8 m^4_{\pi}}{4m^2_{\pi}+\vec{q}\,^2 }\bigg) L(q)\Bigg\}
\\
=   \frac{-\vec{q}\,^4}{16 \pi^2}  \Bigg\{   \frac{-7}{12}  \mathcal{R} - \frac{7}{36} 
-\frac{7}{6} \ln{\bigg(\frac{m_{\pi}}{\mu}\bigg)} -\frac{7}{6} L(q)\Bigg\}+  \mathcal{O}(\chi^4)
\end{multline}

Finally, perfoming the calculations along the lines given above, one finds that the $I^{(5)}_{box}$ integral is given by
\begin{multline}
 I^{(5)}_{box}  =   \frac{-1}{16 \pi^2} \vec{q}\,^2 \Bigg\{ \bigg(   \frac{5}{2}m^2_{\pi} - \frac{1}{4} \vec{q}\,^2\bigg)\mathcal{R} - \frac{1}{12}\vec{q}\,^2 +  3 m^2_{\pi} 
\\ +\bigg(  2 m^2_{\pi}  -\frac{1}{6}\vec{q}\,^2 \bigg) \ln{\bigg(\frac{m_{\pi}}{\mu}\bigg)} +\bigg(  \frac{7}{3}\vec{q}\,^2 - \frac{8}{3}m^2_{\pi}  \\ \hfill 
+ \frac{8 m^4_{\pi}}{4m^2_{\pi}+\vec{q}\,^2 }\bigg) L(q)\Bigg\}
\\
=\frac{- \vec{q}\,^4}{16 \pi^2} \Bigg\{   \frac{-1}{4} \mathcal{R} - \frac{1}{12}
 -\frac{1}{6}  \ln{\bigg(\frac{m_{\pi}}{\mu}\bigg)} + \frac{7}{3}  L(q)\Bigg\} + \mathcal{O}(\chi^4)
\end{multline} 

\subsubsection{$ (I^{(6)}_{box})_{in}$}
We encounter this integral in $B^* \Bar{B} \rightarrow B \Bar{B}^*$, $B \Bar{B} \rightarrow B^* \Bar{B}^*$  and in the subsequent $B^{(*)} B^{(*)}$ counterparts. The $(I^{(6)}_{box})_{in}$ integral is given by,
\begin{multline}
    (I^{(6)}_{box})_{in}  = i\int \frac{d^4l}{(2\pi)^4} \\ \frac{\epsilon_{ijs}  \epsilon_{nrm} (q_2)_j (q_1)_s (q_2)_r(q_1)_m  }  {[l_0-i\epsilon]^2 \big[q_2^2 -m^2_{\pi}+i\epsilon\big] \big[q_1^2-m^2_{\pi}+i\epsilon\big]  }.
\end{multline}\\
Repeating the same steps as in the earlier cases gives
\begin{multline}
    (I^{(6)}_{box})_{in} =  \frac{-3}{4}\mu^{4-D} \int_{0}^{1}dx
    \int \frac{d^{D-1}\vec{l}}{(2\pi)^{D-1}}\epsilon_{ijs}  \epsilon_{nrm}\\
    \frac{  [l_j+q_j(x-1)] (l_s+q_s s)[l_r+q_r(x-1)](l_m+q_mx)}    {\big[\vec{l}^2 + \vec{q}\,^2 x (1-x)+ m^2_{\pi}\big]^{5/2} }.
\end{multline}
Due to anti-symmetry of the Levi-Civita tensor, most of the numerator vanishes,
\begin{multline}
   (I^{(6)}_{box})_{in}  =  \frac{-3}{4}\mu^{4-D} \int_{0}^{1}dx
  \\  \int \frac{d^{D-1}\vec{l}}{(2\pi)^{D-1}}\frac{\epsilon_{ijs}  \epsilon_{nrm} (-l_s q_j)(-l_m q_r)}    {\big[\vec{l}^2 + \vec{q}\,^2 x (1-x)+ m^2_{\pi}\big]^{5/2} }.
\end{multline}
Going to $(D-1)$-dimensional spherical coordinates and using $l_i l_j \rightarrow \frac{\Vec{l}^2 \, \delta_{ij}}{D-1}$, we get
\begin{multline}
    (I^{(6)}_{box})_{in} = \frac{-3\sqrt{\pi}}{(4\pi)^{D/2}} \frac{\mu ^{4-D}}{\Gamma(\frac{D-1}{2})}
    \int_{0}^{1}dx (\delta_{in}\vec{q}\,^2- q_i q_n)
   \\ \int_{0}^{\infty} dl \frac{\frac{l^D}{D-1}}
   {\big[\vec{l}^2 + \vec{q}\,^2 x (1-x)+ m^2_{\pi}\big]^{5/2} }.
\end{multline}
Doing the $l$- and $x$- integration and inserting $D=4-\xi$, one finally finds 
\begin{multline}
    (I^{(6)}_{box})_{in} = \frac{-1}{16 \pi^2} (\delta_{in}\vec{q}\,^2- q_i q_n) \\ \Bigg\{ - \mathcal{R} + 1  -2L(q) -2\ln{\bigg(\frac{m_{\pi}}{\mu}\bigg)} \Bigg\}.
\end{multline}

\subsubsection{$ (I^{(7)}_{box})_{ikn}$}
We encounter this  integral in $B^* \Bar{B} \rightarrow B^* \Bar{B}^*$, $B \Bar{B}^* \rightarrow B^* \Bar{B}^*$ and in the subsequent $B^{(*)} B^{(*)}$ counterparts . The $(I^{(7)}_{box})_{ikn}$ integral is given by,
\begin{multline}
    (I^{(7)}_{box})_{ikn}= i \int \frac{d^4l}{(2\pi)^4} \\
    \frac{ \big( (q_2)_k \, (q_1)_i - (q_2)_i \,(q_1)_k \big)  \epsilon_{num}    (q_2)_u (q_1)_m
    }  {[l_0-i\epsilon]^2  \big[q_2^2 -m^2_{\pi}\big] \big[q_1^2-m^2_{\pi}\big]  }
\end{multline}
Introducing Feynman parameters,  executing the $l^0$-integration, and shifting $\vec{l} \rightarrow \vec{l}+ \vec{p}$ such that $\vec{q_1} = \vec{p}- \vec{l} \rightarrow -\vec{l}$ and  $\vec{q_2} = \vec{p'}- \vec{l} \rightarrow -\vec{l} +\vec{q}$ with $\vec{q}= \vec{p'}-\vec{p}$:
\begin{multline}
   (I^{(7)}_{box})_{ikn} =  \frac{-3}{4} \int_{0}^{1}dx
    \int \frac{d^3\vec{l}}{(2\pi)^3} \\
    \frac{ \big( (l-q)_k \, (l)_i - (l-q)_i \,(l)_k \big)  \epsilon_{num}    (l-q)_u (l)_m
    }  {\big[\vec{l}^2 +(-2 \vec{l} \cdot \vec{q} + \vec{q}\,^2)x + m^2_{\pi}\big]^{5/2} }.
\end{multline}
All terms proportional to $\epsilon_{num}l_u l_m$ or $\epsilon_{num}q_u q_m$ and every term of odd power in $l$ vanishes due to symmetry resulting in
\begin{multline}
   (I^{(7)}_{box})_{ikn} =  \frac{-3}{4}\mu^{4-D} \int_{0}^{1}dx
   \\  \int \frac{d^{D-1}\vec{l}}{(2\pi)^{D-1}}
    \frac{ \big( l_k \, q_i - l_i \,q_k \big)  \epsilon_{num} 
    \big( - l_m q_u \big)
    }  {\big[\vec{l}^2 +(-2 \vec{l} \cdot \vec{q} + \vec{q}\,^2)x + m^2_{\pi}\big]^{5/2} }.
\end{multline}
Solving the remaining integral like in the earlier integral yields
\begin{multline}
     (I^{(7)}_{box})_{ikn}=\frac{-1}{16 \pi^2} (\epsilon_{nku}q_u q_i
-\epsilon_{niu}q_u q_k) \\
\Bigg\{  -\mathcal{R} + 1  -2L(q) -2\ln{\bigg(\frac{m_{\pi}}{\mu}\bigg)} \Bigg\}.
\end{multline}
\section{Partial Wave Projectors}
\label{PWP}
In this section, we have presented the complete set of partial wave projectors used for calculating the potentials
\cite{Baru:2019xnh}. In what follows, 
$\epsilon$ refers to the polarization vector of the heavy vector meson and $\vec n= \vec p/p$.

\begin{equation}
  P(B\Bar{B}(^1 S_0))= 1   
\end{equation}
\begin{equation}
  P(B^* \Bar{B}^*(^1 S_0))= \frac{1}{\sqrt{3}} \big( \vec{\epsilon_1} \cdot \vec{\epsilon_2}\big)   
\end{equation}
\begin{equation}
P(B^* \Bar{B}^*(^5 D_0))= - \sqrt{\frac{3}{8}} S_{xy} v_{xy}    
\end{equation}

\begin{equation}
      P(B \Bar{B}^*(^3 S_1))_x = \epsilon_{2,x}
\end{equation}
\begin{equation}
    P(B^* \Bar{B}(^3 S_1))_h = \epsilon_{1,h} 
\end{equation}
\begin{equation}
  P(B \Bar{B}^*(^3 D_1))_x =- \frac{3}{\sqrt{2}} \epsilon_{2,y}v_{xy}   
\end{equation}
\begin{equation}
  P(B^* \Bar{B}(^3 D_1))_x =- \frac{3}{\sqrt{2}} \epsilon_{1,y}v_{xy}   
\end{equation}
\begin{equation}
 P(B^* \Bar{B}^*(^5 D_1))_h =- \frac{\sqrt{3}}{2}i  \epsilon_{hxj}S_{xy}v_{jy}   
\end{equation}
\begin{equation}
    P(B^* \Bar{B}^*(^3 S_1))_x =A_x 
\end{equation}
\begin{equation}
  P(B^* \Bar{B}^*(^3 D_1))_h =- \frac{3}{\sqrt{2}} A_x v_{hx}   
\end{equation}
\begin{equation}
    P(B \Bar{B}(^1 D_2))_{xy} = - \sqrt{\frac{15}{2}}v_{xy} 
\end{equation}
\begin{equation}
     P(B \Bar{B}^*(^3 D_2))_{zy} = - \frac{\sqrt{5}}{2}i \epsilon_{2,h} \big(\epsilon_{zhx}v_{xy}+\epsilon_{yhx}v_{xz}\big)
\end{equation}
\begin{equation}
  P(B^* \Bar{B}^*(^5 S_2))_{xy} = \frac{1}{2} S_{xy}  
\end{equation}
\begin{equation}
    P(B^* \Bar{B}^*(^1 D_2))_{xy} = - \sqrt{\frac{5}{2}} \big( \vec{\epsilon_1} \cdot \vec{\epsilon_2}\big)  v_{xy}
\end{equation}
 \begin{equation}
 P(B^* \Bar{B^*}(^5 D_2))_{zy} = - \sqrt{\frac{45}{56}} \big( S_{zx}v_{xy}+S_{yx}v_{xz} -\frac{2}{3}\delta_{xy} S_{lx}v_{lx}  \big)
\end{equation}
\begin{equation}
     P(B^* \Bar{B}^*(^5 G_2))_{zw}=  \sqrt{\frac{175}{32}} S_{xy} v_{xyzw}
\end{equation}

where,
\begin{equation}
 v_{xy} =n_x n_y-\frac{1}{3}\delta_{xy}   
\end{equation}
\begin{equation}
    S_{xy} = \epsilon_{1,x}\epsilon_{2,y}+\epsilon_{1,y}\epsilon_{2,x} -\frac{2}{3} \delta_{xy} \big( \vec{\epsilon_{1}} \cdot \vec{\epsilon_{2}}\big)
\end{equation}
\begin{equation}
 A_x = \frac{i}{\sqrt{2}} \epsilon_{xyz} \epsilon_{1,y} \epsilon_{2,z}   
\end{equation}
\begin{multline}
        v_{xyzw}= n_x n_y n_z n_w -\frac{1}{7} \big( n_x n_y \delta_{zw} +n_x n_z \delta_{yw}\\ 
        + n_x n_w \delta_{yz} + n_y n_z \delta_{xw}+
        n_y n_w \delta_{xz}+ n_z n_w \delta_{xy} \big) 
        \\+ \frac{1}{35} \big( \delta_{xy}\delta_{zw} + \delta_{xz}\delta_{yw} + \delta_{xw}\delta_{yz} \big)
\end{multline}

The projectors are normalised as:
\begin{equation}
    \int \frac{d\Omega_n}{4 \pi} P^{\dagger} (\alpha, \vec{n})  P(\alpha, \vec{n} )= 2J+1
\end{equation}

\medskip

\section{The effective potentials}
\label{alleffpotentials}
In this section, all the OPE and TPE potentials to $B^{(*)} \Bar{B}^{(*)} \rightarrow B^{(*)} \Bar{B}^{(*)}$ and $B^{(*)} B^{(*)} \rightarrow B^{(*)} B^{(*)}$ scattering are presented.
\begin{widetext}
     
 \subsection{\texorpdfstring{$\bm{ B \bar{B} \rightarrow B \bar{B}}$}{} }
\begin{equation}
    V_{\rm OPE} (B \Bar{B} \rightarrow B \Bar{B}) =0 
\end{equation}
\begin{equation}
\begin{split}
     V_{\rm TPE} (B \Bar{B} \rightarrow B \Bar{B}) 
     &=  \frac{1}{ 2 f^4_{\pi}} (\vec{\tau_1}\cdot \vec{\tau_2} )\bigg(   I_{fb} - \frac{g^4}{2} I^{(2)}_{box}  \bigg)\\
     &= \frac{\vec{q}\,^2}{16 \pi^2 f^4_{\pi}} \big(\vec{\tau_1}\cdot \vec{\tau_2} \big) \Bigg\{ 
     \mathcal{R} \bigg[ \frac{23}{48}  g^4   
     + \frac{1}{24} \bigg] 
    +  \bigg(\frac{5}{144}   g^4 - \frac{5}{72}  \bigg)  \\
     &+  \bigg( \frac{23}{24} g^4 + \frac{1}{12} \bigg) \ln{\bigg(\frac{m_{\pi}}{\mu}\bigg)}  
     + L(q)  \bigg( \frac{23}{24} g^4 + \frac{1}{12} \bigg)\Bigg\}
\end{split}
\end{equation}
\subsection{\texorpdfstring{$\bm{B \Bar{B} \rightarrow B^* \Bar{B}^*}$}{} }
\begin{equation}
    V_{\rm OPE} (B \Bar{B} \rightarrow B^* \Bar{B}^*) = - \frac{g^2 }{ 4 f^2_{\pi} } (\vec{\tau_1} \cdot \vec{\tau_2}) (\epsilon^*_{1',k} \epsilon^*_{2',n})
    \frac{q_k q_n}{\Vec{q}\,^2+m^2_{\pi}}
\end{equation}

\begin{equation}
    \begin{split}
  V_{\rm TPE} (B \Bar{B} \rightarrow B^* \Bar{B}^*)&=  \frac{3}{128 \pi^2 f^4_{\pi}} g^4 (\epsilon^*_{1',k}  \epsilon^*_{2',n}) 
     \big(\delta_{kn} \vec{q}\,^2- q_k\, q_n \big) \Bigg\{  -\mathcal{R} + 1  - 2L(q) - 2\ln{\bigg(\frac{m_{\pi}}{\mu}\bigg)} \Bigg\}
    \end{split}
\end{equation}
\subsection{\texorpdfstring{$\bm{B^* \Bar{B} \rightarrow B^* \Bar{B}}$}{} }
\begin{equation}
    V_{\rm OPE} (B^* \Bar{B} \rightarrow B^* \Bar{B}) =0 
\end{equation}
\begin{equation}
\begin{split}
   V_{\rm TPE} (B^* \Bar{B} \rightarrow B^* \Bar{B}) &= (\epsilon^*_{1'} \cdot \epsilon_{1}) V_{\rm TPE} (B \Bar{B} \rightarrow B \Bar{B})
\end{split}
\end{equation}
\subsection{\texorpdfstring{$\bm{B^* \Bar{B} \rightarrow B \Bar{B}^*}$}{} }
\begin{flalign}
    \begin{split}
        V_{\rm OPE} (B^* \Bar{B} \rightarrow B \Bar{B}^*) &  = - \frac{g^2 }{ 4f^2_{\pi} } (\vec{\tau_1} \cdot \vec{\tau_2}) (\epsilon^*_{2',n} \epsilon_{1,i})
    \frac{q_iq_n}{\Vec{q}\,^2+m^2_{\pi}}
    \end{split}
\end{flalign}
\begin{equation}
\begin{split}
 V_{\rm TPE} (B^* \Bar{B} \rightarrow B \Bar{B^*}) 
     &=  \frac{3}{128 \pi^2 f^4_{\pi}} g^4 (\epsilon^*_{2',n} \epsilon_{1,i}) (\delta_{in}\vec{q}\,^2- q_i q_n)\Bigg\{  -\mathcal{R} + 1  -2L(q) -2\ln{\bigg(\frac{m_{\pi}}{\mu}\bigg)} \Bigg\}
\end{split}
\end{equation}

\subsection{\texorpdfstring{$\bm{B^* \Bar{B} \rightarrow B^* \Bar{B}^*}$}{} }
\begin{equation}
    \begin{split}
  V_{\rm OPE} (B^* \Bar{B} \rightarrow B^* \Bar{B}^*) &=  i \frac{g^2 }{ 4f^2_{\pi} } (\vec{\tau_1} \cdot \vec{\tau_2})  \epsilon_{ikr} ( \epsilon_{1,i} \epsilon^*_{1',k}  \epsilon^*_{2',n})
  \frac{q_r q_n}{\Vec{q}\,^2+m^2_{\pi}}
    \end{split}
\end{equation}
\begin{equation}
    \begin{split}
      V_{\rm TPE} (B^* \Bar{B} \rightarrow B^* \Bar{B}^*) &=  i\, \frac{3 }{128 \pi^2 f^4_{\pi}} g^4  (\epsilon_{1,i} \epsilon^*_{1',k} \epsilon^*_{2',n} ) 
 \, \big( \epsilon_{nku} q_u q_i - \epsilon_{niu} q_u q_k \big) \\
 &\Bigg\{  -\mathcal{R} + 1  -2L(q) -2\ln{\bigg(\frac{m_{\pi}}{\mu}\bigg)} \Bigg\}
    \end{split}
\end{equation}

\subsection{\texorpdfstring{$\bm{B \Bar{B}^* \rightarrow B^* \Bar{B}^*}$}{} }
\begin{equation}
    \begin{split}
        V_{\rm OPE} (B \Bar{B}^* \rightarrow B^* \Bar{B}^*) &=  i \frac{g^2 }{4 f^2_{\pi} } (\vec{\tau_1} \cdot \vec{\tau_2})  \epsilon_{lns} ( \epsilon_{2,l} \epsilon^*_{1',k}  \epsilon^*_{2',n})
  \frac{q_k q_s}{\Vec{q}\,^2+m^2_{\pi}}
    \end{split}
\end{equation}
\begin{equation}
    \begin{split}
  V_{\rm TPE} (B \Bar{B}^* \rightarrow B^* \Bar{B}^*) 
     &=  i\, \frac{3 }{128 \pi^2 f^4_{\pi}} g^4 (\epsilon_{2,l} \epsilon^*_{1',k} \epsilon^*_{2',n} ) 
 \, \big( \epsilon_{knu} q_u q_l - \epsilon_{klu} q_u q_n \big) \\
 &\Bigg\{ - \mathcal{R} + 1  -2L(q) -2\ln{\bigg(\frac{m_{\pi}}{\mu}\bigg)} \Bigg\}
    \end{split}
\end{equation}

\subsection{\texorpdfstring{$\bm{B^* \Bar{B}^* \rightarrow B^* \Bar{B}^*}$}{} }
\begin{equation}
    \begin{split}
        V_{\rm OPE} (B^* \Bar{B}^* \rightarrow B^* \Bar{B}^*) & =  \frac{g^2 }{4 f^2_{\pi} } (\vec{\tau_1} \cdot \vec{\tau_2})
   \epsilon_{ikr} \epsilon_{lns} (\epsilon_{1,i} \epsilon^*_{1',k}\, \epsilon_{2,l} \epsilon^*_{2',n}) \frac{q_r q_s}{\Vec{q}\,^2+m^2_{\pi}}
    \end{split}
\end{equation}
\begin{equation}
\begin{split}
 V_{\rm TPE}  (B^* \Bar{B}^* \rightarrow B^* \Bar{B}^*)   &= \frac{\vec{q}\,^2}{16 \pi^2 f^4_{\pi}} \big(\vec{\tau_1}\cdot \vec{\tau_2} \big) ( \epsilon_{1} \cdot \epsilon^*_{1'}) (\epsilon_{2} \cdot \epsilon^*_{2'})   \Bigg\{ 
     \mathcal{R} \bigg[\frac{23}{48}    g^4   
     + \frac{1}{24}  \bigg] 
    + \bigg(\frac{5}{144}   g^4 - \frac{5}{72}  \bigg)
    \\ &+ \ln{\bigg(\frac{m_{\pi}}{\mu}\bigg)}   \bigg( \frac{23}{24} g^4 + \frac{1}{12} \bigg)  
     + L(q)   \bigg( \frac{23}{24} g^4 + \frac{1}{12} \bigg)
    \Bigg\}\\ &
     +  \frac{ \vec{q}\,^2 g^4}{128 \pi^2 f^4_{\pi}} ( \epsilon_{1,i} \epsilon^*_{1',k} \epsilon_{2,l}  \epsilon^*_{2',n})
    \big(  \delta_{in}\delta_{kl} -\delta_{il}\delta_{kn} \big) 
     \Bigg\{ 
       \frac{-7}{3}    \mathcal{R}  - 5 \ln{\bigg(\frac{m_{\pi}}{\mu}\bigg)}  - \frac{15}{2} 
     L(q)   \Bigg\} \\
     &+   \frac{ g^4}{16 \pi^2 f^4_{\pi}} ( \epsilon_{1,i} \epsilon^*_{1',k} \epsilon_{2,l}  \epsilon^*_{2',n}) \big[ q_k q_n\delta_{il}-q_k q_l\delta_{in}+q_i q_l\delta_{kn}-q_i q_n\delta_{kl}\big] \\
     &\times \frac{3}{16  } \Bigg\{ 
     2     \mathcal{R}  -2   
     +4 \ln{\bigg(\frac{m_{\pi}}{\mu}\bigg)}    +4  L(q)
     \Bigg\} 
\end{split}
\end{equation}

\subsection{\texorpdfstring{$\bm{B \Bar{B}^* \rightarrow B \Bar{B}}$}{} }
\begin{equation}
    \begin{split}
        V_{\rm OPE} (B \Bar{B}^* \rightarrow B \Bar{B})
        &=V_{\rm TPE} (B \Bar{B}^* \rightarrow B \Bar{B}) =0
    \end{split}
\end{equation}

\subsection{\texorpdfstring{$\bm{B B \rightarrow B B}$}{} }
\begin{equation}
    V_{\rm OPE} (B B \rightarrow B B) =0 
\end{equation}

\begin{equation}
\begin{split}
     V_{\rm TPE} (B B \rightarrow B B) &=  \frac{1}{ 2 f^4_{\pi}} (\vec{\tau_1}\cdot \vec{\tau_2} )\bigg(  - I_{fb} + g^2 I_{tr}- \frac{g^4}{2} I^{(2)}_{box}  \bigg)\\
     &= \frac{\vec{q}\,^2}{16 \pi^2 f^4_{\pi}} \big(\vec{\tau_1}\cdot \vec{\tau_2} \big) \Bigg\{ 
     \mathcal{R} \bigg[ \frac{23}{48}   g^4 -\frac{5}{24}   g^2   
     - \frac{1}{24}  \bigg] 
  + \bigg(\frac{5}{144}   g^4 +\frac{13}{72}g^2 + \frac{5}{72}  \bigg) 
    \\ & + \ln{\bigg(\frac{m_{\pi}}{\mu}\bigg)}   \bigg( \frac{23}{24} g^4 -\frac{5}{12} g^2 - \frac{1}{12} \bigg) 
     + L(q)  \bigg( \frac{23}{24} g^4 - \frac{5}{12} g^2 - \frac{1}{12} \bigg)
     \Bigg\}
\end{split}
\end{equation}

\subsection{\texorpdfstring{$\bm{B B \rightarrow B^*  B^*}$}{} }
\begin{equation}
V_{\rm OPE} (B B \rightarrow B^*  B^*) =  \frac{g^2 }{4 f^2_{\pi} } (\vec{\tau_1} \cdot \vec{\tau_2}) (\epsilon^*_{1',k}\epsilon^*_{2',n})
    \frac{q_k q_n}{\Vec{q}\,^2+m^2_{\pi}}
\end{equation}

\begin{equation}
    \begin{split}
  V_{\rm TPE} (B B \rightarrow B^*  B^*)&=    \frac{3}{128 \pi^2 f^4_{\pi}} g^4 (\epsilon^*_{1',k}  \epsilon^*_{2',n}) 
     \big(\delta_{kn} \vec{q}\,^2- q_k\, q_n \big) \Bigg\{  -\mathcal{R} + 1  - 2L(q) - 2\ln{\bigg(\frac{m_{\pi}}{\mu}\bigg)} \Bigg\}
    \end{split}
\end{equation}

\subsection{\texorpdfstring{$\bm{B^* B\rightarrow B^* B}$}{} }
\begin{equation}
    V_{\rm OPE} (B^* B\rightarrow B^* B) =0 
\end{equation}
\begin{equation}
\begin{split}
   V_{\rm TPE} (B^* B\rightarrow B^* B) &=   (\epsilon^*_{1'} \cdot \epsilon_{1})  V_{\rm TPE} (B B\rightarrow B B) 
\end{split}
\end{equation}

\subsection{\texorpdfstring{$\bm{B^* B \rightarrow B B^*}$}{} }
\begin{flalign}
    \begin{split}
        V_{\rm OPE} (B^* B \rightarrow B B^*) &=
         \frac{g^2}{4 f^2_{\pi} } (\vec{\tau_1} \cdot \vec{\tau_2}) (\epsilon^*_{2',n} \epsilon_{1,i})
    \frac{q_iq_n}{\Vec{q}\,^2+m^2_{\pi}}
    \end{split}
\end{flalign}
\begin{equation}
\begin{split}
 V_{\rm TPE} (B^* B \rightarrow B B^*) &=  \frac{3}{128 \pi^2 f^4_{\pi}} g^4 (\epsilon^*_{2',n} \epsilon_{1,i}) (\delta_{in}\vec{q}\,^2- q_i q_n)\Bigg\{  -\mathcal{R} + 1  -2L(q) -2\ln{\bigg(\frac{m_{\pi}}{\mu}\bigg)} \Bigg\}
\end{split}
\end{equation}

\subsection{\texorpdfstring{$\bm{B^* B \rightarrow B^* B^*}$}{} }
\begin{equation}
    \begin{split}
  V_{\rm OPE} (B^* B \rightarrow B^* B^*) &=       -i  \frac{g^2 }{4 f^2_{\pi} } (\vec{\tau_1} \cdot \vec{\tau_2})  \epsilon_{ikr} ( \epsilon_{1,i} \epsilon^*_{1',k}  \epsilon^*_{2',n})
  \frac{q_r q_n}{\Vec{q}\,^2+m^2_{\pi}}
    \end{split}
\end{equation}
\begin{equation}
    \begin{split}
      V_{\rm TPE} (B^* B \rightarrow B^* B^*) &=  i\, \frac{3 }{128 \pi^2 f^4_{\pi}} g^4  (\epsilon_{1,i} \epsilon^*_{1',k} \epsilon^*_{2',n} ) 
 \, \big( \epsilon_{nku} q_u q_i - \epsilon_{niu} q_u q_k \big) \\
 &\Bigg\{  -\mathcal{R} + 1  -2L(q) -2\ln{\bigg(\frac{m_{\pi}}{\mu}\bigg)} \Bigg\}
    \end{split}
\end{equation}

\subsection{\texorpdfstring{$\bm{B {B^*} \rightarrow B^* {B^*}}$}{} }
\begin{equation}
    \begin{split}
        V_{\rm OPE} (B {B^*} \rightarrow B^*  {B^*}) &= -i \frac{g^2 }{4 f^2_{\pi} } (\vec{\tau_1} \cdot \vec{\tau_2})  \epsilon_{lns} ( \epsilon_{2,l} \epsilon^*_{1',k}  \epsilon^*_{2',n})
  \frac{q_k q_s}{\Vec{q}\,^2+m^2_{\pi}}
    \end{split}
\end{equation}

\begin{equation}
    \begin{split}
  V_{\rm TPE} (B {B^*} \rightarrow B^* {B^*}) &=  i\, \frac{3 }{128 \pi^2 f^4_{\pi}} g^4 (\epsilon_{2,l} \epsilon^*_{1',k} \epsilon^*_{2',n} ) 
 \, \big( \epsilon_{knu} q_u q_l - \epsilon_{klu} q_u q_n \big) \\
 &\Bigg\{ - \mathcal{R} + 1  -2L(q) -2\ln{\bigg(\frac{m_{\pi}}{\mu}\bigg)} \Bigg\}
    \end{split}
\end{equation}

\subsection{\texorpdfstring{$\bm{B^* B^* \rightarrow B^* B^*}$}{} }
\begin{equation}
\begin{split}
        V_{\rm OPE} (B^* B^* \rightarrow B^* B^*) &= -  \frac{g^2 }{4 f^2_{\pi} } (\vec{\tau_1} \cdot \vec{\tau_2})
   \epsilon_{ikr} \epsilon_{lns} (\epsilon_{1,i} \epsilon^*_{1',k}\, \epsilon_{2,l} \epsilon^*_{2',n}) \frac{q_r q_s}{\Vec{q}\,^2+m^2_{\pi}}
\end{split}
\end{equation}
\begin{equation}
    \begin{split}
         V_{\rm TPE}  (B^* B^* \rightarrow B^* B^*)  
     &= \frac{\vec{q}\,^2}{16 \pi^2 f^4_{\pi}} \big(\vec{\tau_1}\cdot \vec{\tau_2} \big) ( \epsilon_{1} \cdot \epsilon^*_{1'}) (\epsilon_{2} \cdot \epsilon^*_{2'})   \Bigg\{ 
     \mathcal{R} \bigg[ \frac{23}{48}   g^4 -\frac{5}{24} g^2   - \frac{1}{24}  \bigg] 
    + \bigg(\frac{5}{144}   g^4 +\frac{13}{72}g^2 + \frac{5}{72}  \bigg) 
    \\& + \ln{\bigg(\frac{m_{\pi}}{\mu}\bigg)}   \bigg( \frac{23}{24} g^4 -\frac{5}{12} g^2 - \frac{1}{12} \bigg) 
     + L(q)  \bigg( \frac{23}{24} g^4 - \frac{5}{12} g^2 - \frac{1}{12} \bigg)
    \Bigg\}
     \\ &+  \frac{ \vec{q}\,^2  g^4}{128 \pi^2 f^4_{\pi}} ( \epsilon_{1,i} \epsilon^*_{1',k} \epsilon_{2,l}  \epsilon^*_{2',n})
    \big(  \delta_{in}\delta_{kl} -\delta_{il}\delta_{kn} \big) 
    \Bigg\{ 
       \frac{-7}{3}    \mathcal{R}
     - 5\ln{\bigg(\frac{m_{\pi}}{\mu}\bigg)}  - \frac{15}{2} 
     L(q)   \Bigg\} \\
     &+   \frac{ g^4}{16 \pi^2 f^4_{\pi}} ( \epsilon_{1,i} \epsilon^*_{1',k} \epsilon_{2,l}  \epsilon^*_{2',n}) \big[ q_k q_n\delta_{il}-q_k q_l\delta_{in}+q_i q_l\delta_{kn}-q_i q_n\delta_{kl}\big] \\
     &\times \frac{3}{16   }\Bigg\{ 
     2     \mathcal{R}  -2   
     +4 \ln{\bigg(\frac{m_{\pi}}{\mu}\bigg)}    +4  L(q)
     \Bigg\} 
    \end{split}
\end{equation}

\subsection{\texorpdfstring{$\bm{B B^* \rightarrow B B}$}{} }
\begin{equation}
    \begin{split}
        V_{\rm OPE} (B B^* \rightarrow B B)
        &=V_{\rm TPE} (B B^* \rightarrow B B) =0
    \end{split}
\end{equation}

\section{Partial wave decomposition}
\label{PWDforrest}
Here, we present the partial wave projected potentials for the rest of the channels, apart from  with $J^{PC}=0^{++}$, which is given in the main text.
\subsubsection{$\mathbf{J^{PC}=1^{++}}$}
\begin{equation}\label{Eq:VCT1++}
 V^{1^{++},I}_{\rm CT}=   
    \begin{pmatrix} 
	  \mathcal{C}_d^I+  \mathcal{C}_f^I +( \mathcal{D}_d^I + \mathcal{D}_f^I) (p^2+p'^2) & -\mathcal{D}^I_{SD} p'^2 & -\sqrt{3} \mathcal{D}^I_{SD} p'^2 \\
		-\mathcal{D}^I_{SD} p'^2 & 0 & 0  \\
		-\sqrt{3} \mathcal{D}^I_{SD} p^2 & 0 & 0
	\end{pmatrix}
\end{equation}
\begin{equation}
 V^{1^{++}}_{\rm OPE}=   -  \frac{g^2}{4 f^2_{\pi}} (\vec{\tau_1}\cdot \vec{\tau_2})
    \begin{pmatrix} 
	 \frac{1}{3}Q_2 & \frac{1}{\sqrt{18}}(3Q_{n'}-Q_2) & \frac{7}{18\sqrt{6}}(3Q_{n'}-Q_2)  \\
	\frac{1}{\sqrt{18}}(3Q_{n}-Q_2) & -\frac{1}{6} (V_{\rm OPE}^{1^{++}})_{22} & 0 \\
		\frac{7}{18\sqrt{6}}(3Q_{n}-Q_2) & 0 &
  \frac{5}{162} (Q_{x^2}+Q_x)
	\end{pmatrix}
\end{equation}
 \begin{equation}
     (V_{\rm OPE}^{1^{++}})_{22}= 3Q_{n'} +3Q_n -9Q_x -Q_2
 \end{equation}

\begin{equation}
 V^{1^{++}}_{\rm TPE}=   \frac{ 1}{16 \pi^2 f^4_{\pi}}
    \begin{pmatrix} 
	(V_{\rm TPE}^{1^{++}})_{11} &\frac{3}{8\sqrt{2}} \, g^4 (V_{\rm TPE}^{1^{++}})_{12}  & \frac{9}{4\sqrt{24}} \, g^4 (V_{\rm TPE}^{1^{++}})_{13}  \\
	\frac{3}{8\sqrt{2}} \, g^4 (V_{\rm TPE}^{1^{++}})_{21} & (V_{\rm TPE}^{1^{++}})_{22} & \frac{3 \sqrt{3}}{8} \, g^4 (V_{\rm TPE}^{1^{++}})_{23}  \\
		\frac{9}{4\sqrt{24}} \, g^4 (V_{\rm TPE}^{1^{++}})_{31}   & \frac{3 \sqrt{3}}{8} \, g^4 (V_{\rm TPE}^{1^{++}})_{32} & (V_{\rm TPE}^{1^{++}})_{33}
	\end{pmatrix}
\end{equation}
where,
\begin{equation}
(V_{\rm TPE}^{1^{++}})_{11}= \Bar{S}_0 + \frac{g^4 }{4}\bigg\{ (p'\,^2+ p^2) \bigg[- \mathcal{R}+1- 2 \ln{\bigg(\frac{m_{\pi}}{\mu}\bigg)} \bigg] -2 R_2(p',p) \bigg\}
\end{equation}
\begin{equation}
    (V_{\rm TPE}^{1^{++}})_{12}= (V_{\rm TPE}^{1^{++}})_{13}= \frac{2}{3} (p'\,^2) \bigg[- \mathcal{R}+ 1- 2 \ln{\bigg(\frac{m_{\pi}}{\mu}\bigg)} \bigg] +\frac{2}{3} R_2(p',p) - 2 R_{n'}(p',p)
\end{equation}
\begin{equation}
 (V_{\rm TPE}^{1^{++}})_{21}= (V_{\rm TPE}^{1^{++}})_{31}= \frac{2}{3} (p^2) \bigg[- \mathcal{R}+ 1- 2 \ln{\bigg(\frac{m_{\pi}}{\mu}\bigg)} \bigg] +\frac{2}{3} R_2(p',p) - 2 R_{n}(p',p)
\end{equation}
\begin{equation}
(V_{\rm TPE}^{1^{++}})_{22}= \Bar{S}_2 + g^4 \frac{3}{8} \bigg\{ R_{0}(p',p)- 3R_{2x}(p',p) + 3R_{x}(p',p)- R_{n'}(p',p)-R_{n}(p',p)+ \frac{1}{3} R_{2}(p',p)  \bigg\}
\end{equation}

\begin{equation}
(V_{\rm TPE}^{1^{++}})_{23}=(V_{\rm TPE}^{1^{++}})_{32}= \frac{2}{3} R_{2}(p',p)+ 3R_{x}(p',p) - R_{n'}(p',p)- R_{n}(p',p)-R_{x^2}(p',p)
\end{equation}
\begin{multline}
    (V_{\rm TPE}^{1^{++}})_{33}=7 \Bar{S}_2+ \frac{g^4}{4} \Bigg\{ \frac{45}{2} R_{2}(p',p)- \frac{135}{2}R_{x^2}(p',p)  
    - \frac{7}{12} \bigg( 15 R_{n} (p',p) + 15 R_{n'}(p',p) 
   +8 R_2(p',p)- 45 R_{x}(p',p) \\-39 R_{x^2}(p',p) \bigg)  \Bigg\}
\end{multline}

\subsubsection{$\mathbf{J^{PC}=1^{+-}}$}
\begin{equation}\label{Eq:VCT1+-}
 V^{1^{+-},I}_{\rm CT}=   
    \begin{pmatrix} 
 \mathcal{C}^I_d+ \mathcal{D}^I_d  (p^2+p'^2) & \mathcal{D}^I_{SD} p'\,^2 & \mathcal{C}^I_f+ \mathcal{D}^I_f  (p^2+p'^2) & \mathcal{D}^I_{SD} p'\,^2 \\
		\mathcal{D}^I_{SD} p^2 & 0 & \mathcal{D}^I_{SD} p^2 & 0  \\
		\mathcal{C}^I_f+ \mathcal{D}^I_f  (p^2+p'^2) & \mathcal{D}^I_{SD} p'\,^2 & \mathcal{C}^I_d+ \mathcal{D}^I_d  (p^2+p'^2) & \mathcal{D}^I_{SD} p'\,^2 \\
		\mathcal{D}^I_{SD} p^2  & 0 & \mathcal{D}^I_{SD} p^2  & 0
	\end{pmatrix}
\end{equation}
\begin{equation}
\begin{split}
 V^{1^{+-}}_{\rm OPE} &=   - \frac{g^2}{4f^2_{\pi}} (\vec{\tau_1}\cdot \vec{\tau_2})\\
    &\times \begin{pmatrix} 
	- \frac{1}{3}Q_2 &  \frac{1}{\sqrt{18}}(3Q_{n'}-Q_2)  & \frac{2}{3}Q_2 & \frac{1}{\sqrt{18}}(3Q_{n'}-Q_2)  \\
	\frac{1}{\sqrt{18}}(3Q_{n}-Q_2) & -\frac{1}{6} (V_{\rm OPE}^{1^{+-}})_{22} & \frac{1}{\sqrt{18}}(3Q_{n}-Q_2) &  (V_{\rm OPE}^{1^{+-}})_{24}  \\
		\frac{2}{3}Q_2 & \frac{1}{\sqrt{18}}(3Q_{n'}-Q_2) & - \frac{1}{3}Q_2 & 
		\frac{1}{\sqrt{18}}(3Q_{n'}-Q_2)\\
		\frac{1}{\sqrt{18}}(3Q_{n}-Q_2) &   (V_{\rm OPE}^{1^{+-}})_{24}& 
		\frac{1}{\sqrt{18}}(3Q_{n}-Q_2) & -\frac{1}{6}  (V_{\rm OPE}^{1^{+-}})_{22}
	\end{pmatrix}
	\end{split}
\end{equation}

where,
\begin{equation}
    (V_{\rm OPE}^{1^{+-}})_{22} =3Q_n(p',p) +3Q_{n'} (p',p)-9Q_x (p',p) -Q_{x^2} (p',p)
\end{equation}
\begin{equation}
    (V_{\rm OPE}^{1^{+-}})_{24} =\frac{1}{2}Q_n (p',p)+\frac{1}{2}Q_{n'} (p',p)-\frac{3}{2}Q_x (p',p) + \frac{3}{2}Q_{x^2}(p',p)-\frac{2}{3}Q_2(p',p)
\end{equation}

\begin{equation}
    \begin{split}
 V^{1^{+-}}_{\rm TPE}&=  \frac{1}{16 \pi^2 f^4_{\pi}}
     \begin{pmatrix} 
     (V_{\rm TPE}^{1^{+-}})_{11} & \frac{-3}{8\sqrt{2}} \, g^4 (V_{\rm TPE}^{1^{+-}})_{12} & (2 g^4) (V_{\rm TPE}^{1^{+-}})_{13} &  \frac{-3}{8\sqrt{2}} \, g^4 (V_{\rm TPE}^{1^{+-}})_{14} \\
	\frac{-3}{8\sqrt{2}} \, g^4 (V_{\rm TPE}^{1^{+-}})_{21} & (V_{\rm TPE}^{1^{+-}})_{22}  &  \frac{-3}{8\sqrt{2}} \, g^4 (V_{\rm TPE}^{1^{+-}})_{23} &   \frac{1}{4} g^4 (V_{\rm TPE}^{1^{+-}})_{24} \\
		(2 g^4) (V_{\rm TPE}^{1^{+-}})_{31}  &  \frac{-3}{8\sqrt{2}} \, g^4 (V_{\rm TPE}^{1^{+-}})_{32} & (V_{\rm TPE}^{1^{+-}})_{33} & \frac{1}{16\sqrt{2}} \, g^4 (V_{\rm TPE}^{1^{+-}})_{34}\\
		 \frac{-3}{8\sqrt{2}} \, g^4 (V_{\rm TPE}^{1^{+-}})_{41} & \frac{1}{4} g^4 (V_{\rm TPE}^{1^{+-}})_{42}  & \frac{1}{16\sqrt{2}} \, g^4 (V_{\rm TPE}^{1^{+-}})_{43} & (V_{\rm TPE}^{1^{+-}})_{44}
	\end{pmatrix}       
    \end{split}
\end{equation}
where,
\begin{equation}
    (V_{\rm TPE}^{1^{+-}})_{11}=  \Bar{S}_0 - \frac{g^4}{4} \bigg\{ (p'\,^2+ p^2) \bigg[- \mathcal{R}+ 1- 2 \ln{\bigg(\frac{m_{\pi}}{\mu}\bigg)} \bigg] -2 R_2(p',p) \bigg\}
\end{equation}
\begin{equation}
    (V_{\rm TPE}^{1^{+-}})_{12}= (V_{\rm TPE}^{1^{+-}})_{14}=(V_{\rm TPE}^{1^{+-}})_{32}=  \frac{2}{3} (p'\,^2) \bigg[- \mathcal{R}+1- 2 \ln{\bigg(\frac{m_{\pi}}{\mu}\bigg)} \bigg] +\frac{2}{3} R_2(p',p) - 2 R_{n'}(p',p)
\end{equation}
\begin{equation}
    (V_{\rm TPE}^{1^{+-}})_{13}=  (V_{\rm TPE}^{1^{+-}})_{31}= \frac{1}{4}(p'\,^2+ p^2) \bigg[- \mathcal{R}+1- 2 \ln{\bigg(\frac{m_{\pi}}{\mu}\bigg)} \bigg] -2 R_2(p',p) 
\end{equation}
\begin{equation}
    (V_{\rm TPE}^{1^{+-}})_{21}=  (V_{\rm TPE}^{1^{+-}})_{23}=(V_{\rm TPE}^{1^{+-}})_{41}= \frac{2}{3} (p^2) \bigg[- \mathcal{R}+1- 2 \ln{\bigg(\frac{m_{\pi}}{\mu}\bigg)} \bigg] +\frac{2}{3} R_2(p',p) - 2 R_{n}(p',p)
\end{equation}
\begin{equation}
(V_{\rm TPE}^{1^{+-}})_{22}= \Bar{S}_2 - g^4 \frac{3}{8} \bigg\{ R_{0}(p',p)- 3R_{2x}(p',p) + 3R_{x}(p',p)- R_{n'}(p',p)-R_{n}(p',p)+ \frac{1}{3} R_{2}(p',p)  \bigg\}
\end{equation}
\begin{equation}
(V_{\rm TPE}^{1^{+-}})_{24}= (V_{\rm TPE}^{1^{+-}})_{42}=   \frac{9}{4} R_{x^2}(p',p)- \frac{27}{4}R_{x}(p',p) + \frac{9}{4}R_{n'}(p',p) + \frac{9}{4}R_{n}(p',p)-\frac{3}{2} R_{2}(p',p) 
\end{equation}
\begin{equation}
    (V_{\rm TPE}^{1^{+-}})_{33}=\Bar{S}_0- \frac{g^4}{8} \Bigg\{- 2 (p^2+p'\,^2) \mathcal{R}+ 2 (p^2+p'\,^2)- 4(p^2+p'\,^2)\ln{\bigg(\frac{m_{\pi}}{\mu}\bigg)} - 4 R_{2}(p',p)  \Bigg\}
\end{equation}
\begin{equation}
     (V_{\rm TPE}^{1^{+-}})_{34} = 2p'\,^2  \bigg[2 \mathcal{R}- 2 +4 \ln{\bigg(\frac{m_{\pi}}{\mu}\bigg)} \bigg] +4 \bigg[ 3 R_{n'}(p',p)- R_2(p',p) \bigg] 
\end{equation}
\begin{equation}
     (V_{\rm TPE}^{1^{+-}})_{43} = 2p^2  \bigg[2 \mathcal{R}-2 + 4 \ln{\bigg(\frac{m_{\pi}}{\mu}\bigg)} \bigg]  + 4 \bigg[ 3 R_{n}(p',p)- R_2(p',p) \bigg] 
\end{equation}

\begin{multline}
    (V_{\rm TPE}^{1^{+-}})_{44}= -\Bar{S}_2+ \frac{g^4}{16} \Bigg\{ \frac{15}{2} R_{2}(p',p) - \frac{45}{2}R_{x^2}(p',p) 
   + \frac{21}{2} \bigg( 2R_{n} (p',p)+ 2R_{n'}(p',p) 
    -\frac{8}{3} R_2(p',p) - 6 R_{x}(p',p) \\+ 6 R_{x^2}(p',p) \bigg) \Bigg\}
\end{multline}
\subsubsection{$\mathbf{J^{PC}=2^{++}}$}
\begin{equation}\label{Eq:VCT2++}
    V^{2^{++},I}_{\rm CT}= 
\begin{pmatrix} 
0 & 0 &   - \sqrt{\frac{3}{5}}\mathcal{D}^I_{SD} p^2  &  0 & 0 & 0  \\
	0 & 0 & -\frac{3}{\sqrt{5}}\mathcal{D}^I_{SD} p^2 &  0 & 0 & 0  \\
	 - \sqrt{\frac{3}{5}}\mathcal{D}^I_{SD} p'\,^2 & -\frac{3}{\sqrt{5}}\mathcal{D}^I_{SD} p'\,^2 &  \mathcal{C}^I_d+  \mathcal{C}^I_f +( \mathcal{D}^I_d + \mathcal{D}^I_f) (p^2+p'^2) &-\frac{1}{\sqrt{5}}\mathcal{D}^I_{SD} p'\,^2 &  \sqrt{\frac{7}{5}}\mathcal{D}^I_{SD} p'\,^2 & 0\\
		0 &   0  & -\frac{1}{\sqrt{5}}\mathcal{D}^I_{SD} p^2 &  0 & 0 & 0 \\
		0 & 0 &  \sqrt{\frac{7}{5}}\mathcal{D}^I_{SD} p^2 & 0 & 0 &0\\ 
	 0 & 0 & 0 & 0 & 0 &0
	\end{pmatrix}
\end{equation}

\begin{multline}
 V^{2^{++}}_{\rm OPE}=    -\frac{g^2}{4 f^2_{\pi}} (\vec{\tau_1}\cdot \vec{\tau_2})\\
      \times
    \begin{pmatrix} 
0 & 0 & - \frac{1}{\sqrt{30}}(3Q_{n'}-Q_2) & \frac{1}{\sqrt{12}}(3Q_{x^2}-Q_2) & - \frac{1}{\sqrt{21}} K_{15} & - \frac{3}{\sqrt{560}} K_{16}  \\
	0 & 0 & \frac{1}{\sqrt{20}}(3Q_{n'}-Q_2) &  0 & \frac{1}{\sqrt{504}} K_{25} & -\frac{1}{\sqrt{1400}} K_{26}  \\
	- \frac{1}{\sqrt{30}}(3Q_{n}-Q_2) & \frac{1}{\sqrt{20}}(3Q_n-Q_2) & - \frac{1}{12}Q_2 & 
		\frac{1}{\sqrt{90}}(3Q_n-Q_2) & - \sqrt{\frac{7}{90}}(3Q_{n}-Q_2) & 0\\
		\frac{1}{\sqrt{12}}(3Q_{x^2}-Q_2) &   0  & \frac{1}{\sqrt{90}}(3Q_{n'}-Q_2) & \frac{1}{6}(3Q_{x^2}-Q_2) & \frac{1}{\sqrt{63}} K_{45} & \frac{1}{8 \sqrt{35}} K_{46}\\
		-\frac{1}{\sqrt{21}} K_{51} & \frac{1}{\sqrt{504}} K_{52} & -\sqrt{\frac{7}{90}} (3Q_{n'}-Q_2) & \frac{1}{\sqrt{63}} K_{54} & \frac{3}{14} K_{55} & -\frac{1}{56\sqrt{5}} K_{56}\\ 
	- \frac{3}{\sqrt{560}} K_{61} & -\frac{1}{\sqrt{1400}} K_{62} & 0 & \frac{1}{8 \sqrt{35}} K_{64} & -\frac{1}{56\sqrt{5}} K_{65} & \frac{1}{28} K_{66}
	\end{pmatrix}
\end{multline}

where, 
\begin{equation}
  K_{51}= K_{15} = 3 Q_{x^2}-9Q_x+3Q_x+3Q_{n'}-2Q_2  
\end{equation}
\begin{equation}
  K_{61} = 35 Q_{n'x^2}-Q_{x^2}+20Q_x-5Q_{n'}+2Q_n+Q_2  
\end{equation}
\begin{equation}
 K_{16} = 35 Q_{n x^2}-Q_{x^2}+20Q_x-5Q_{n} + 2Q_{n'}+Q_2   
\end{equation}
\begin{equation}
    K_{52} =K_{25}= 18 Q_{ x^2}-18 Q_{x} + 6Q_{n'} + 6Q_{n}+ 8 Q_2
\end{equation}
\begin{equation}
    K_{62} = 35 Q_{n'x^2}-5 Q_{x^2}- 20Q_x - 5Q_{n'} + 2Q_n + Q_2
\end{equation}
\begin{equation}
    K_{26} = 35 Q_{nx^2}-5 Q_{x^2}- 20Q_x - 5Q_{n} + 2Q_{n'} + Q_2
\end{equation}
\begin{equation}
K_{54} = K_{45}=  3 Q_{x^2} - Q_{x} + 3Q_{n'} + 3Q_{n} - 2Q_{2} \end{equation}
\begin{equation}
K_{64} = 35 Q_{n'x^2}-35 Q_{x^2}- 140 Q_x - 5Q_{n'} + 2Q_n + Q_2  \end{equation}
\begin{equation}
K_{46} = 35 Q_{nx^2}-35 Q_{x^2}- 140 Q_x - 5Q_{n} + 2Q_{n'} + Q_2 \end{equation}
\begin{equation}
 K_{55} = -8 Q_{ x^2} +11 Q_{x} - 6(Q_{n'}+Q_n) + 7 Q_2   
\end{equation}
\begin{equation}
 K_{65} = 35 Q_{n'x^2}-5 Q_{x^2}+ 20Q_x - 5Q_{n'} + 2Q_n + Q_2   
\end{equation}
\begin{equation}
    K_{56} = 35 Q_{nx^2}-5 Q_{x^2}+ 20Q_x - 5Q_{n} + 2Q_{n'} + Q_2
\end{equation}
\begin{equation}
 K_{66}= 245 Q_{x^3}- 105 (Q_{n'x^2} +Q_{nx^2}) + 15 Q_{x^2} + 15 (Q_{n'} +Q_{n}) +5 Q_x -3Q_2     
\end{equation}

\begin{multline}
    V^{2^{++}}_{\rm TPE} =   \frac{1}{16 \pi^2 f^4_{\pi}}\\
    \times \scalemath{0.9}{ 
    \begin{pmatrix} 
(V_{\rm TPE}^{2^{++}})_{11} & 0 &  \frac{3\sqrt{3}}{8\sqrt{10}}\,g^4(V_{\rm TPE}^{2^{++}})_{13}  &  \frac{\sqrt{3}}{8}\,g^4(V_{\rm TPE}^{2^{++}})_{14} & \frac{3\sqrt{3}}{4\sqrt{7}}\,g^4(V_{\rm TPE}^{2^{++}})_{15} & \frac{1}{32\sqrt{105}}\,g^4(V_{\rm TPE}^{2^{++}})_{16}  \\
	0 & 0 & \frac{9}{8\sqrt{10}} \,g^4(V_{\rm TPE}^{2^{++}})_{23}&  0 &  \frac{-9}{8\sqrt{14}} \,g^4(V_{\rm TPE}^{2^{++}})_{25} & \frac{-3}{8\sqrt{70}} \,g^4(V_{\rm TPE}^{2^{++}})_{26}  \\
	\frac{3\sqrt{3}}{8\sqrt{10}}\,g^4(V_{\rm TPE}^{2^{++}})_{31} & \frac{9}{8\sqrt{10}} \,g^4(V_{\rm TPE}^{2^{++}})_{32} & (V_{\rm TPE}^{2^{++}})_{33} & \frac{-1}{16\sqrt{10}}\,g^4(V_{\rm TPE}^{2^{++}})_{34} &
		\frac{\sqrt{7}}{16\sqrt{10}}\,g^4(V_{\rm TPE}^{2^{++}})_{35} &  0\\
		\frac{\sqrt{3}}{8}\,g^4(V_{\rm TPE}^{2^{++}})_{41} &   0  & \frac{-1}{16\sqrt{10}}\,g^4(V_{\rm TPE}^{2^{++}})_{43} & (V_{\rm TPE}^{2^{++}})_{44} & \frac{1}{16\sqrt{7}}\,g^4(V_{\rm TPE}^{2^{++}})_{45} & \frac{-3}{16\sqrt{35}}\,g^4(V_{\rm TPE}^{2^{++}})_{46}\\
		\frac{3\sqrt{3}}{4\sqrt{7}}\,g^4(V_{\rm TPE}^{2^{++}})_{51} & \frac{-9}{8\sqrt{14}} \,g^4(V_{\rm TPE}^{2^{++}})_{52}& \frac{\sqrt{7}}{16\sqrt{10}}\,g^4(V_{\rm TPE}^{2^{++}})_{53} &  \frac{1}{16\sqrt{7}}\,g^4(V_{\rm TPE}^{2^{++}})_{54} & (V_{\rm TPE}^{2^{++}})_{55} &  \frac{3}{224 \sqrt{5}}\,g^4(V_{\rm TPE}^{2^{++}})_{56}\\ 
 \frac{1}{32\sqrt{105}}\,g^4(V_{\rm TPE}^{2^{++}})_{61} & \frac{-3}{8\sqrt{70}} \,g^4(V_{\rm TPE}^{2^{++}})_{62} & 0 & \frac{-3}{16\sqrt{35}}\,g^4(V_{\rm TPE}^{2^{++}})_{64} & \frac{3}{224 \sqrt{5}}\,g^4(V_{\rm TPE}^{2^{++}})_{65}& (V_{\rm TPE}^{2^{++}})_{66} 
	\end{pmatrix}
 }
\end{multline}

where,
\begin{equation}
(V_{\rm TPE}^{2^{++}})_{11}= \Bar{S}_2(p',p)
\end{equation}
\begin{equation}
   (V_{\rm TPE}^{2^{++}})_{13}=(V_{\rm TPE}^{2^{++}})_{23}= \frac{2}{3} p^2  \bigg[-\mathcal{R}+ 1- 2 \ln{\bigg(\frac{m_{\pi}}{\mu}\bigg)} \bigg]+\frac{2}{3}R_{2}(p',p) -2R_{n}(p',p)
\end{equation}
\begin{equation}
(V_{\rm TPE}^{2^{++}})_{14}= (V_{\rm TPE}^{2^{++}})_{41}=   3 R_{x^2}(p',p)- 9 R_{2x}(p',p) + 3 R_{0}(p',p) - R_{2}(p',p) 
\end{equation}
\begin{equation}
(V_{\rm TPE}^{2^{++}})_{15}=(V_{\rm TPE}^{2^{++}})_{25}= (V_{\rm TPE}^{2^{++}})_{51}= (V_{\rm TPE}^{2^{++}})_{52}=   \frac{2}{3} R_{2}(p',p)+3  R_{x}(p',p) -  R_{n}(p',p) - R_{n'}(p',p) -R_{x^2}(p',p) 
\end{equation}
\begin{equation}
    (V_{\rm TPE}^{2^{++}})_{16}= 63 R_{n'x^2}(p',p)-90  R_{x^2}(p',p)- 360 R_{x}(p',p)-90 R_{n'}(p',p) + 36 R_{n}(p',p)+18 R_{2}(p',p)  
\end{equation}
\begin{equation}
(V_{\rm TPE}^{2^{++}})_{26}= 35 R_{n'x^2}(p',p)-5  R_{x^2}(p',p)- 2 R_{x}(p',p)-5 R_{n'}(p',p) + 2 R_{n}(p',p)+ R_{2}(p',p)  
\end{equation}
\begin{equation}
    (V_{\rm TPE}^{2^{++}})_{31}= (V_{\rm TPE}^{2^{++}})_{32}= \frac{2}{3} p'\,^2  \bigg[-\mathcal{R}+ 1- 2 \ln{\bigg(\frac{m_{\pi}}{\mu}\bigg)} \bigg]+\frac{2}{3}R_{2}(p',p) -2R_{n'}(p',p)
\end{equation}
\begin{equation}
    (V_{\rm TPE}^{2^{++}})_{33}=\Bar{S}_0+ \frac{g^4}{8} \Bigg\{- 2 (p^2+p'\,^2) \mathcal{R}+ 2 (p^2+p'\,^2)- 4(p^2+p'\,^2)\ln{\bigg(\frac{m_{\pi}}{\mu}\bigg)}- 4 R_{2}(p',p)  \Bigg\}
\end{equation}
\begin{equation}
     (V_{\rm TPE}^{2^{++}})_{34} = (V_{\rm TPE}^{2^{++}})_{35} = 2p'\,^2  \bigg[2 \mathcal{R}- 2 + 4 \ln{\bigg(\frac{m_{\pi}}{\mu}\bigg)} \bigg]  + 4 \bigg[ 3 R_{n'}(p',p)- R_2(p',p) \bigg]  
\end{equation}
\begin{equation}
     (V_{\rm TPE}^{2^{++}})_{43} = (V_{\rm TPE}^{2^{++}})_{53} = 2p^2  \bigg[2 \mathcal{R}- 2 + 4 \ln{\bigg(\frac{m_{\pi}}{\mu}\bigg)} \bigg] + 4 \bigg[ 3 R_{n}(p',p)- R_2(p',p) \bigg]  
\end{equation}
\begin{equation}
    (V_{\rm TPE}^{2^{++}})_{44}= \Bar{S}_2+ \frac{g^4}{8} \Bigg\{  \frac{45}{2}R_{x^2}(p',p) - \frac{15}{2} R_{2}(p',p) 
    +\frac{7}{2} R_0(p',p)-
    \frac{21}{2}  R_{2x} (p',p)  \Bigg\}
\end{equation}
\begin{equation}
     (V_{\rm TPE}^{2^{++}})_{45} = (V_{\rm TPE}^{2^{++}})_{54}=  - \frac{7}{2}\bigg[ 3 R_{n}(p',p)+ 3R_{n'}(p',p)  - 9 R_{x}(p',p) +3 R_{x^2}(p',p) 
    - 2 R_2(p',p)  \bigg] 
\end{equation}
\begin{multline}
     (V_{\rm TPE}^{2^{++}})_{46} = (V_{\rm TPE}^{2^{++}})_{56}=   - \frac{7}{2}\bigg[ 35 R_{n'x^2}(p',p) -5 R_{n'}(p',p) - 20 R_{x}(p',p) + R_{2}(p',p) + 2 R_{n}(p',p)
    -5 R_{x^2}(p',p)   \bigg] 
\end{multline}

\begin{multline}
    (V_{\rm TPE}^{2^{++}})_{55}= \Bar{S}_2+ \frac{g^4}{16} \Bigg\{ \frac{15}{2} R_{2}(p',p)  - \frac{45}{2}R_{x^2}(p',p)
   - \frac{1}{4} \bigg( -9 R_{n} (p',p) -9 R_{n'}(p',p) -51 R_{x^2}(p',p)
    +27  R_{x}(p',p) \\+20 R_2(p',p) \bigg) \Bigg\}
\end{multline}
\begin{equation}
(V_{\rm TPE}^{2^{++}})_{61}= 63 R_{nx^2}(p',p)-90  R_{x^2}(p',p)- 360 R_{x}(p',p)-90 R_{n}(p',p) + 36 R_{n'}(p',p)+18 R_{2}(p',p)  
\end{equation}
\begin{equation}
(V_{\rm TPE}^{2^{++}})_{62}= 35 R_{nx^2}(p',p)-5  R_{x^2}(p',p)- 2 R_{x}(p',p)-5 R_{n}(p',p) + 2 R_{n'}(p',p)+ R_{2}(p',p)  
\end{equation}

\begin{multline}
     (V_{\rm TPE}^{2^{++}})_{64} = (V_{\rm TPE}^{2^{++}})_{65}=   - \frac{7}{2}\bigg[ 35 R_{nx^2}(p',p) -5 R_{n}(p',p) - 20 R_{x}(p',p) + R_{2}(p',p) + 2 R_{n'}(p',p)
    -5 R_{x^2}(p',p)   \bigg]  
\end{multline}
\begin{multline}
    (V_{\rm TPE}^{2^{++}})_{66}= \Bar{S}_4- \frac{15 g^4}{32} \bigg[ 35 R_{x^4} (p',p)- 30 R_{x^2}(p',p) +3 R_{2}(p',p) \bigg] 
     +\frac{3g^4}{112} \Bigg\{
   - \frac{7}{2} \bigg( 105 R_{n'x^2} (p',p) + 105 R_{n x^2} (p',p)\\
   -15 R_{n'}(p',p) - 15 R_{n}(p',p)
    +45  R_{x}(p',p) +3 R_2(p',p) -15 R_{x^2}(p',p) -245 R_{nn'x^3}(p',p)\bigg)  \Bigg\}
\end{multline}
and
\begin{multline}
     \Bar{S}_4 = \int^{1}_{-1} \frac{dx}{2} \frac{(35x^4 - 30 x^2+3)}{8} (\vec{\tau_1}\cdot \vec{\tau_2} ) \vec{q}\,^2 \Bigg\{    \mathcal{R} \bigg[\frac{23}{48}   g^4 
     + \frac{1}{24}  \bigg] 
    +  \bigg(\frac{5}{144}   g^4 - \frac{5}{72}  \bigg)  
     + \ln{\bigg(\frac{m_{\pi}}{\mu}\bigg)}  \bigg( \frac{23}{24} g^4 + \frac{1}{12} \bigg)
      \\+ L(q)   \bigg( \frac{23}{24} g^4 + \frac{1}{12} \bigg) \Bigg\}
\end{multline}

All the $Q$ and $R$ functions mentioned above are functions of $p$ and $p'$ (as in $Q(p',p)$), but for simplicity reasons the $(p',p)$ was avoided.

\section{Calculation of integrals of partial wave decomposition}
\label{PWDint}
The calculations of integrals from partial wave decomposition are shown here.
Specifically, we present the integration of $Q$, $R$, $S$ and $T$ terms encountered in the partial wave decomposition here. 
The following notation is used below: 
 $|\vec{p'}|= p'$ , $|\vec{p}|= p$,$|\vec{q}|= q $, $\vec{n}= \vec{p}/p$, $\vec{n'}= \vec{p'}/p'$. In addition, $\vec{q}\,^2 = {p'}^2+p^2-2p'px$ and $\Vec{n'}\cdot \vec{n}$ and
\begin{equation}
    \vec{n}\cdot \vec{q}= p'x-p= \frac{{p'}^2-p^2-q^2}{2p}
\end{equation}
\begin{equation}
    \vec{n'}\cdot \vec{q}= p'-px= \frac{{p'}^2-p^2+q^2}{2p'}
\end{equation}

\subsection{Q-integrals}
\begin{equation}
  Q_2(p',p)= \int_{-1}^{1} \frac{dx}{2} \frac{\vec{q}\,^2}{\vec{q}\,^2+m^2_{\pi}} 
    = 1  +  \mathcal{O}(\chi^4)
\end{equation}
\begin{equation}
 \begin{split}
    Q_n(p',p)&= \int_{-1}^{1} \frac{dx}{2} \frac{(\vec{n}\cdot\vec{q})^2}{\vec{q}\,^2+m^2_{\pi}}= 1 - \frac{{p'}^2 + p^2  }{4p^2}+ \frac{({p'}^2 - p^2 )^2 }{8p'p^3} 
    \arctanh{\bigg( \frac{2p' p}{{p'}^2 + p^2+ m^2_{\pi}}\bigg)}+  \mathcal{O}(\chi^4)
 \end{split}
\end{equation}
\begin{equation}
 \begin{split}
    Q_{n'}(p',p)&= \int_{-1}^{1} \frac{dx}{2} \frac{(\vec{n'}\cdot\vec{q})^2}{\vec{q}\,^2+m^2_{\pi}}
    = 1 - \frac{{p'}^2 + p^2 }{4{p'}^2}+ \frac{(-{p'}^2 + p^2 )^2 }{8{p'}^3p} 
    \arctanh{\bigg( \frac{2p' p}{{p'}^2 + p^2+ m^2_{\pi}}\bigg)} +  \mathcal{O}(\chi^4)
 \end{split}
\end{equation}
\begin{multline}
    Q_x(p',p)= \int_{-1}^{1} \frac{dx}{2} \frac{(\vec{n'}\cdot\vec{q})(\vec{n}\cdot\vec{q})x}{\vec{q}\,^2+m^2_{\pi}}
    = \frac{5}{12} - \frac{{p'}^4 + p^4  }{8{p'}^2p^2}+ \frac{\big( ({p'}^2 - p^2)^2 \big) ({p'}^2 + p^2 ) }{16{p'}^3p^3} 
     \arctanh{\bigg( \frac{2p' p}{{p'}^2 + p^2+ m^2_{\pi}}\bigg)} +  \mathcal{O}(\chi^4)
\end{multline}
\begin{equation}
 \begin{split}
    Q_{x^2}(p',p)&= \int_{-1}^{1} \frac{dx}{2} \frac{(\vec{q}\,^2 x^2)}{\vec{q}\,^2+m^2_{\pi}}
    = \frac{1}{3}+  \mathcal{O}(\chi^4)
 \end{split}
\end{equation}
\begin{multline}
    Q_{nx^2}(p',p)= \int_{-1}^{1} \frac{dx}{2} \frac{(\vec{n}\cdot\vec{q})^2 x^2}{\vec{q}\,^2+m^2_{\pi}}
    =  \frac{ p^4(5{p'}^2 )-p^6  -{p'}^6 }{16{p'}^2p^4}+ \frac{- {p'}^4}{48 {p}^2 {p'}^2} 
    + \frac{(p^4-{p'}^4)^2}{32{p'}^3p^5}  \arctanh{\bigg( \frac{2p' p}{{p'}^2 + p^2+ m^2_{\pi}}\bigg)}+  \mathcal{O}(\chi^4)
\end{multline}
\begin{multline}
    Q_{n'x^2}(p',p)= \int_{-1}^{1} \frac{dx}{2} \frac{(\vec{n'}\cdot\vec{q})^2 x^2}{\vec{q}\,^2+m^2_{\pi}}
    =  \frac{ {p'}^4(5{p}^2 )-{p'}^6  -{p}^6 }{16{p}^2{p'}^4}+ \frac{- {p}^4}{48 {p'}^2 {p}^2}
    + \frac{({p'}^4-{p}^4)^2}{32{p'}^5p^3}  \arctanh{\bigg( \frac{2p' p}{{p'}^2 + p^2+ m^2_{\pi}}\bigg)}+  \mathcal{O}(\chi^4)
\end{multline}
\begin{multline}
    Q_{x^3}(p',p)= \int_{-1}^{1} \frac{dx}{2} \frac{(\vec{n'}\cdot\vec{q})(\vec{n}\cdot\vec{q}) x^3}{\vec{q}\,^2+m^2_{\pi}}
    =  \frac{ -2{p'}^6 - p^4 ( 2 {p'}^2)
    } {48{p'}^4p^2} 
    + \frac{ 59 {p'}^2}{240{p'}^2} + \frac{ - p^8  - {p'}^8}{32 {p'}^4 p^4}\\
    + \frac{(({p'}^2-{p}^2))({p'}^2+p^2 )^3}
    {64{p'}^5p^5}  \arctanh{\bigg( \frac{2p' p}{{p'}^2 + p^2+ m^2_{\pi}}\bigg)}+  \mathcal{O}(\chi^4)
\end{multline}
\subsection{R-integrals}
Since  $\vec{q}\,^2 = {p'}^2+p^2-2p'px$, we can substitute $x$ inside the $R$-integrals:
\begin{equation}
    \frac{dx}{2}= - \frac{q}{2p'p}dq
\end{equation}
In the following, $q$ is relabeled as $\rho$ to avoid ambiguity between the transferred momentum and the integration variable. The limits of integration are:
\begin{equation}
    x_b =1\rightarrow \rho_b=p'-p
\end{equation}
\begin{equation}
    x_a =-1\rightarrow \rho_a=p'+p
\end{equation}
The $R$-integrals are now written as,

\begin{equation}
 R_0 (p',p) = \int_{-1}^{1} \frac{dx}{2} L(q)= \frac{1}{2p'p}\int_{p'-p}^{p'+p} d\rho \, \rho  L(\rho) = -\frac{1}{2} +
 \frac{1}{4p'p} \bigg[\rho^2 L(\rho) \bigg]_{p'-p}^{p'+p} +  \mathcal{O}(\chi^4)
\end{equation}
\begin{equation}
    R_2 (p',p) = \int_{-1}^{1} \frac{dx}{2} \vec{q}^2 L(q) = \frac{1}{2p'p}\int_{p'-p}^{p'+p} d\rho \, \rho^3  L(\rho)
    = \frac{-1}{8p'p} \bigg[ \frac{\rho^4}{4}  -\rho^4 L(\rho) \bigg]_{p'-p}^{p'+p} +  \mathcal{O}(\chi^4)
\end{equation}
\begin{equation}
    R_4 (p',p) = \int_{-1}^{1} \frac{dx}{2} \vec{q}^4 L(q)
    = \frac{1}{12p'p} \bigg[-\frac{\rho^6}{6}  + \rho^6 L(\rho) \bigg]_{p'-p}^{p'+p} +  \mathcal{O}(\chi^4)
\end{equation}
\begin{equation}
    R_6 (p',p) = \int_{-1}^{1} \frac{dx}{2} \vec{q}^6 L(q)
    = -\frac{1}{16p'p} \bigg[\frac{\rho^8}{8} 
    -\rho^8 L(\rho)  \bigg]_{p'-p}^{p'+p} +  \mathcal{O}(\chi^4)
\end{equation}
\begin{equation}
    R_8 (p',p) = \int_{-1}^{1} \frac{dx}{2} \vec{q}^8 L(q) = 
     \frac{1}{8p'p} \bigg[\frac{-\rho^{10}}{25} 
    + \frac{2\rho^{10}}{5}  L(\rho)  \bigg]_{p'-p}^{p'+p} +  \mathcal{O}(\chi^4)
\end{equation}
\begin{equation}
    R_{10} (p',p) = \int_{-1}^{1} \frac{dx}{2} \vec{q}^{10} L(q)
    = \frac{1}{2 p'p} \bigg[\frac{ -\rho^{12} }{144} 
    + \frac{\rho^{12}}{12} L(\rho)   \bigg]_{p'-p}^{p'+p} +  \mathcal{O}(\chi^4)
\end{equation}
\begin{multline}
   \begin{split}
        R_n (p',p) = \int_{-1}^{1} \frac{dx}{2} (\vec{n}\cdot\vec{q})^2 L(q) &= \frac{1}{8p'p^3}\int_{p'-p}^{p'+p} d\rho  \big(\rho^4-2\rho^2({p'}^2-p^2)+({p'}^2-p^2)^2\big) \rho \cdot L(\rho)\\
    &= \frac{1}{4p^2} \Big(R_4(p',p)-2({p'}^2-p^2)R_2(p',p)+({p'}^2-p^2)^2 R_0(p',p)  \Big)+  \mathcal{O}(\chi^4)
   \end{split}
\end{multline}
\begin{equation}
 R_{n'} (p',p) = \int_{-1}^{1} \frac{dx}{2} (\vec{n'}\cdot\vec{q})^2 L(q)
 =  \frac{1}{4p'\,^2} \Big(R_4(p',p)+2({p'}^2-p^2)R_2(p',p)+({p'}^2-p^2)^2 R_0(p',p)  \Big) +  \mathcal{O}(\chi^4)
\end{equation}

\begin{multline}
    R_x (p',p) = \int_{-1}^{1} \frac{dx}{2} (\vec{n'}\cdot\vec{q})(\vec{n}\cdot\vec{q})x L(q) 
    = \frac{1}{8{p'}^2p^2} \Big(R_6(p',p)-({p'}^2+p^2)R_4(p',p)-({p'}^2-p^2)^2 R_2(p',p)\\
    +({p'}^2-p^2)^2 ({p'}^2+p^2) R_0(p',p)  \Big)+  \mathcal{O}(\chi^4)
\end{multline}
\begin{equation}
\begin{split}
    R_{x^2} (p',p) &= \int_{-1}^{1} \frac{dx}{2} \vec{q}\,^2 x^2 L(q)  
    = \frac{1}{4{p'}^2p^2} \Big(R_6(p',p)-2({p'}^2+p^2)R_4(p',p)+ ({p'}^2+p^2)^2 R_2(p',p) \Big) +  \mathcal{O}(\chi^4)
\end{split}
\end{equation}
\begin{equation}
\begin{split}
    R_{2x} (p',p) &= \int_{-1}^{1} \frac{dx}{2}  x^2 L(q)  
    = \frac{1}{4{p'}^2p^2} \Big(R_4(p',p)-2({p'}^2+p^2)R_2(p',p)+ ({p'}^2+p^2)^2 R_0(p',p) \Big)+  \mathcal{O}(\chi^4)
\end{split}
\end{equation}
\begin{multline}
    R_{x^4} (p',p) = \int_{-1}^{1} \frac{dx}{2}  x^4 \vec{q}\,^2 L(q)  
    = \frac{1}{16{p'}^4p^4} \Big(R_{10}(p',p) - 4({p'}^2+p^2) R_8(p',p)+ 6 ({p'}^2+p^2) R_6(p',p) \\-  4({p'}^2+p^2)^3 R_4(p',p) + ({p'}^2+p^2)^4 R_2(p',p) \Big) +  \mathcal{O}(\chi^4)
\end{multline}
\begin{multline}
    R_{nx^2} (p',p) = \int_{-1}^{1} \frac{dx}{2} (\vec{n}\cdot\vec{q}) x^2 L(q)  
    = \frac{1}{16{p'}^2p^4} \Big(R_8 (p',p)-4{p'}^2 R_6(p',p) + (6{p'}^4-2p^4)R_4(p',p) \\ + 4{p'}^2({p'}^4-p^4) R_2(p',p) + ({p'}^4-p^4)^2 R_0(p',p) \Big)+  \mathcal{O}(\chi^4)
\end{multline}

\begin{multline}
    R_{n'x^2} (p',p) = \int_{-1}^{1} \frac{dx}{2} (\vec{n'}\cdot\vec{q}) x^2 L(q)   
    = \frac{1}{16{p'}^2p^4} \Big(R_8(p',p)-4{p}^2 R_6(p',p) + (6{p}^4-2{p'}^4)R_4(p',p) \\+ 4{p}^2({p}^4-{p'}^4) R_2(p',p)  + ({p}^4-{p'}^4)^2 R_0(p',p) \Big) +  \mathcal{O}(\chi^4)
\end{multline}
\begin{multline}
    R_{n n'x^3} (p',p) = \int_{-1}^{1} \frac{dx}{2} (\vec{n'}\cdot\vec{q})(\vec{n}\cdot\vec{q})x^3 L(q)   
    = \frac{1}{16{p'}^2p^4} \Big(R_{10}(p',p) - 3 ({p'}^2+p^2) R_8(p',p) \\+ 2 (p^4+ 4{p'}^2 p^2 + {p'}^4) R_6(p',p) 
    +2 (p^6 - 3{p'}^2 p^4 -3{p'}^4 p^2  + {p'}^6) R_4(p',p) - 3 ({p}^4- {p'}^4)^2 R_2(p',p) 
   \\ + ({p}^2-{p'}^2)^2 ({p}^2+{p'}^2)^4 R_0(p',p)   \Big)+  \mathcal{O}(\chi^4)
\end{multline}

$L(\rho)$ is given by,
\begin{equation}
L(\rho)= \frac{\sqrt{4m^2_{\pi}+ \rho^2}}{\rho} \ln{\bigg(\frac{\sqrt{4m^2_{\pi}+ \rho^2}+\rho}{2 m_{\pi}}\bigg)} \end{equation}
\subsection{$\mathbf{\Bar{S}}$-integrals and S-integrals}
The $\mathrm{\Bar{S}}$-integrals can be written in general as,

\begin{multline}
  \Bar{S}_k(p',p)=  \int_{-1}^{1} \frac{dx}{2} P_k(x)  \big(\vec{\tau_1}\cdot \vec{\tau_2} \big) \vec{q}\,^2 \Bigg\{  
     \mathcal{R} \bigg[\frac{23}{48}   g^4   
     + \frac{1}{24}  \bigg] 
    +  \bigg(\frac{5}{144}   g^4 - \frac{5}{72}  \bigg)  
     + \ln{\bigg(\frac{m_{\pi}}{\mu}\bigg)}  \bigg( \frac{23}{24} g^4 + \frac{1}{12} \bigg)
      + L(q)   \bigg( \frac{23}{24} g^4 + \frac{1}{12} \bigg) \Bigg\}
\end{multline}

where $P_k(x)$ denotes the $k$-th Legendre polynomial. 
\begin{multline}
  \Bar{S}_0(p',p)=\big(\vec{\tau_1}\cdot \vec{\tau_2} \big) 
  \Bigg\{ \mathcal{R}  ({p'}^2 +p^2) \bigg( \frac{23}{24}g^4  +\frac{1}{24}\bigg)  
  +({p'}^2 +p^2) \bigg( \frac{5}{144}g^4  -\frac{5}{72}\bigg)    
+({p'}^2 +p^2) \bigg( \frac{23}{24}g^4 +\frac{1}{12}\bigg) 
\ln{\bigg(\frac{m_{\pi}}{\mu}\bigg)} \\
+R_2(p',p) \bigg(\frac{23}{24}g^4 +\frac{1}{12} \bigg)       \Bigg\}
\end{multline} 
In the case of $\Bar{S}_2(p',p)$ and $\Bar{S}_4(p',p)$, with the exception of $L(q)$ terms, every other term vanishes after the $x$-integration.
\begin{multline}
\Bar{S}_2(p',p)=\big(\vec{\tau_1}\cdot \vec{\tau_2} \big) \int_{-1}^{1} \frac{dx}{2} \bigg(\frac{3x^2-1}{2}\bigg) L(q)    \vec{q}\,^2 \bigg( \frac{23}{24} g^4 + \frac{1}{12} \bigg)
\\= \frac{\big(\vec{\tau_1}\cdot \vec{\tau_2} \big) }{8{p'}^2p^2} \Bigg\{
\Big( 3 R_6 -6({p'}^2+p^2)R_4 + \big(3 ({p'}^2+p^2)^2 -4{p'}^2p^2 \big) R_2 \Big)\bigg( \frac{23}{24} g^4 + \frac{1}{12} \bigg)
    \Bigg\} 
\end{multline}  

\begin{multline}
\Bar{S}_4(p',p)=\big(\vec{\tau_1}\cdot \vec{\tau_2} \big) \int_{-1}^{1} \frac{dx}{2} \bigg(\frac{35x^4 -30x^2+3}{8}\bigg) L(q)   \vec{q}\,^2 \bigg( \frac{23}{24} g^4 + \frac{1}{12} \bigg)
\\= \frac{\big(\vec{\tau_1}\cdot \vec{\tau_2} \big) }{128{p'}^4p^4} \Bigg\{
\bigg[ \Big(35 p^8 +35 {p'}^8 +20 p^6 {p'}^2+18 p^4 {p'}^4 +20 p^2 {p'}^6\Big) R_2 + (210 p^4+300 p^2 {p'}^2  +210 
   {p'}^4 ) R_6  \\- \left(140 p^2 +140{p'}^2\right) R_8  - \left( 140 p^6+180 p^4 {p'}^2+180 p^2 {p'}^4+140  {p'}^6\right) R_4  +35 R_{10} \bigg]\bigg( \frac{23}{24} g^4 + \frac{1}{12} \bigg) \Bigg\}
\end{multline}  
 The $S$-integrals are given by,
\begin{multline}
  S_k(p',p)=  \int_{-1}^{1} \frac{dx}{2} P_k(x)  \big(\vec{\tau_1}\cdot \vec{\tau_2} \big) \vec{q}\,^2  \Bigg\{ 
     \mathcal{R} \bigg[ \frac{23}{48}  g^4 -\frac{5}{24}   g^2    - \frac{1}{24}  \bigg] 
     +  \bigg(\frac{5}{144}   g^4 +\frac{13}{72}g^2 + \frac{5}{72}  \bigg) \\
     + \ln{\bigg(\frac{m_{\pi}}{\mu}\bigg)}   \bigg( \frac{23}{24} g^4 -\frac{5}{12} g^2 - \frac{1}{12} \bigg) 
     + L(q)   \bigg( \frac{23}{24} g^4 - \frac{5}{12} g^2 - \frac{1}{12} \bigg)   \Bigg\}
     \label{eq:BBTint}
\end{multline} 
Starting with $S_0(p',p)$,
\begin{multline}
  S_0(p',p)=\big(\vec{\tau_1}\cdot \vec{\tau_2} \big) \Bigg\{ 
  \mathcal{R}   ({p'}^2 +p^2) \bigg( \frac{23}{24}g^4  -\frac{5}{24} g^2 -\frac{1}{24}\bigg) 
  +({p'}^2 +p^2) \bigg(\frac{5}{144}   g^4 +\frac{13}{72}g^2 + \frac{5}{72}  \bigg) \\   
+({p'}^2 +p^2) \bigg( \frac{23}{24} g^4 -\frac{5}{12} g^2 - \frac{1}{12} \bigg) 
\ln{\bigg(\frac{m_{\pi}}{\mu}\bigg)} 
+R_2 \bigg( \frac{23}{24} g^4 - \frac{5}{12} g^2 - \frac{1}{12} \bigg)  \Bigg\}
\end{multline} 
Similar to the earlier case ($\Bar{S}_2(p',p)$, $\Bar{S}_4(p',p)$), only the $L(q)$ terms contribute in $S_2(p',p)$ and $S_4(p',p)$.
\begin{multline}
S_2(p',p)=\big(\vec{\tau_1}\cdot \vec{\tau_2} \big)\int_{-1}^{1} \frac{dx}{2} \bigg(\frac{3x^2-1}{2}\bigg) L(q)  \vec{q}\,^2 \bigg( \frac{23}{24} g^4 - \frac{5}{12} g^2 - \frac{1}{12} \bigg)
       \\ 
= \frac{\big(\vec{\tau_1}\cdot \vec{\tau_2} \big)}{8{p'}^2p^2} \Bigg\{
\Big( 3 R_6 -6({p'}^2+p^2)R_4 + \big(3 ({p'}^2+p^2)^2 -4{p'}^2p^2 \big) R_2 \Big) \bigg( \frac{23}{24} g^4 - \frac{5}{12} g^2 - \frac{1}{12} \bigg)
    \Bigg\}    
\end{multline}

\begin{multline}
S_4(p',p)=\big(\vec{\tau_1}\cdot \vec{\tau_2} \big)\int_{-1}^{1} \frac{dx}{2} \bigg(\frac{35x^4 -30x^2+3}{8}\bigg) L(q)   \vec{q}\,^2 \bigg( \frac{23}{24} g^4 - \frac{5}{12} g^2 - \frac{1}{12} \bigg)
    \\   
= \frac{\big(\vec{\tau_1}\cdot \vec{\tau_2} \big)}{128{p'}^4p^4} \Bigg\{
\bigg[ \Big(35 p^8 +35 {p'}^8 +20 p^6 {p'}^2+18 p^4 {p'}^4 +20 p^2 {p'}^6\Big) R_2 + (210 p^4+300 p^2 {p'}^2  +210 
   {p'}^4 ) R_6 -  \left(140 p^2 +140{p'}^2\right) R_8 \\- \left( 140 p^6+180 p^4 {p'}^2+180 p^2 {p'}^4+140  {p'}^6\right) R_4  +35 R_{10} \bigg]\bigg( \frac{23}{24} g^4 - \frac{5}{12} g^2 - \frac{1}{12} \bigg) \Bigg\}
\end{multline}

\end{widetext}

\end{document}